%% file: ms_arxiv.tex
\documentclass[10pt,preprint]{aastex}
\newcommand\Teff{\mbox{$T_{\rm eff}$}}

\begin{document}

\shortauthors{Metchev et al.}
\shorttitle{Weather on Other Worlds. II.}

\title{WEATHER ON OTHER WORLDS. II. SURVEY RESULTS: SPOTS ARE UBIQUITOUS ON L AND T DWARFS}

\author{Stanimir A.\ Metchev\altaffilmark{1,2}, Aren Heinze\altaffilmark{2}, D\'{a}niel Apai\altaffilmark{3,4}, Davin Flateau\altaffilmark{4}, Jacqueline Radigan\altaffilmark{5}, Adam Burgasser\altaffilmark{6},
Mark S.~Marley\altaffilmark{7}, \'{E}tienne Artigau\altaffilmark{8}, Peter Plavchan\altaffilmark{9}}
\and
\author{Bertrand Goldman\altaffilmark{10}}

\altaffiltext{1}{The University of Western Ontario, Department of Physics and Astronomy, 1151 Richmond Avenue, London, ON N6A 3K7, Canada, smetchev@uwo.ca}
\altaffiltext{2}{Stony Brook University, Department of Physics and Astronomy, Stony Brook, 100 Nicolls Road, NY 11794--3800, USA}
\altaffiltext{3}{The University of Arizona, Department of Astronomy, 933 N.\ Cherry Avenue, Tucson, AZ 85721, USA}
\altaffiltext{4}{The University of Arizona, Department of Planetary Sciences and Lunar and Planetary Laboratory, 1629 E University Blvd, Tucson, AZ 85721, USA}
\altaffiltext{5}{Space Telescope Science Institute, 3700 San Martin Drive, Baltimore, MD 21218, USA}
\altaffiltext{6}{University of California San Diego, Center for Astrophysics and Space Science, 9500 Gilman Drive, Mail Code 0424, La Jolla, CA 92093, USA}
\altaffiltext{7}{NASA Ames Research Center, MS-245-3, Moffett Field, CA 94035, USA}
\altaffiltext{8}{Universit\'{e} de Montr\'{e}al, D\'{e}partement de Physique and Observatoire du Mont M\'{e}gantic, C.P.\ 6128, Succ.\ Centre-Ville, Montr\'{e}al, QC, H3C 3J7, Canada}
\altaffiltext{9}{Missouri State University, Department of Physics, Astronomy and Materials Science, 901 S.\ National Avenue, Springfield, MO 65897, USA}
\altaffiltext{10}{Max-Planck-Institut f\"{u}r Astronomie, K\"{o}nigstuhl 17, 69117, Heidelberg, Germany}

\begin{abstract}
We present results from the {\it Weather on Other Worlds} {\it Spitzer} Exploration Science program to investigate photometric variability in L and T dwarfs, usually attributed to patchy clouds.  We surveyed 44 L3--T8 dwarfs, spanning a range of $J-K_s$ colors and surface gravities.  We find that 14/23 ($61\%_{-20\%}^{+17\%}$, 95\% confidence) of our single L3--L9.5 dwarfs are variable with peak-to-peak amplitudes between 0.2\% and 1.5\%, and 5/16 ($31\%_{-17\%}^{+25\%}$) of our single T0--T8 dwarfs are variable with amplitudes between 0.8\% and 4.6\%.  After correcting for sensitivity, we find that $80\%_{-27\%}^{+20\%}$ of L dwarfs vary by $\geq$0.2\%, and $36\%_{-17\%}^{+26\%}$ of T dwarfs vary by $\geq$0.4\%.  Given viewing geometry considerations, we conclude that photospheric heterogeneities causing $>$0.2\% 3--5~$\micron$ flux variations are present on virtually all L dwarfs, and probably on most T dwarfs.  A third of L dwarf variables show irregular light curves, indicating that L dwarfs may have multiple spots that evolve over a single rotation.  Also, approximately a third of the periodicities are on time scales $>$10~h, suggesting that slowly-rotating brown dwarfs may be common.  We observe an increase in the maximum amplitudes over the entire spectral type range, revealing a potential for greater temperature contrasts in T dwarfs than in L dwarfs.  We find a tentative association (92\% confidence) between low surface gravity and high-amplitude variability among L3--L5.5 dwarfs. Although we can not confirm whether lower gravity is also correlated with a higher incidence of variables, the result is promising for the characterization of directly imaged young extrasolar planets through variability.

\end{abstract}

\keywords{brown dwarfs -- stars: low-mass -- stars: rotation -- stars: starspots -- stars: variables: general -- techniques: photometric}

\section{INTRODUCTION}

\subsection{Clouds in L- and T-type Atmospheres: Theoretical and Observational Perspectives}

The atmospheres of ultra-cool ($>$M7) dwarfs are distinct from those of warmer stars because their effective temperatures span the condensation points of various chemical compounds.  Their spectral energy distributions (SEDs) are affected by condensate opacities arising from a multi-layer structure of cloud decks with distinct compositions.  At high temperatures and pressures these include refractory compounds (e.g., oxides, silicates, etc; collectively referred to as ``dust''), while at lower temperatures and pressures the condensates consist of more volatile compounds (e.g., alkali salts, water; e.g., \citealt{fegley_lodders96, ackerman_marley01, morley_etal12}).   

The framework of cloudy models has been very successful at describing L and T dwarf atmospheres.  While prior dusty and dust-free models were able to reproduce some of the gross characteristics of early L-type and late T-type atmospheres, respectively \citep[e.g., the DUSTY and COND Phoenix models;][]{allard_etal01}, the more nuanced picture of photospheric condensate cloud formation \citep{ackerman_marley01, cooper_etal03, helling_etal06, allard_etal12} provides a more accurate representation of the spectra, colors, and chemical compositions across the L and T dwarf sequence \citep{cushing_etal08, stephens_etal09}.  The cloud model has also been able to explain the re-appearance of photospheric condensate opacity in the form of salt (KCl) clouds in the coldest T and early Y dwarfs \citep{burgasser_etal10, morley_etal12}.  A cloud or sedimentation prescription is now considered a fundamental parameter in characterizing brown dwarf atmospheres, along with effective temperature, surface gravity, metallicity, and vertical mixing.  

Despite these gains, 1D cloudy model atmospheres do not encapsulate the complexity of cloud structures observed on Solar System giant planets, which are dominated by bands, jets, spots, and storms.  \citet{ackerman_marley01} proposed that such heterogeneous cloud structures may be present in brown dwarf atmospheres, leading to rotationally modulated flux variations and explaining many of the unusual characteristics of the L dwarf/T dwarf transition \citep{burgasser_etal02, marley_etal10}.  Thus, \citet{gelino_marley00} inferred that rotationally-induced modulations in Jupiter's light curve caused by its bright, clear equatorial ``5~$\micron$ hot spots'' would create as much as 20\% peak-to-peak variability at 
4.78~$\micron$.  The patchy cloud paradigm postulates similar heterogeneous atmospheric structures on L and T dwarfs, most prominently at the L-to-T transition, where dust clouds break up as they rain out of the visible atmosphere.  Evidence in support of cloud break-up is independently provided by detections of $J$-band flux reversals in double brown dwarfs with component spectral types near the L/T transition \citep{burgasser_etal06, burgasser_etal13, liu_etal06, looper_etal08b}.  The cooler, early- to mid-T component in such binaries is brighter at $J$ band than the warmer late-L to early-T component because the relative lack of molecular opacity at $J$ band reveals much deeper, hotter layers in a dust-free atmosphere.  The presence of a patchy cloud cover on L and at least early T dwarfs now forms an integral part of our understanding of ultra-cool atmospheres, as dramatically revealed by \citet{crossfield_etal14} through Doppler imaging of one of the two nearest brown dwarfs, Luhman 16B \citep[a.k.a., WISE J104915.57--531906.1B][]{luhman13}.

Other mechanisms for surface brightness heterogeneities are also possible.  Notably,  M and early-L dwarfs are known to have elevated magnetospheric activity \citep{schmidt_etal07, hallinan_etal08, berger_etal10, west_etal11}.  The coupling of magnetic fields with the atmosphere could result in either hot or cold spots, and may be difficult to distinguish from cloud structures with similar temperature differentials.  There is evidence from radio observations that flaring activity extends even into the T dwarfs \citep{route_wolszczan12}.  However, in general the neutral atmospheres of L and T dwarfs are considered too electrically resistive to support magnetic starspots \citep{mohanty_etal02, gelino_etal02, chabrier_kuker06}.

More recently, temperature fluctuations driven by deep atmospheric instabilities \citep{robinson_marley14} or by  jet or eddy circulation in the stratified layer above the convective zone \citep{zhang_showman14} have also been invoked as sources of brightness heterogeneities on brown dwarfs.  The former process may be important in late T dwarfs, where the radiative time scale is long enough so that the temperature fluctuations dissipate on time scales longer than a minute.  The \citeauthor{zhang_showman14} scenario might offer an alternative to patchy clouds, even if it may  also drive the creation of cloud heterogeneities. Still, \citet{radigan_etal12} and \citet{apai_etal13} show that temperature fluctuations alone can not account for the observed color and spectral variations of brown dwarfs near the L/T transition.



Like patchy clouds, these other mechanisms are also expected to produce rotationally modulated brightness variations.  Given the presence of condensates in ultra-cool atmospheres, clouds likely contribute to the observed variations regardless of the underlying mechanism that generates them.   

\subsection{Detecting Patchy Clouds through Periodic Variability}

Numerous attempts to detect periodic flux variations in L and T dwarfs have been made over the past 14 years.  The majority of these have been ground-based, with detection thresholds of 10--100 milli-magnitudes (mmag) on time scales of tens of minutes to weeks (e.g., \citealt{bailer-jones_mundt01, gelino_etal02, clarke_etal02, clarke_etal08, enoch_etal03, koen_etal05, goldman05, khandrika_etal13, girardin_etal13, wilson_etal14}).  The detections or evidence for periodicity have often been marginal, and in some cases unconfirmed in subsequent observations \citep[e.g.,][]{clarke_etal03, clarke_etal08, koen04, koen13a, radigan14}.  A notable exception early-on was the detection of a 1.8~h photometric period in the L2 dwarf Kelu-1 with a 1.1\% peak-to-peak amplitude at 860~nm \citep{clarke_etal02}.  While subsequent $I$-band observations did not confirm the periodicity \citep{clarke_etal03}, the same period did recur in the H$\alpha$ line intensity \citep{clarke_etal03} and in subsequent $g^\prime$ observations \citep{littlefair_etal06}.

A very careful discussion of the caveats involved in the photometric monitoring for low-amplitude variability of faint targets is presented in \citet{koen13a}, who stresses the need for an accurate understanding of seeing fluctuations.  \citeauthor{koen13a}'s summary analysis of a decade-long optical monitoring campaign of 125 ultracool dwarfs finds evidence for variability in 19 objects, for an overall variability fraction of 15\%.  The majority (17) of the variables have spectral types $\leq$L5, although two T dwarfs also show variations in mean $I$-band flux from one observing run to the next.

Cooler brown dwarfs---with spectral types $>$L5 ($\Teff\lesssim1500$~K)---might be expected to show greater variability because of the greater abundance of condensates in their atmospheres, and because of the anticipated peak in the silicate cloud disruption rate at the L-to-T spectral type transition.  However, cooler brown dwarfs are also fainter, have stronger intrinsic water absorption, and so pose greater challenges for precision photometry from the ground. 

\citet{artigau_etal09} were the first to detect highly significant, periodic, and repeatable variations in a $>$L5 dwarf: the T2.5 dwarf SIMP J013656.5+093347 ($\Delta J=50$~mmag, $P=2.4$ hr; later confirmed in \citealt{apai_etal13} and \citealt{metchev_etal13}). These were strongly suggestive of patchiness in the cloud cover, and the differing $J$- and $K_{\rm s}$-band variability amplitudes offered a new window into the vertical structure of a brown dwarf atmosphere.  More recently, \citet[][henceforth, RLJ14]{radigan_etal14} completed the most comprehensive and sensitive ground-based variability survey of L and T dwarfs, detecting highly significant periodic $J$-band modulations in nine out of 57 objects (16\% variability fraction).  The success of the \citeauthor{artigau_etal09} and RLJ14 campaigns was a direct result of the intensive monitoring of individual objects over entire nights. The RLJ14 campaign reveals that $J$-band variability is enhanced both in frequency and amplitude at the L-to-T spectral type transition, as expected from the disruption of silicate clouds.

Space-based photometric monitoring programs do not face the same difficulties as ground-based programs, and can attain much higher precision.  In a pilot variability study with the {\it Spitzer Space Telescope}, \citet{morales_calderon_etal06} detected tentative [4.5]-band variations in two of their three L dwarf targets.  However, non-confirmations of the variations in the IRAC [8.0] band prevented them from ruling out instrumental effects in the light curves.  More recently, \citet[][hereafter: B14]{buenzli_etal14} conducted a 22-target 1.1--1.7~$\micron$ grism spectroscopy survey with the {\it Hubble Space Telescope (HST)}, and found convincing 
evidence for variations in at least six (27\%) brown dwarfs.
B14's results demonstrate that detectable variability exists beyond the L/T transition, and that in fact low-level heterogeneities may be a frequent characteristic of L and T type atmospheres.  

The evidence for variability across the L and T domains finds further support in the recent ground-based work of \citet{khandrika_etal13} and \citet{wilson_etal14}.  However, the shorter monitoring periods (less than one rotation) in these studies and the poorer photometric precision leave open the possibility that the variations may be correlated with time-variable telluric water absorption \citep[e.g.,][]{artigau06} or seeing \citep{koen13a}.  Thus, a re-analysis of the \citet{wilson_etal14} data by \citet[][henceforth, R14]{radigan14} confirms only three of the claimed 11 new variables.  Henceforth, we rely on the set of RLJ14, R14, and B14 analyses as references for the near-IR (1.1--1.7~$\micron$) variability properties of L and T dwarfs.

\subsection{{\it Weather on Other Worlds:} a {\it Spitzer} Exploration Science Program} 

The ensemble of empirical evidence to date indicates that variability is not unusual in L and T dwarfs, with overall variability frequencies between 16\%--27\% at optical and near-IR wavelengths.  Clearly, these detections are subject to significant incompleteness, either because of the relatively poorer photometric sensitivity of ground-based observations, or because of the generally shorter monitoring periods in past ground- or space-based observations.  In all likelihood, low-amplitude and/or long-period variables are missing from the existing surveys.

The {\it Spitzer Warm Mission} offers an opportunity to study brown dwarf variability at high precision, high cadence, and over unprecedentedly long uninterrupted intervals.  It also offers a distinct set of wavelengths from those employed in ground-based and {\it HST} surveys to date, which give complementary information on the atmospheres and cloud structures of brown dwarfs.

We carried out a comprehensive brown dwarf precision monitoring campaign with {\it Spitzer} as an Exploration Science Program (GO 80179, PI: S.\ Metchev).  The principal goal of the {\it Weather on Other Worlds} program was to trace the emergence and then decline of the cloud disruption phenomenon at the transition between dusty L-type and dust-free T-type atmospheres.  A secondary goal was to trace the dependence of cloud disruption on surface gravity (a proxy for youth) and $J-K_s$ color (a proxy for dustiness) for insights into the cloud structures of young and dusty directly imaged extrasolar planets.

First results from our {\it Spitzer} campaign, including a detection of the lowest amplitude ultra-cool variable reported to date (DENIS-P~J1058.7--1548 [L3]) and a description of our data analysis methods, were reported in \citet{heinze_etal13}.  Here we present the results from the entire survey.  Future publications will address in detail aspects of the survey, including L dwarf variability, irregular and long-term variability, results from observations over a broader, 0.7--5~$\micron$ wavelength range, and theoretical interpretations of the global variability trends in the context of atmospheric phenomena.

In this paper we first present the sample (Sec.~\ref{sec_sample}), observing strategy (Sec.~\ref{sec_observations}), and data analysis methods (Sec.~\ref{sec_data_reduction}) for the {\it Weather on Other Worlds} campaign.  We then present results on the variability frequency, amplitudes, and periods of L3--T8 dwarfs (Sec.~\ref{sec_variability}).
We discuss the prevalence of spots on L and T dwarfs and draw comparisons to previous surveys at complementary wavelengths (Sec.~\ref{sec_spots}).  The principal findings are summarized in Section~\ref{sec_conclusion}. 

\section{SAMPLE SELECTION
\label{sec_sample}}


Our survey sample was designed to test for the presence of photometric variations across a representative range of effective temperatures, surface gravities, and atmospheric dust content.  Our targets were selected to be bright, with IRAC channel 1 (3.6~$\micron$, [3.6]) or channel 2 (4.5~$\micron$, [4.5]) magnitudes brighter than 14.5~mag, to optimize sensitivity to small-amplitude variations.  

We included 44 targets---25 L dwarfs and 19 T dwarfs---spanning the L3--T8 spectral type range, including dwarfs with blue, median, or red $J-K_s$ colors at similar spectral subtypes 
(Table~\ref{tab_sample}).  
Objects earlier than L3 were excluded to avoid contamination with magnetospheric activity, common among earlier-type dwarfs \citep{schmidt_etal07, berger_etal10, west_etal11}.  Nevertheless, we incorporated a control object for recognizing potential activity-induced photometric effects: the radio emitting L3.5 dwarf 2MASSW J0036159+182110 \citep{berger02} which is also a known irregular optical variable \citep{maiti07, lane_etal07, koen13a}.  As this object was added to our program deliberately because of its known variability and radio emission, we do not include it in our statistical considerations.

Seven of our targets were chosen because they showed observational evidence for low or moderately low surface gravities, suggesting ages less than 500~Myr.  Six of these were in the L3--L5 spectral type range, where they could be compared to an approximately equal number of targets with higher gravities.  The deliberate inclusion of low-gravity objects biases our sample of L3--L5 dwarfs, although some would have been included anyway on account of their very red $J-K_s$ colors.  The influence of low surface gravity on variability is discussed in Section~\ref{sec_low_gravity}.

After the sample selection was complete, four of our targets turned out to be resolved $<$0$\farcs5$ binaries.
As we can not separate these with the {\it Spitzer} point-spread function (PSF), we exclude them from our statistical analysis, although we do present results on them for completeness.  Known and additional suspected binaries are  discussed in Section~\ref{sec_binaries}.  

Henceforth, whenever we consider the statistical properties of L3--T8 dwarf variability, we limit our analysis to the sample of 39 objects---23 L dwarfs and 16 T dwarfs---none of which have been spatially resolved into tight binaries, and none of which were a priori known to be magnetically active.  We will refer to this sample as the ``unresolved sample.''  


\input{table1.tex}

Our entire sample was selected blindly with regard to previously detected variability.  After the sample was finalized, two of the objects were recognized as known variables:  2MASS J11263991--5003550 (L4.5; RLJ14) and 2MASS J22282889--4310262 \citep[T6;][]{clarke_etal08, buenzli_etal12}.  More recently, 2MASS~J08251968+2115521 (L7.5) was also identified as a variable in B14. 

\subsection{Known and Candidate Binaries
\label{sec_binaries}}

Unresolved binaries can display unusual properties for their composite spectral types that are not necessarily representative of isolated single objects.  Therefore, care needs to be taken to treat these properly in statistical studies.  There are four known $<$0$\farcs$5 binaries in our sample that are unresolved by the {\it Spitzer} PSF (FWHM = $1\farcs7$), all noted in Table~\ref{tab_sample}.  
Lacking accurate spectroscopic and photometric information for some of these, we have chosen to exclude all resolved tight binaries from our statistical analysis.


There are two additional candidate binaries,
both identified in the literature through the spectral decomposition technique \citep{burgasser07}.  The method entails fitting the spectrum of an unresolved L or T dwarf with combinations of L + T dwarf spectroscopic templates to test whether the target may be an unresolved binary with components of disparate spectral types.  Some candidate spectral binaries identified in this manner have subsequently been separated with high angular resolution observations.  Examples include two of the tight binaries included in our sample: SDSS J205235.31--160929.8 \citep[T1 composite;][]{chiu_etal06} and SDSS J151114.66+060742.9 \citep[T2 composite;][]{albert_etal11}, both identified as strong spectral binary candidates by \citet{burgasser_etal10b}, and both subsequently resolved with laser guide-star (LGS) adaptive optics (AO) imaging \citep{stumpf_etal11, bardalez_gagliuffi_etal15}.  For SDSS J205235.31--160929.8 we adopt the T1 + T2.5 component spectral types determined photometrically by \citet{stumpf_etal11}.  For SDSS J151114.66+060742.9 we adopt the tentative L5.5 + T5 component spectral decomposition of \citet{burgasser_etal10b}, with a more accurate spectroscopic characterization of the system expected in \citet{bardalez_gagliuffi_etal15}.

While candidate binaries identified through spectral decomposition have often been separated in multiple components, should they remain unresolved in high-angular resolution observations, they are potential candidates for variability.  Rather than representing two distinct brown dwarf components, the composite spectrum may instead be revealing the two-temperature nature of the photosphere of a single object: e.g., through a combination of regions with thick clouds and regions with thin clouds \citep{apai_etal13}.  Such is the case of the T1.5 dwarf 2MASS J21392676+0220226 \citep{burgasser_etal06b}, suggested as a strong L8.5 + T3.5 spectral binary candidate by \citet{burgasser_etal10b}, but identified as a $J$-band variable \citep{radigan_etal12} that is unresolved in {\it HST} images \citep{apai_etal13}, and exhibits no radial velocity variations \citep{khandrika_etal13}.

We gauge whether the two remaining candidate spectral binaries in our sample may contain multiple components by checking archival high-angular resolution observations from the {\it HST}\footnote{https://archive.stsci.edu/hst/}, NASA Keck\footnote{http://www2.keck.hawaii.edu/koa/public/koa.php}, and ESO VLT\footnote{http://archive.eso.org} archives.  Archival {\it HST}/WFC3 observations exist for both: 2MASS J09490860--1545485 \citep[T2, a weak T1 + T2 candidate;][]{burgasser_etal10b} and 2MASS J13243559+6358284 (T2.5, a strong L8 + T3.5 candidate; \citealt{burgasser_etal10b}; also an L9 + T2 candidate from \citealt{geissler_etal11}).
The {\it HST} images do not resolve the candidate spectral binaries down to $0\farcs10$.
Additional Keck LGS AO observations exist for 2MASS J13243559+6358284, although because of sub-optimal AO correction, the angular resolution is not better than in the {\it HST}/WFC3 images.  

Since neither of the above two candidate spectral binaries are resolved down to $0\farcs10$, we treat them as single objects in our analysis, although radial velocity monitoring would be needed to establish this with confidence \citep[e.g.,][]{burgasser_etal08b}.  Notably, we find that 
the strong binary candidate 2MASS J13243559+6358284 is one of our highest-amplitude variables (Sections~\ref{sec_data_reduction}--\ref{sec_variability}, Table~\ref{tab_results}).  That is, it may parallel the case of the candidate spectral binary-turned-variable 2MASS J21392676+0220226.


Finally, our sample includes as two separate targets the individual components of the known 9$\farcs$4 binary SDSS J141624.08+134826.7 / ULAS J141623.94+134836.3 \citep{scholz10a}.  The components, a blue L6 and a blue T7.5 dwarf, are sufficiently well separated that both can be measured accurately and simultaneously.  The majority of the remaining targets have been observed at high-angular resolution, with LGS AO, or with the {\it HST}, and remain unresolved.  We presume that all of these brown dwarfs are single.




\subsection{Low Surface Gravity Objects
\label{sec_low_gravity}}


We included a sequence of six L3--L5.5 dwarfs with signatures of low surface gravity (i.e., youth) 
and the moderately young ($\sim$500~Myr) T2.5 dwarf HN~PegB \citep{luhman_etal07b} to further investigate the dependence of cloud structure on surface gravity.  One of the low-gravity dwarfs, SDSSp J224953.45+004404.2 (L3$\beta$), is a tight binary, for a total of eight individual low-gravity dwarfs.

The youth of three of these (2MASS J04210718-6306022 [L5$\beta$], 2MASSI J1726000+153819 [L3$\beta$], and 2MASSW J2208136+292121 [L3$\gamma$]) is discussed in \citet{cruz_etal09} and \citet{gagne_etal14}.  \citet{cruz_etal09} spectroscopically classify them as $\beta$- or $\gamma$-type low-gravity objects.  We discuss the remaining low-gravity L3--L5.5 dwarfs below.

2MASS J16154255+4953211 (L4$\beta$) is identified as a possible low-gravity L4 dwarf from optical spectra by \citet{cruz_etal07} and \citet{gagne_etal14}.  Low-gravity features in the near-IR spectrum---weak \ion{K}{1} lines and weak metal-hydride absorption---are also noted by \citet{kirkpatrick_etal08}, who estimate a tentative age of 100~Myr.  \citet{geissler_etal11} further note the similarity of the 0.8--2.4~$\micron$ low resolution spectrum of this object to the known 20--300~Myr-old L3 dwarf G~196--3B \citep{rebolo_etal98}, although tentatively assign it an L6 spectral type based on the 0.8--1.2~$\micron$ continuum. To maintain consistency with the optical spectral type classification for $<$L9 dwarfs, we adopt the \citet{cruz_etal07} L4 spectral type, and $\beta$-class gravity as for other $\sim$100~Myr-old L0--L5 dwarfs (including G~196--3B) in \citet{cruz_etal09}.

2MASS J18212815+1414010 (L4.5) is identified as a peculiarly red L4.5 dwarf by \citet{looper_etal08}.  That study notes a number of spectroscopic features indicating moderately low gravity, including relative weakness in the alkaline and FeH strengths and sharpness of the $H$-band continuum.  \citet{gagne_etal14} conclude that despite its signatures of youth this object does not belong to any known young moving group.   \citet{looper_etal08} mention unusually high atmospheric dust content as an alternate explanation for these traits, although that may also be the result from low surface gravity.  Notably, low gravity likely does not account for similar characteristics observed in another unusually red L dwarf studied by \citet{looper_etal08} in parallel: 2MASS J21481633+4003594 (L6).  \citet{looper_etal08} point to the latter object's high galactic tangential velocity ($v_{\tan} \sim62$~km~s$^{-1}$) as evidence against youth, while noting that the tangential velocity of 2MASS J18212815+1414010 is much lower ($v_{\tan} \sim10$~km~s$^{-1}$), and so fully consistent with youth.  Therefore, we tentatively adopt the low-gravity hypothesis for this object.  2MASS J18212815+1414010 may be somewhat older than the $\sim$100~Myr $\beta$-class objects of \citet{cruz_etal09}, hence we do not assign a gravity class.

SDSSp J224953.45+004404.2 (L3$\beta$) is resolved into a pair of $0\farcs32$ L3 and L5 dwarfs using Keck LGS AO by \citet{allers_etal10}.  They note that the L3 component shows low-gravity features (weak alkali and FeH absorption, strong VO absorption) similar to G~196--3B (L3$\beta$), and conclude that both components are young.  We adopt $\beta$-class gravity, as for G~196--3B, even if \citet{gagne_etal14} do not find an association with any known young moving group.  

It is possible that not all of the above putative low-gravity objects are actually young: as deliberated for 2MASS J18212815+1414010 (L4.5) in \citet{looper_etal08}.  Nonetheless, they all share characteristics linked to low surface gravities in ultra-cool dwarfs: weak alkali lines and metal-hydride bands, enhanced VO absorption, red 1--2.5~$\micron$ continua, and peaked $H$-band spectra.  These are more generally associated with enhanced dust content \citep{looper_etal08, kirkpatrick_etal10}, which can be the result of low surface gravity.  While we refer to this set of objects in the present analysis as being low-gravity, we are likely selecting for a more general dependence on atmospheric dustiness.

\section{OBSERVATIONS
\label{sec_observations}}

We observed all objects in staring mode with {\it Spitzer} in IRAC channels 1 and 2  for a total of 891~h.  The default observing sequence was a 14~h astronomical observing request (AOR) in channel 1, immediately followed by a 7~h AOR in channel 2.  The combined sequence was intended to detect periods up to $\sim$10~h in channel 1, and to then measure the [4.5]/[3.6] variability amplitude ratio as a probe of the temperature gradient among heterogeneous cloud layers.  

For eight of the objects, including the radio-emiting 2MASSW~J0036159+182110, $v \sin i$ measurements from high-dispersion spectroscopy were available from the literature.  We used these to set upper limits on the expected rotation periods, assuming radii equal to Jupiter's.  The maximum rotation periods for these were between 3--6 hrs, and we correspondingly planned shorter---twice the maximum period---AORs in channel 1.  
The AOR durations for each target are included in Table~\ref{tab_sample}.  All exposures were 12~sec long, taken in full-array readout mode.

As the execution of the program commenced in the second half of 2011, the {\it Spitzer} Science Center (SSC) was implementing a novel acquisition peak-up scheme to improve the stability of the telescope pointing over long staring observations.  Previous experience had shown that the telescope can take up to 30--45~min after target acquisition to stabilize its pointing within the boundaries of a $1\farcs2$ pixel.  Pointing that is stable to a fraction of a pixel is necessary to avoid systematic errors arising from variations in sensitivities among pixels or within individual pixels---the latter known as the ``pixel phase effect'' \citep{reach_etal05}.  On the advice of the SSC, we added a 30~min channel 1 ``acquisition'' AOR to the beginning of our staring sequence on each target.  Thirty-four of our targets were observed with the extra 30~min AOR for acquisition.

We further experimented with acquiring our targets on the well-characterized IRAC channel 1 ``sweet spot.''  As of early 2012, the sweet spot was a region approximately one-third of a pixel in area with very well characterized pixel phase: the result of extensive calibration by the SSC.  However, we found that the relatively large uncertainties in the proper motions of our targets, determined at the time exclusively from ground-based parallax programs \citep{tinney_etal03b, vrba_etal04, faherty_etal09}, prevented us from obtaining sufficiently accurate positions to ensure placement on the sweet spot.  Having attempted this mostly unsuccessfully for several targets, we abandoned the approach, and instead opted for positioning near the center of the detector.  A central location allowed better overlap of comparison stars with concurrent ground-based monitoring programs of the same targets.

\begin{figure}
\epsscale{0.7}
\plotone{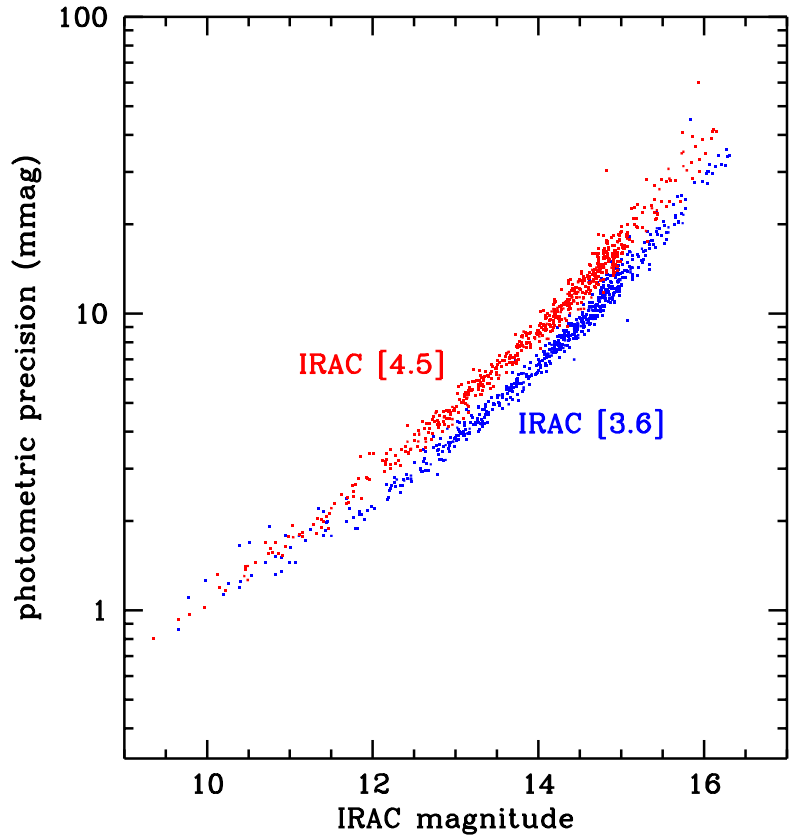}
\figcaption{\footnotesize Photometric precision attained on point sources in the {\it Weather on Other Worlds} program as a function of object brightness at IRAC [3.6]  and [4.5].  
Photometric apertures are optimized individually for each object, with brighter objects generally having larger optimal apertures.   Photometry is binned in 10-point bins, corresponding to a sampling interval of 120 seconds.
\label{fig_photometric_precision}}
\end{figure}

\section{DATA REDUCTION AND VARIABILITY ANALYSIS
\label{sec_data_reduction}}

Our data reduction and analysis approach is presented in detail in the first announcement of results from the {\it Weather on Other Worlds} program \citep{heinze_etal13}.  Here we summarize the steps only briefly, and discuss areas where our analysis has been updated.

\subsection{Photometry and Identification of Variables}
\label{sec_variable_identification}

We use aperture photometry with radii optimized to deliver the lowest RMS scatter in the measured fluxes.  We average down random noise by binning the photometry in 10-image bins, which yields a sampling interval of about 120~s and retains sensitivity to variations on the timescales of interest ($\gtrsim$0.5~h).   We correct for the pixel phase effect by fitting the measured flux of each source as a 2-D quadratic function of position on the detector.

Our limiting precision, determined as the standard deviation of the binned data after the removal of the pixel phase fit, for over 600 stars identified as non-variable in our survey images is shown as a function of magnitude in Figure~\ref{fig_photometric_precision}.   We attained $\approx$20\% better photometric precision at [3.6] than at [4.5] for the same nominal magnitudes.  Consequently, our L dwarf light curves had higher SNR at [3.6].  However, the $[3.6]-[4.5]$ colors for most T dwarfs are sufficiently red that their photometry in the [4.5] band had similar or better precision than at [3.6].

We identify variable sources by creating Lomb-Scargle periodograms of the pixel-phase-corrected data sets 
using a routine from \citet{press_etal92}, in which the periodogram power at angular frequency $\omega$
is defined as:

\begin{equation}
P(\omega) \equiv \frac{1}{2 \sigma^2} \left\{\frac{\left[\sum_j (h_j - \overline{h}) \cos \omega (t_j-\tau) \right]^2}{\sum_j  \cos^2 \omega (t_j-\tau)} + \frac{\left[\sum_j (h_j - \overline{h}) \sin \omega (t_j-\tau) \right]^2}{\sum_j  \sin^2 \omega (t_j-\tau)} \right\}
\label{eq:periodogram01}
\end{equation}

where $\overline{h}$ and $\sigma^2$ are the mean and variance of the data points $h_j$ taken at times $t_j$, and $\tau$ is defined by the relation:

\begin{equation}
\tan(2 \omega \tau) = \frac{\sum_j \sin 2 \omega t_j}{\sum_j \cos 2 \omega t_j}.
\label{eq:periodogram02}
\end{equation}

The range and sampling of periods probed by the periodogram is determined by the oversampling factors in the frequency and time domains (parameters {\sc ofac} and {\sc hifac} in the \citealt{press_etal92} routine).  We set the oversampling factors in the frequency and time domains to 200.0 and 0.2, respectively, which for a typical [3.6] data set results in the investigation of about 7500 distinct periods ranging from 0.36~h to $>$100~h, with a sampling interval that is constant in frequency and has a value of 0.00036 cycles~h$^{-1}$. The false alarm probability (FAP) is $M e^{-P}$, where $P$ is the periodogram power of the highest peak, and $M$ is the number of independent frequencies: equal to the number of data points multiplied by the time-domain oversampling factor {\sc hifac}.  With ten-point binned data, typically $M \sim 360\times0.2 = 72$ at [3.6] and half that value at [4.5].

A similar approach was developed independently and presented in RLJ14.  We find that while a periodogram is most sensitive to sinusoidal signals (that is, they generate the lowest FAP at a given amplitude), it remains a useful means of detecting non-sinusoidal and even non-periodic signals, including linear trends.  Additionally, we determine that although our preliminary method of correcting the pixel phase effect can distort an astrophysical signal, it is extremely unlikely to suppress its periodogram power completely or to prevent the detection of a true variable.

\begin{figure}[ht]
\epsscale{1.0}
\plottwo{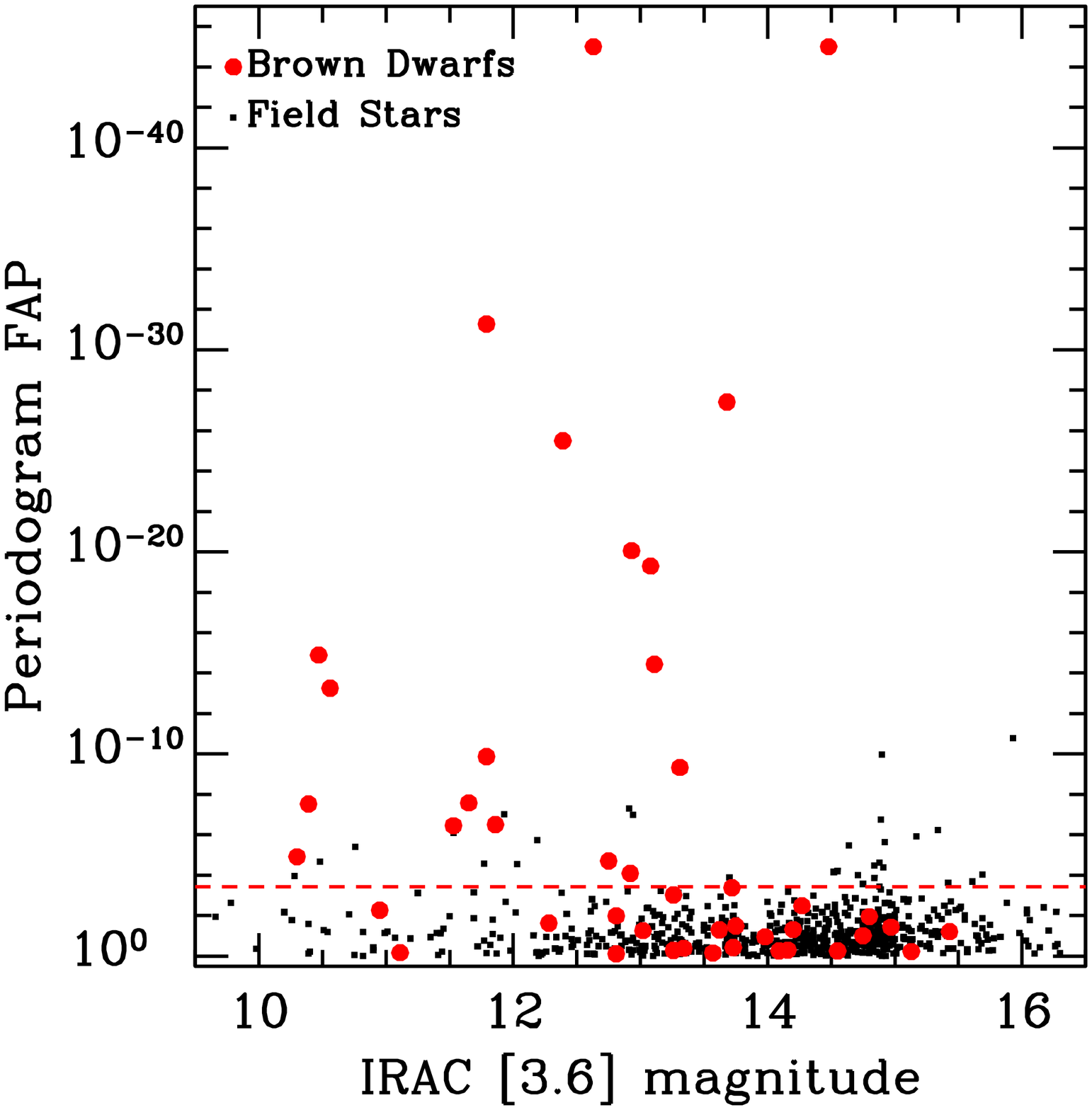}{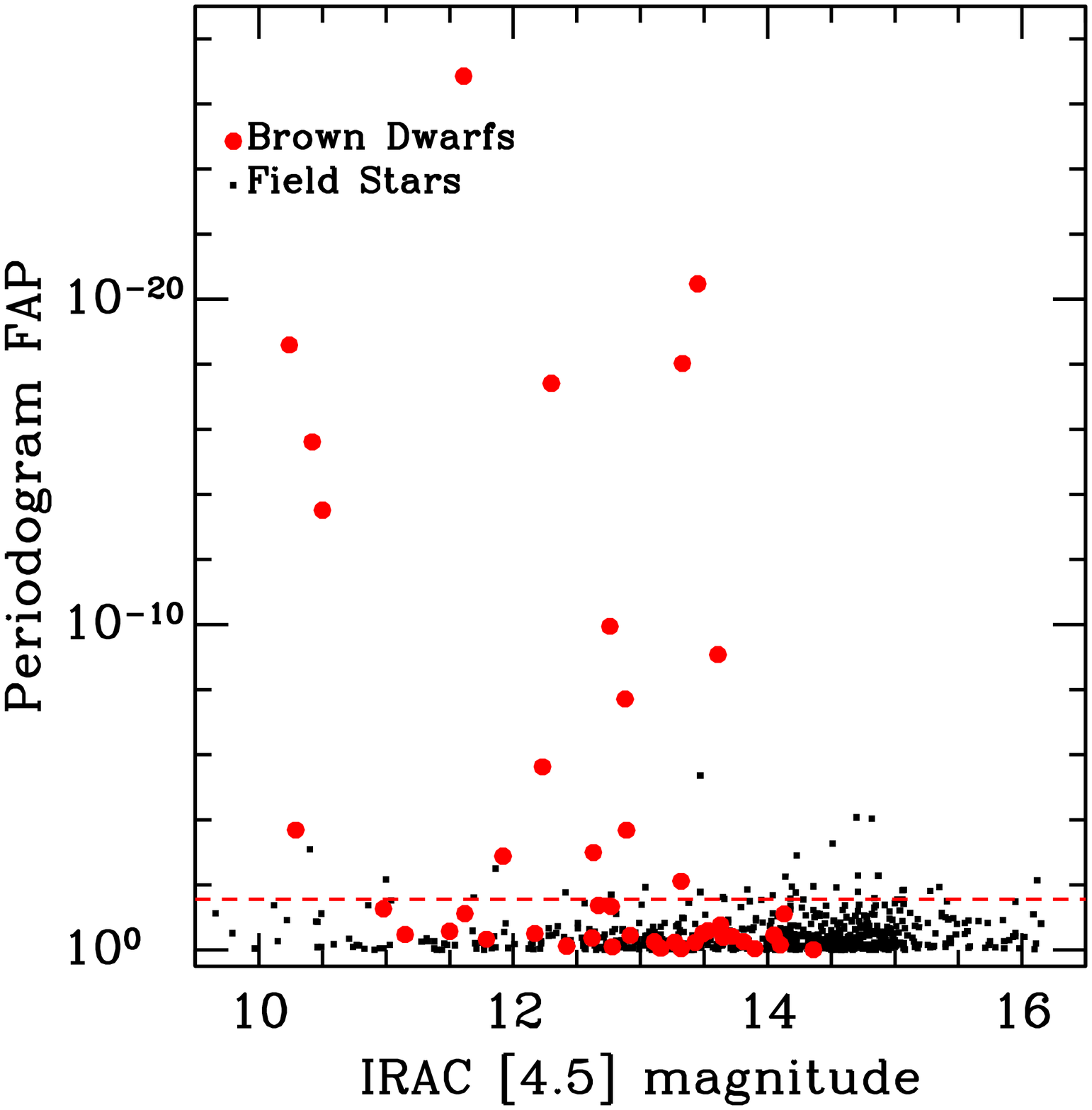}
\figcaption{\footnotesize Periodogram FAPs for the observed L and T dwarfs (large dots) vs.\ comparison stars (small dots) within the same IRAC fields at [3.6] ({\it left}) and [4.5] ({\it right}).  There are 19 probable variables at [3.6], and 16 at [4.5]: all displaying variability greater than that of 95\% of the comparison stars, maked by the horizontal dashed line.  The relative dearth of low-FAP variables among the $\gtrsim$13.5~mag survey targets is likely the result of poorer photometric precision.
\label{fig_fap}}
\end{figure}

The FAP of the strongest peak in the periodogram measures the likelihood that any apparent coherent variations are caused by random noise.  Rather than uncritically accepting all sources with FAP $<$5\% as variables with 95\% confidence, we have performed periodogram analyses on a large number of comparison stars in the fields of our targets to arrive at a robust understanding both of the statistics of IRAC photometry and of the performance of our periodogram-based method for identifying variables.  Excluding obvious variables (e.g., eclipsing binaries and RR Lyrae stars) filtered out by eye, we have [3.6] photometry for 636 field stars and [4.5] photometry for 652 stars, with magnitudes in the same range as our brown dwarfs.  Among these stars, the 5th percentile in the FAP value of the strongest peak in each periodogram is $3.7 \times 10^{-4} = 10^{-3.4}$ at [3.6] and $2.9 \times 10^{-2} = 10^{-1.5}$ at [4.5] (Fig.~\ref{fig_fap}). The fact that these values are both smaller than $5 \times 10^{-2}$ indicates that some of the stars have non-random variations, which may be caused by either low-level astrophysical variability or residual IRAC systematics.  The lower 5th percentile FAP value for [3.6] likely reflects the longer monitoring interval, which produces greater sensitivity to variations regardless of origin. 

\begin{figure}
\plottwo{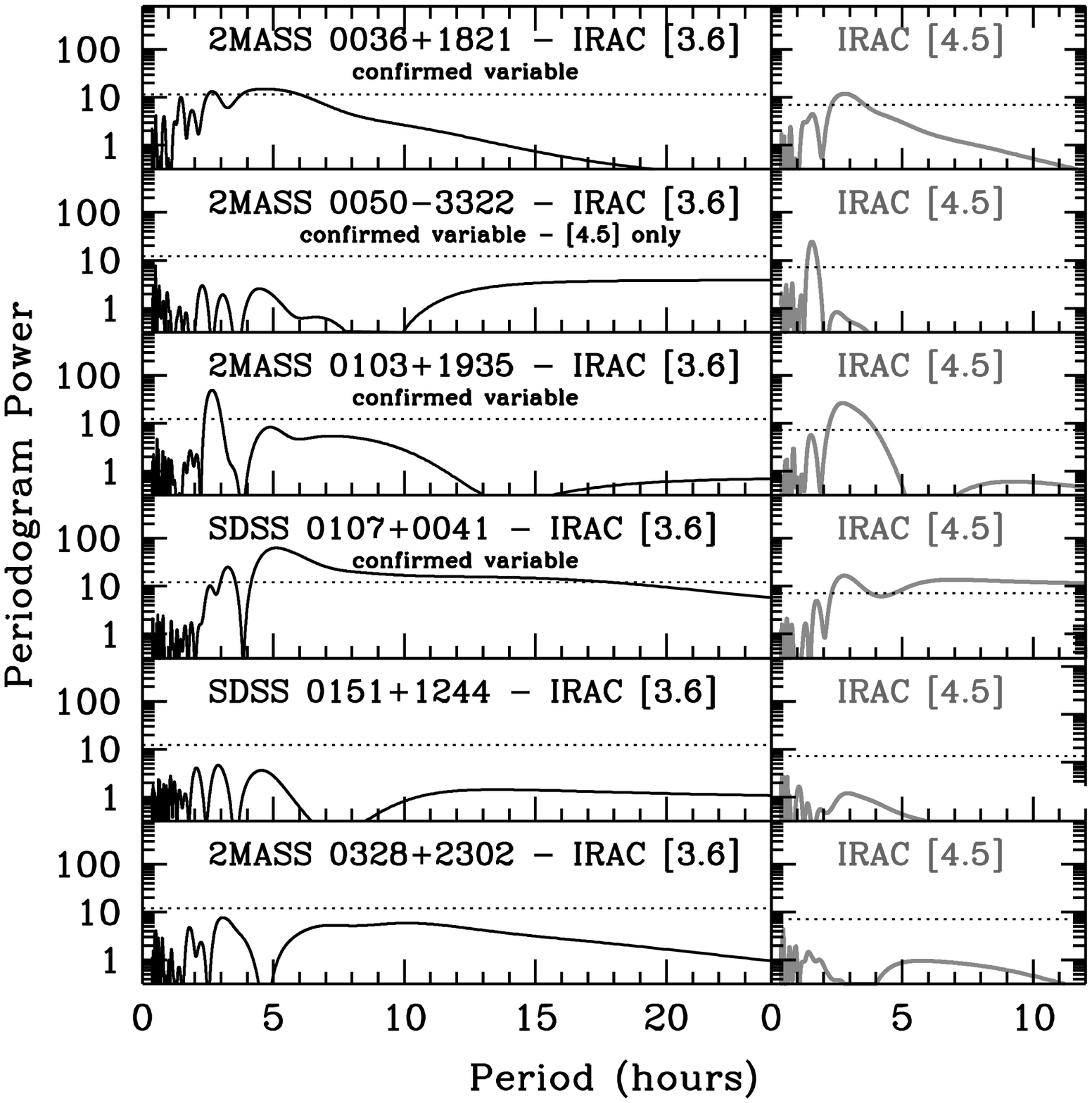}{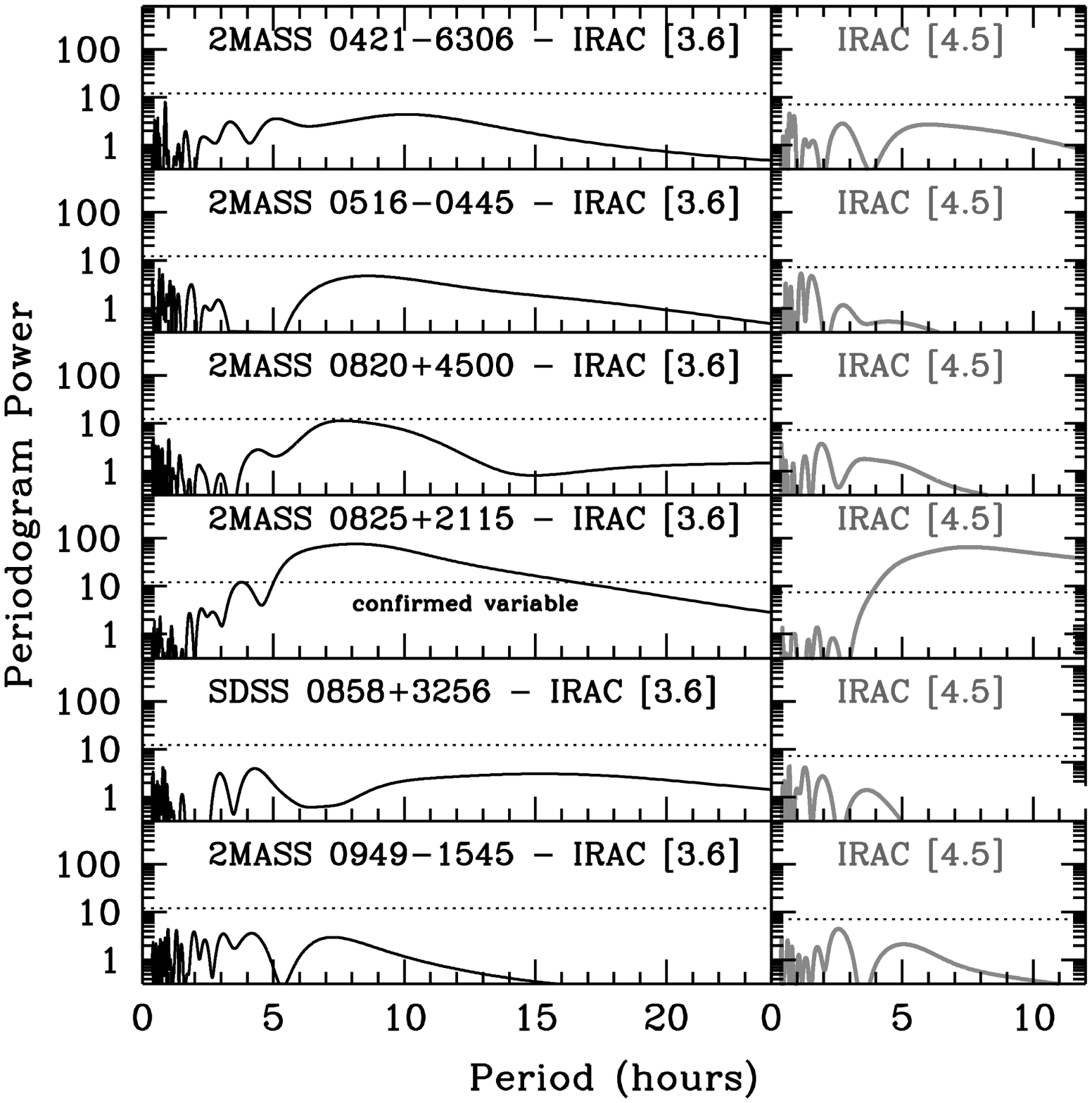} \\
\plottwo{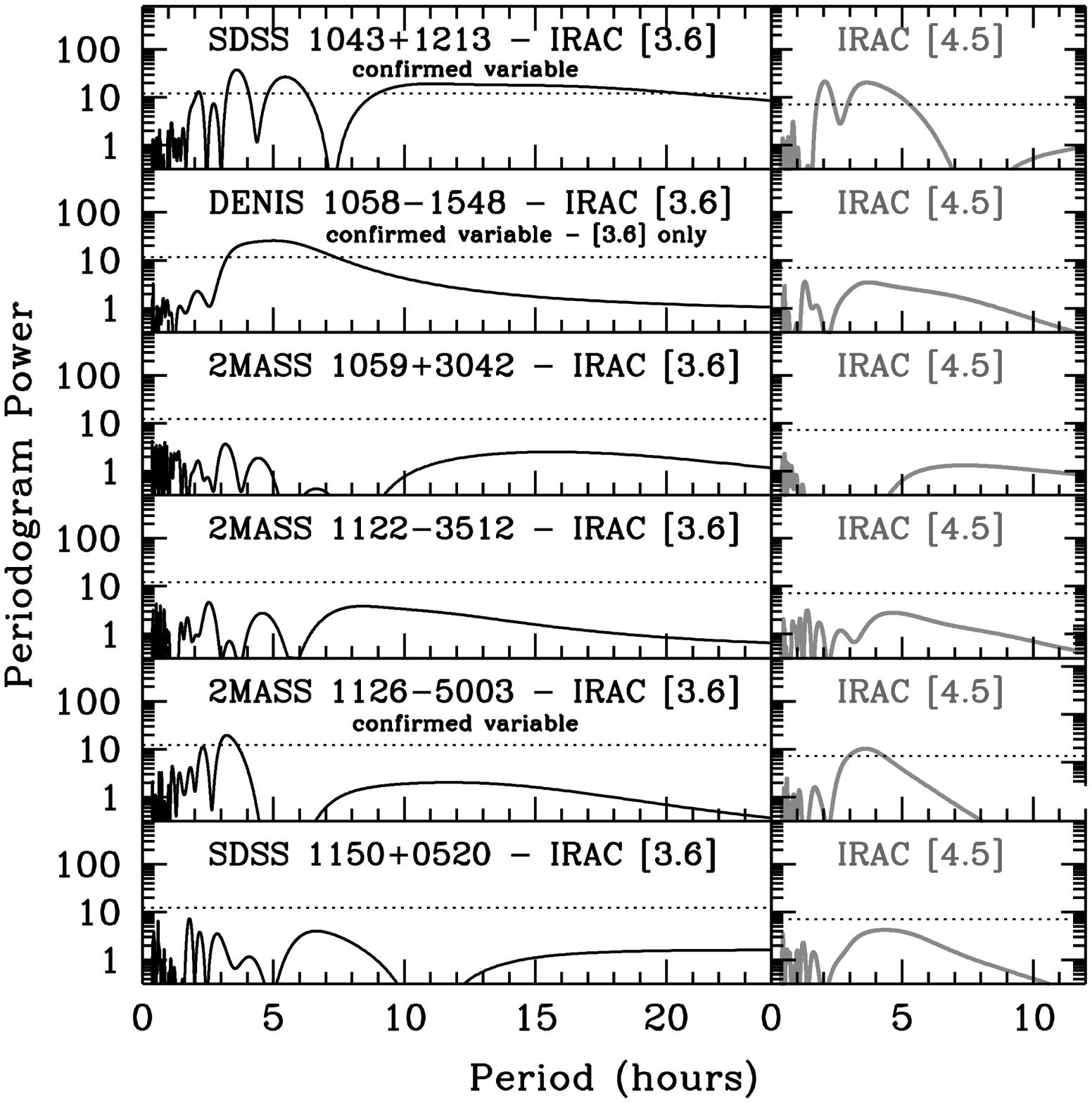}{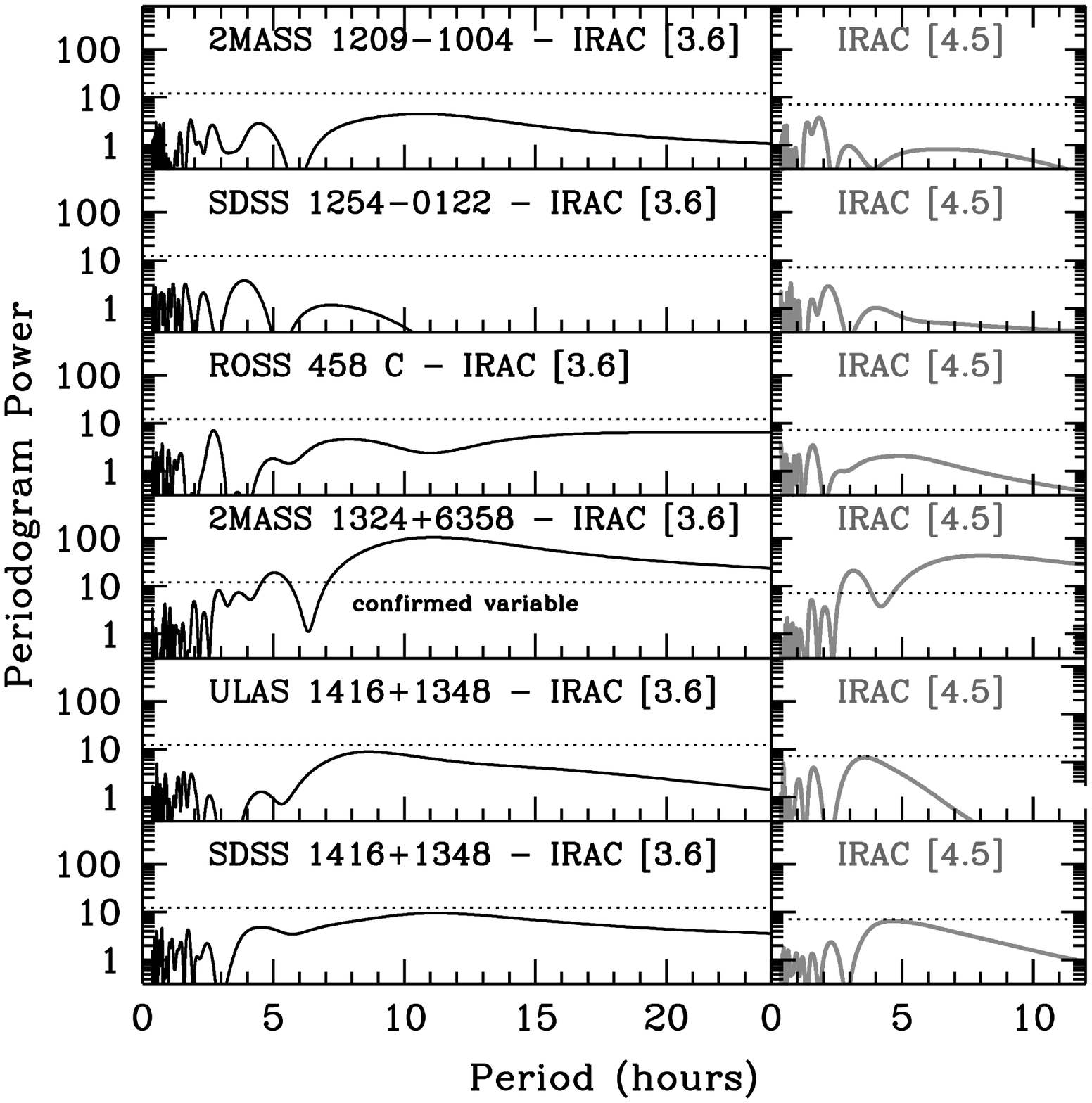}
\figcaption{\footnotesize Periodogram power distributions of the light curves of our objects after the initial pixel phase correction.  The dotted lines correspond to the FAP thresholds determined for each of the [3.6] and [4.5] bands as described in Section~\ref{sec_variable_identification} and Figure~\ref{fig_fap}.  Any object with periodogram power above the threshold at either of the IRAC bands is considered to be variable.
\label{fig_periodograms}}
\end{figure}

We identify as genuine astrophysical variables at the 95\% confidence level all brown dwarfs with FAP values at either band below the corresponding fifth percentile threshold (Fig.~\ref{fig_periodograms}).  Considering these as 95\% confidence thresholds is conservative 
because it implicitly assumes that all of the comparison stars with low FAP values are merely affected by residual systematics, when in fact some of them are probably astrophysical variables in their own right. 

We find a total of 21 variable brown dwarfs, including one binary and the deliberately added magnetically active L3.5 dwarf 2MASSW J0036159+182110), 15 of which are variable in both bands, four are variable only at [3.6], one (2MASS J00501994--3322402; T7) is variable only at [4.5], and one (HN~PegB; T2.5) is variable at [4.5] and has a FAP value on the threshold at [3.6].  Only one of the four known close binary systems (SDSS J151114.66+060742.9 [L5.5 + T5]) is variable, only at [3.6].  We argue in Section~\ref{sec_sdss1511} that the variations are likely associated with the brighter component.

We note that given our $\gtrsim$95\% confidence threshold on variability detection, we expect on average two false-positive identifications of variability in our unresolved sample of 39 L3--T8 dwarfs.  If such exist in our sample, these would most certainly be among the [3.6]-only variables, as dual-band variables are independently confirmed at [3.6] and [4.5], and the one [4.5]-only variable is highly significant.

\begin{figure}
\figurenum{3}
\plottwo{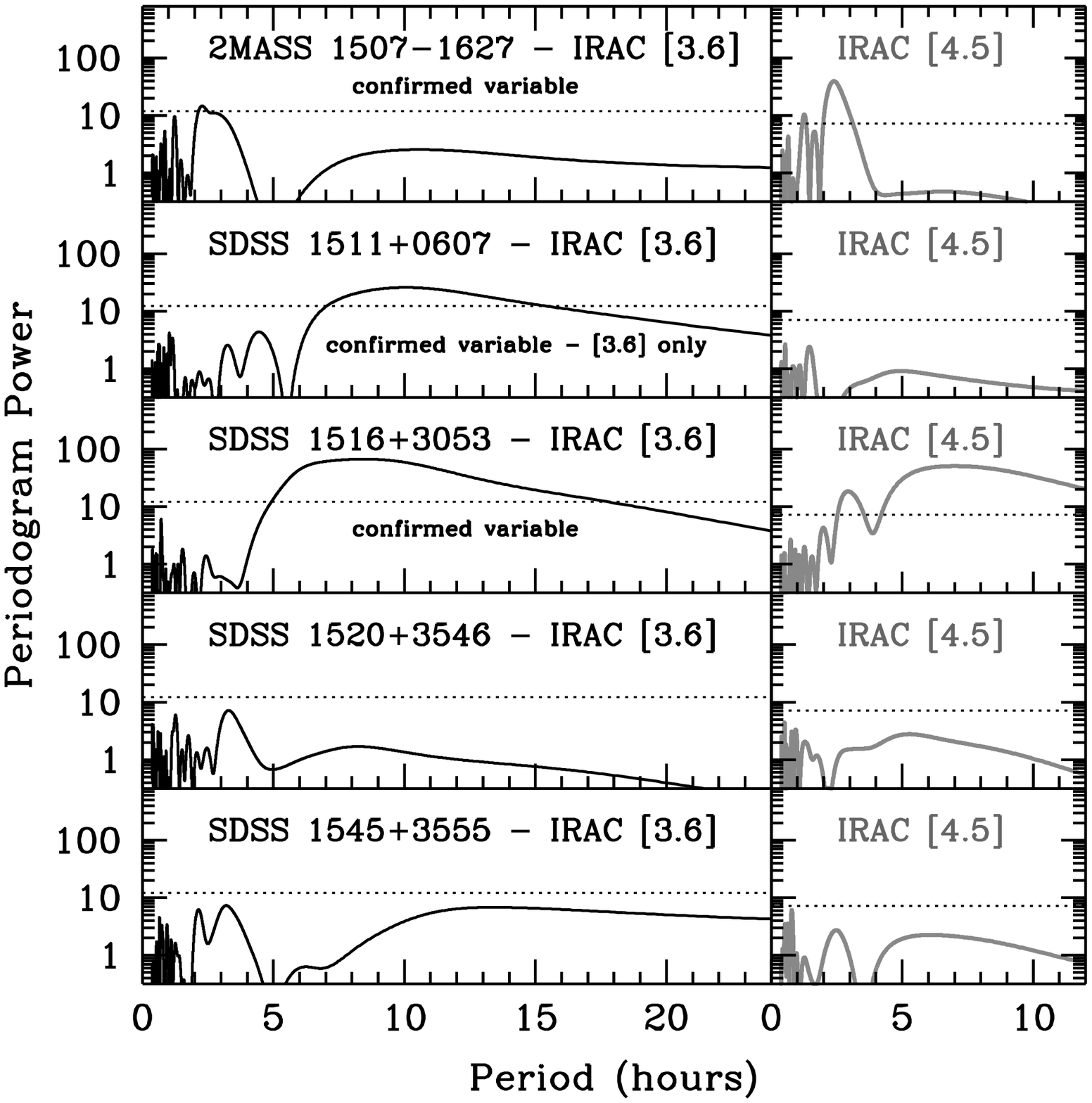}{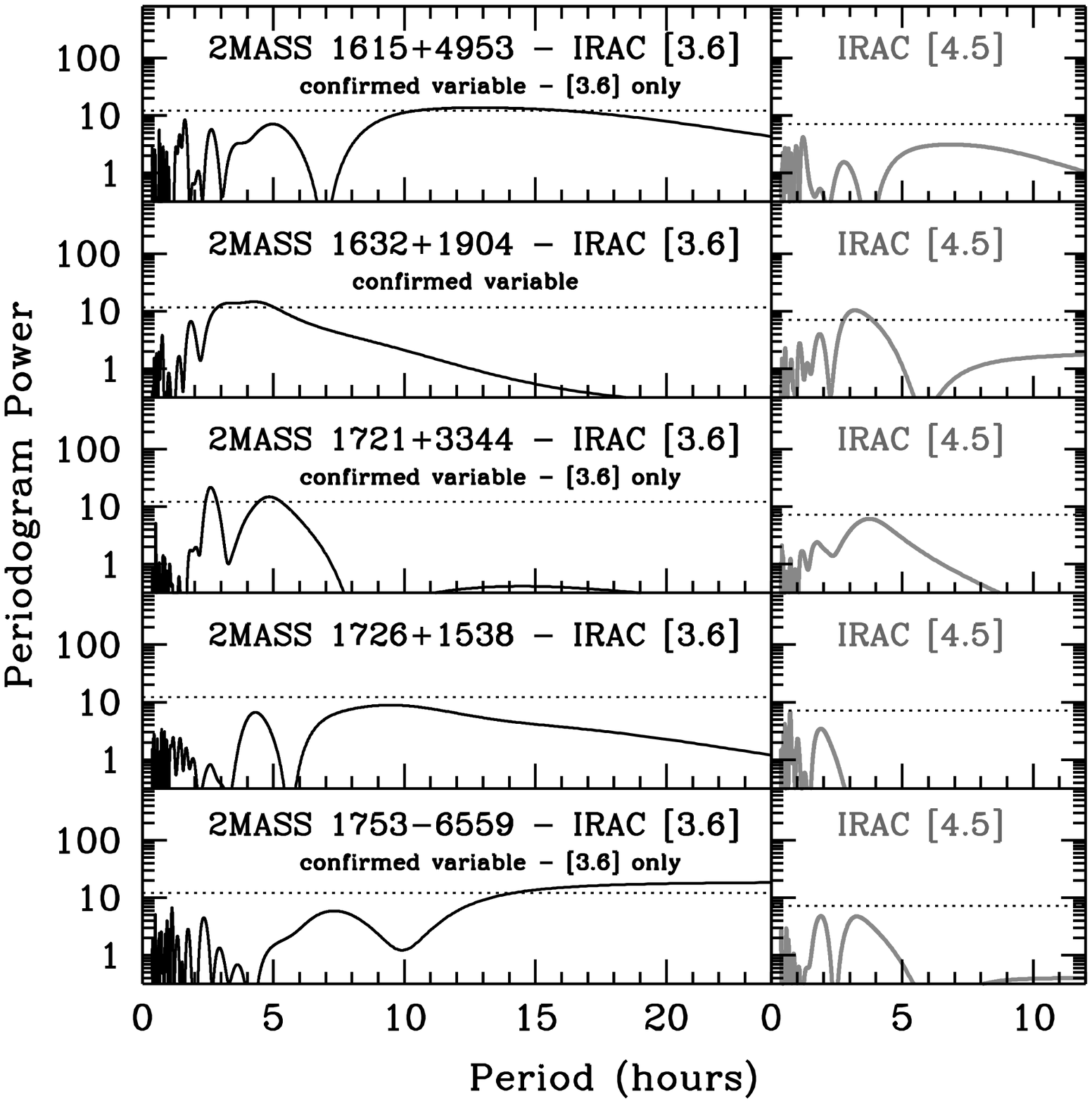} \\
\plottwo{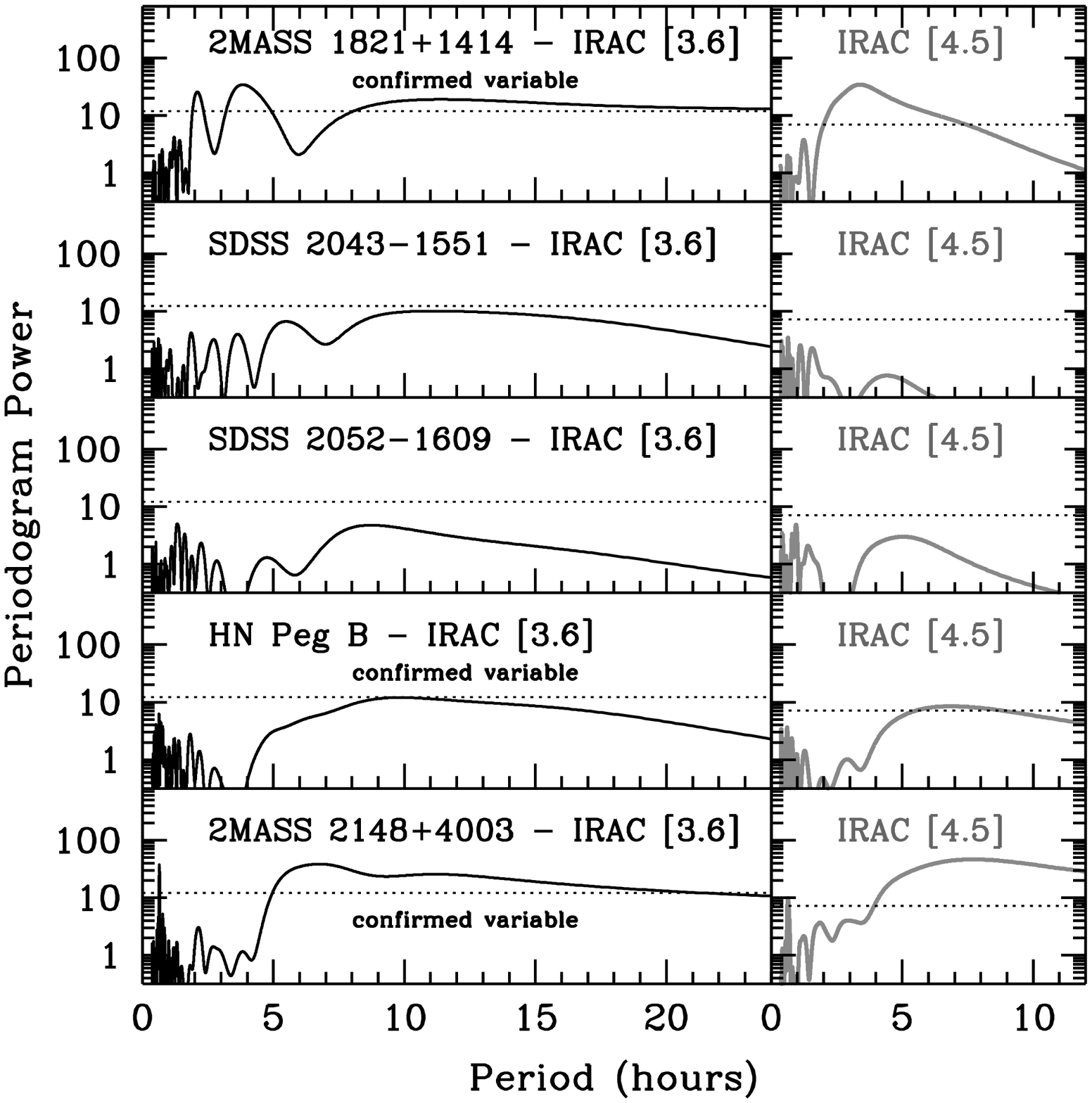}{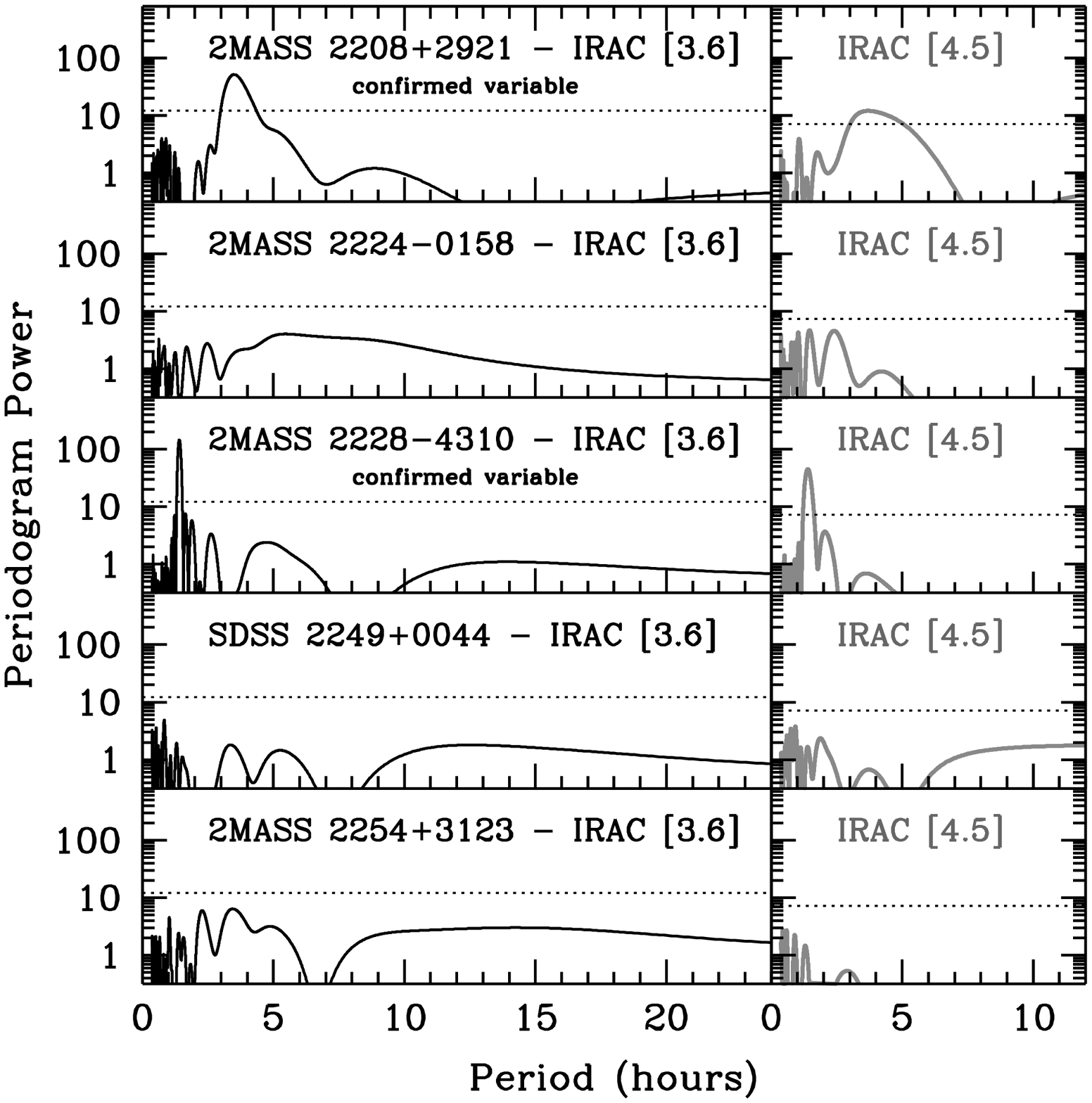}
\figcaption{\footnotesize Continued.}
\end{figure}

\subsection{Fitting for Periods, Amplitudes, and Waveforms}
\label{sec_variability_fitting}

We analyze the photometry of our variable objects by fitting an astrophysical model together with the pixel phase correction, using the simultaneous/iterative least-squares method described in \citet{heinze_etal13}. This fitting procedure removes the distortion of astrophysical variability that the pixel phase correction can impose if it is applied to the photometry independent of an astrophysical fit. We therefore use the results from the combined model plus pixel phase fits, rather than from the initial periodograms, to identify the true periods and amplitudes of our variables. 

Our astrophysical model is a truncated Fourier series:

\begin{equation}
F(t) = 1.0 + \sum_{j=1}^n a_j \sin\left(\frac{2 j \pi t}{P} + \phi_j \right),
\label{eq:fourier01}
\end{equation}

or equivalently, for purposes of linear least-squares fitting:

\begin{equation}
F(t) = 1.0 + \sum_{j=1}^n a_j \sin\left(\frac{2 j \pi t}{P}\right) + b_j \cos\left(\frac{2 j \pi t}{P}\right).
\label{eq:fourier02}
\end{equation}

For each object we set the number of terms $n$ in the series to the smallest value that produces fit residuals consistent with random noise: such that the periodogram of the residuals have an FAP $\geq1\%$ at the strongest peak.
We apply this model independently to the [3.6] and [4.5] data for each object, with no constraint on phasing or common periodicity.  The peak-to-peak amplitudes and uncertainties resulting from these fits are shown in columns $A[3.6]$ and $A[4.5]$ of Table \ref{tab_results}.  Where the period is longer than the monitoring interval in a given band, the true amplitude at that band could be larger than our quoted value.

Once we have fit the photometry from each band individually, we attempt a simultaneous fit to the photometry in both bands, where the period, phase, and waveform (i.e., the relative amplitudes of the Fourier terms) of the astrophysical model are constrained to be the same, but the overall amplitude is allowed to differ.  This ignores possible phase shifts between the two IRAC bands, and such have been reported in a T dwarf over the broader 1--5~$\micron$ wavelength range by \citet{buenzli_etal12}.  
However, phase shifts would not necessarily be expected in our case because both IRAC channels probe very similar atmospheric pressures \citep[e.g., see Fig.~7 in][]{ackerman_marley01}.  We find no evidence for phase shifts among our variables with regular, periodic curves (Fig.~\ref{fig_variable_curves}).  The possibility of a [3.6]-to-[4.5] phase shift is explored in one of our targets with a more complex light curve, SDSSp J010752.33+004156.1 [L8], in forthcoming work by \citet{flateau_etal15}.

\begin{figure}
\plottwo{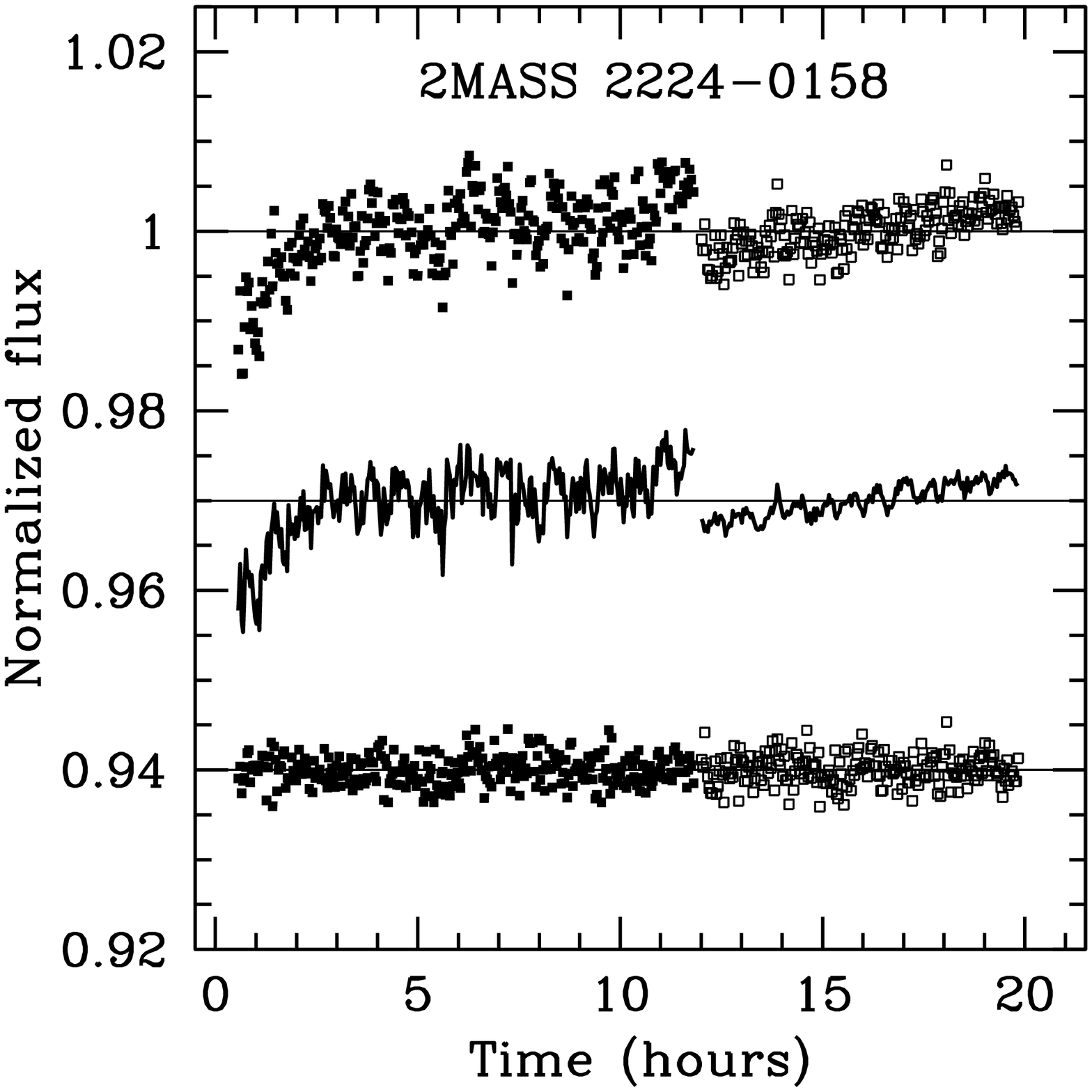}{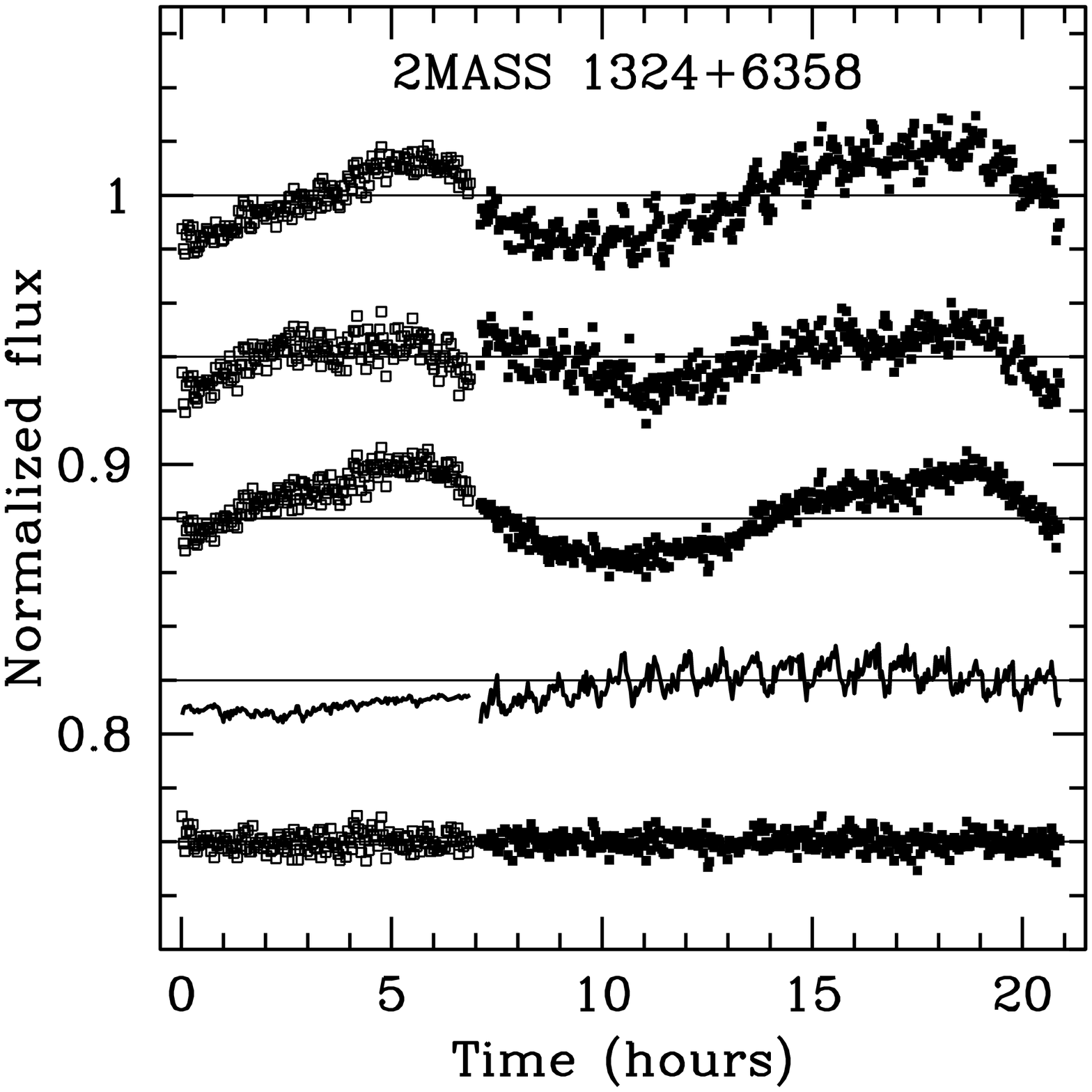}
\figcaption{\footnotesize The effect of pixel phase corrections and Fourier term fitting.  Solid points are [3.6] photometry and open squares are [4.5] photometry.  {\bf Left:} pixel phase correction for the non-variable L4.5 dwarf 2MASS J2224438--015852.  The uncorrected data are shown at the top, the contribution from pixel phase is shown in the middle, and the corrected data are at the bottom: all offset by --0.03 for clarity.  {\bf Right:} pixel phase correction for the variable T2.5 dwarf 2MASS J13243559+6358284. Starting at the top, the five time series are the raw photometry; the photometry after the initial pixel phase correction that did not include an astrophysical model (i.e., the input to our initial periodogram analysis for variable identification in Section~\ref{sec_variable_identification}); the photometry after pixel phase correction combined with a truncated Fourier series astrophysical model (see Section~\ref{sec_variability_fitting}); the final contribution from pixel phase; and the residuals from the final model.  Each time series is offset --0.06 relative to the previous one for clarity.  
\label{fig_ppc_and_model}}
\end{figure}

We perform a finely spaced 2-D grid search over period and [4.5]/[3.6] amplitude ratio, and use our simultaneous/iterative 
least-squares fitting method to determine the dominant waveforms
at each grid point.  The approach is analogous to that used in \citet{heinze_etal13}, but augmented to fit two bands simultaneously.
 Similarly to the earlier method, it also solves simultaneously for the parameters of the pixel phase correction, which are different, and independent, for [3.6] and [4.5].  The outcomes at various steps of our simultaneous pixel phase correction and Fourier term fitting are shown in Figure~\ref{fig_ppc_and_model}: for a non-variable object (2MASSW J2224438--015852; left panel) and for a variable object (2MASS J13243559+6358284; right panel).  The initial pixel phase correction distorts the astrophysical signal in the variable object, but does not remove it or prevent the periodogram analysis from detecting the variability.  By including the astrophysical Fourier model, the final fit accurately determines the pixel phase parameters and removes the distortion.
  


\subsection{Classification of Variables: Regular, Irregular, and Long-Period}
\label{sec_variable_classification}

\begin{figure}
\plottwo{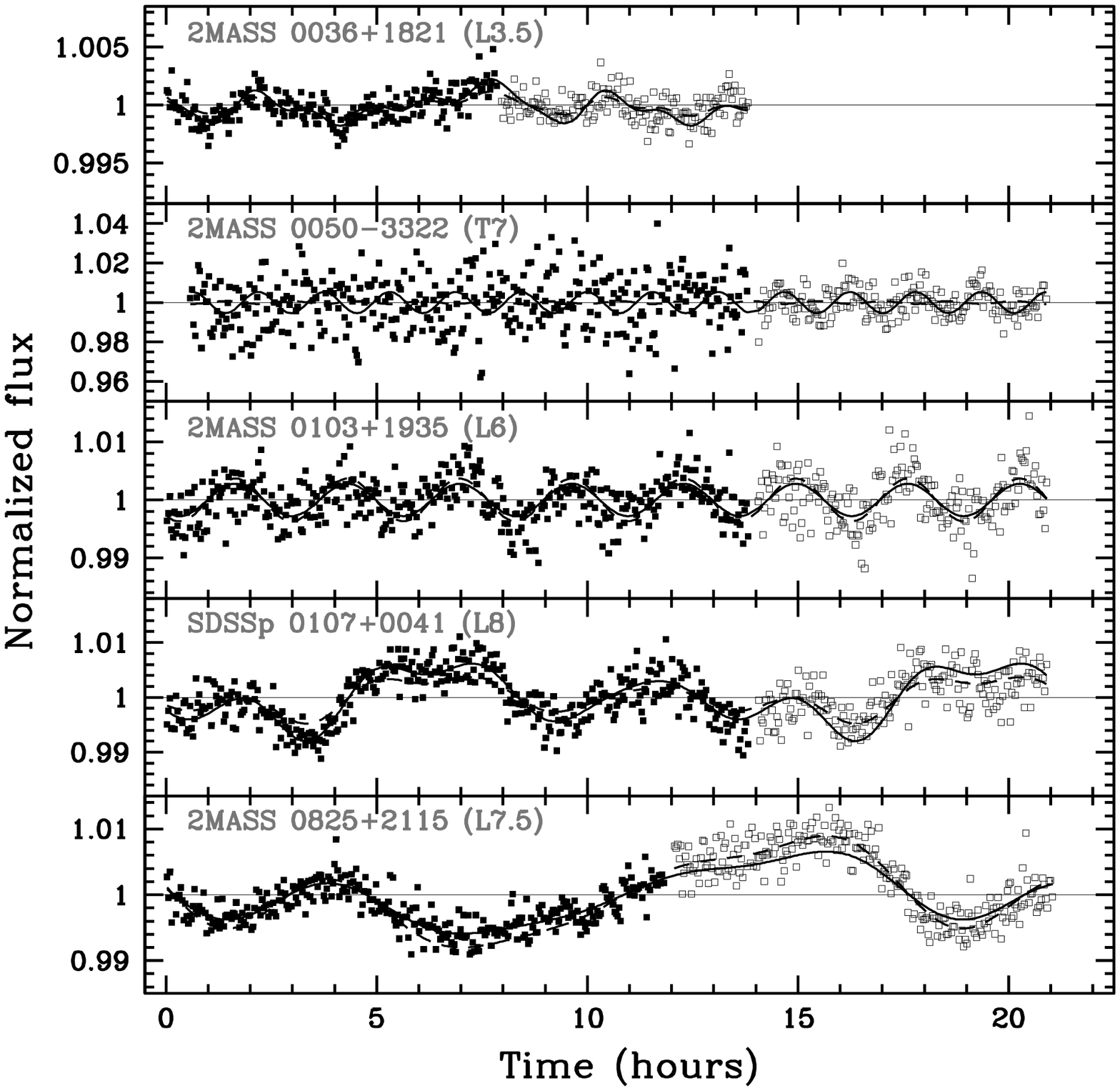}{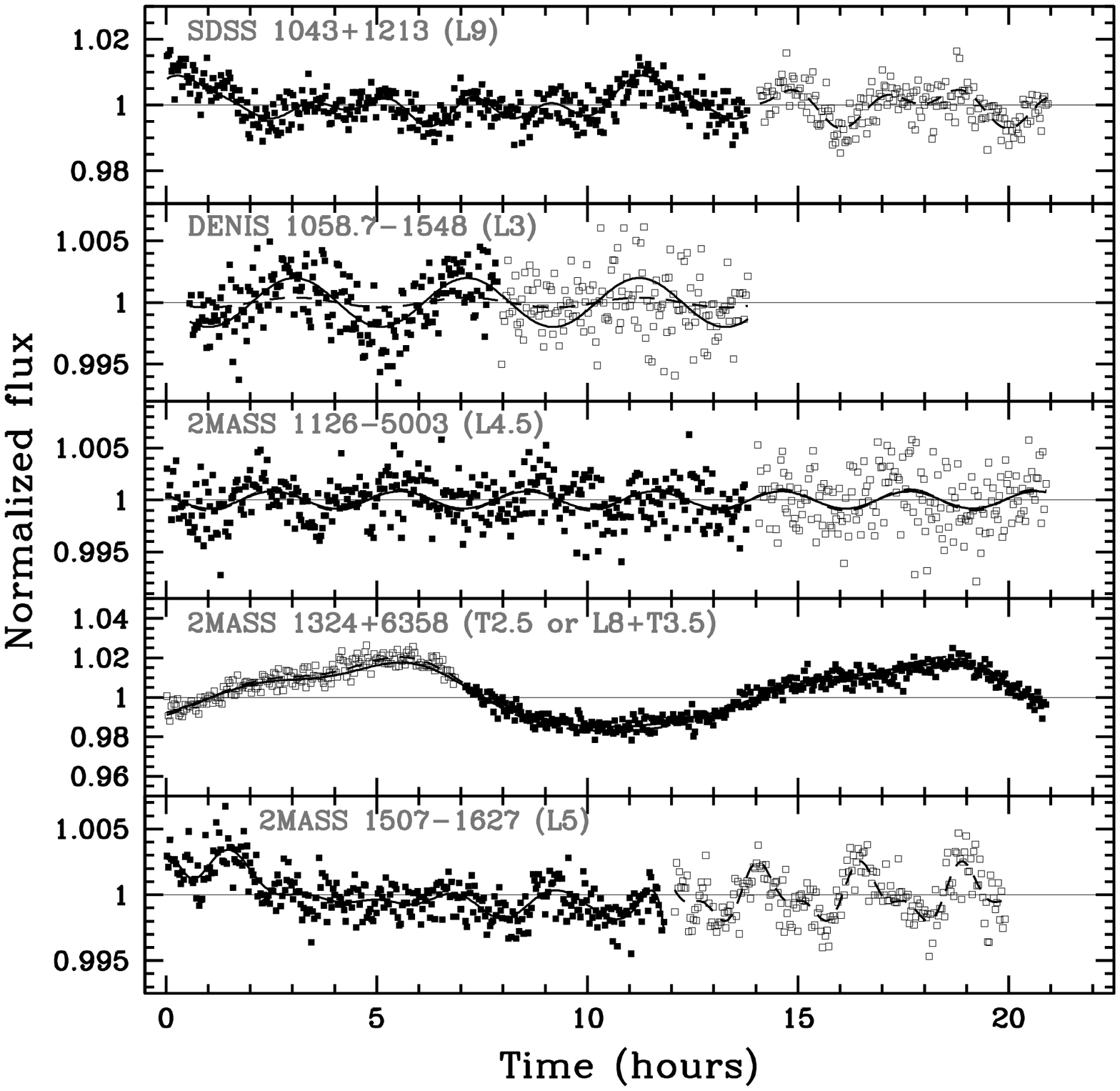} \\
\plottwo{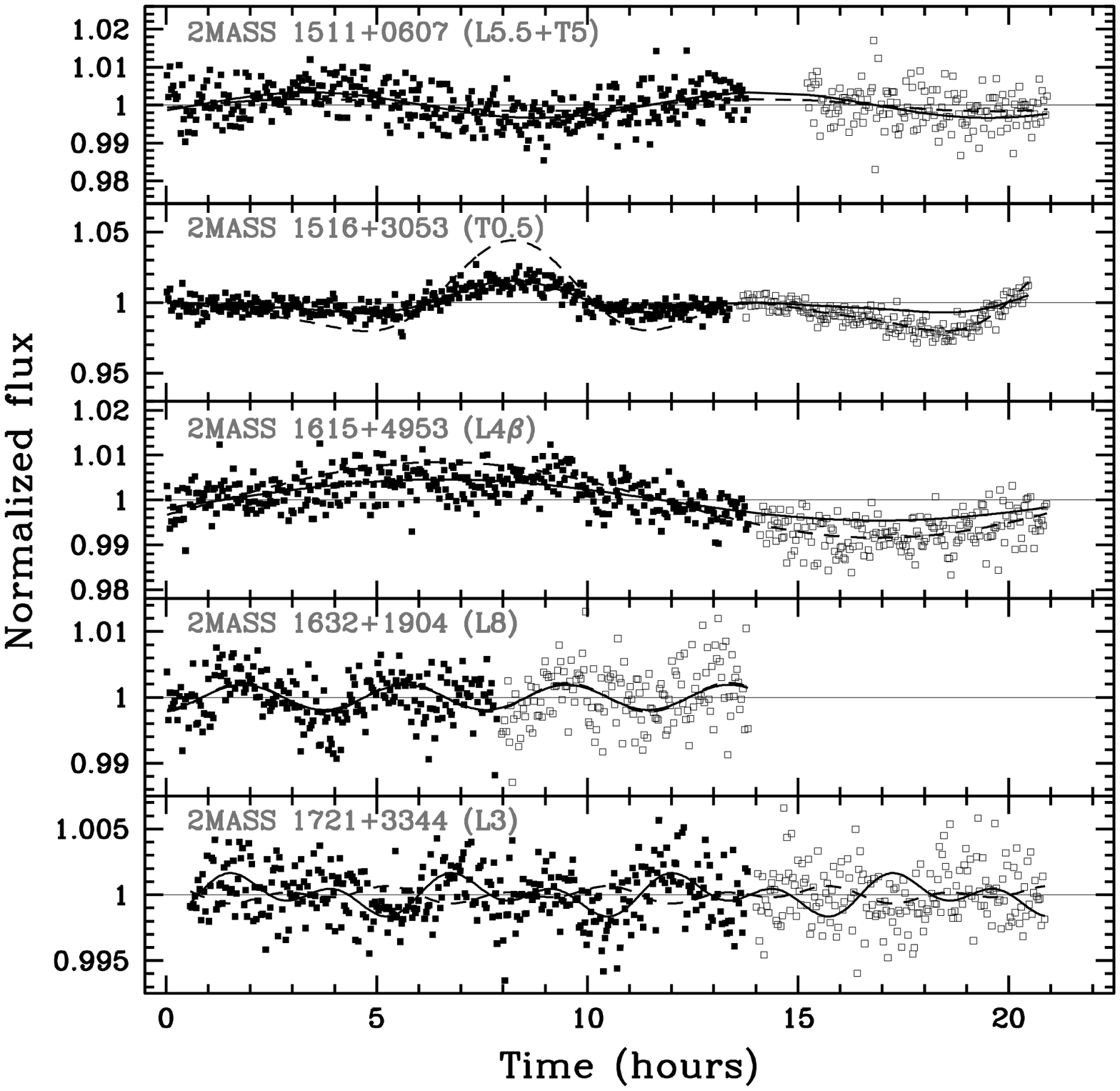}{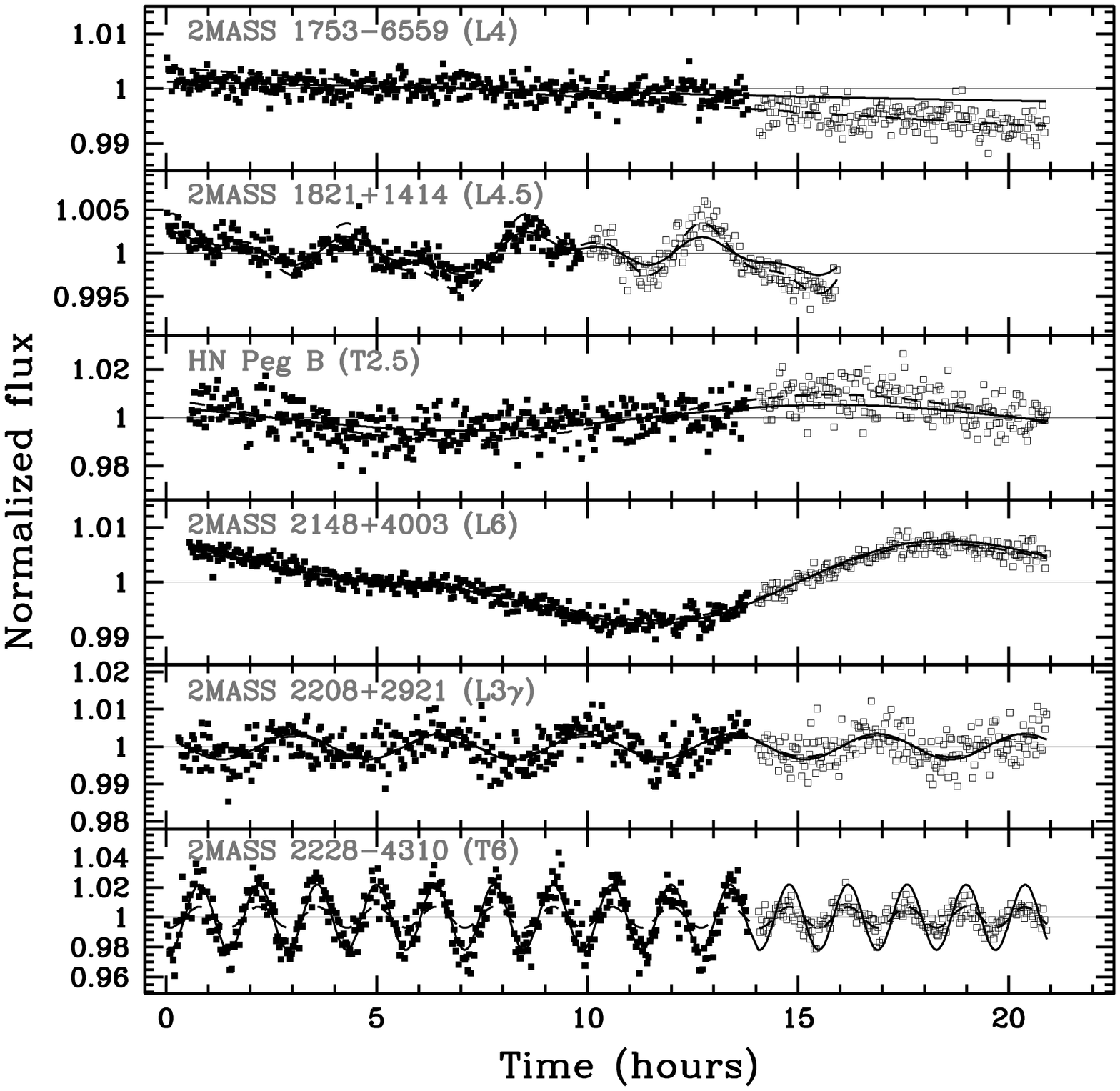}
\figcaption{\footnotesize Normalized {\it Spitzer} IRAC [3.6] (filled symbols) and [4.5] (open symbols) light curves of our 21 variable L and T dwarfs, ordered by R.A.  The fitted curves, solid for [3.6] and dashed for [4.5], are the lowest-order Fourier models that produced satisfactory fits.  The period, phase, and waveform 
are constrained to be the same for [3.6] and [4.5], but the overall amplitudes are permitted to differ.
\label{fig_variable_curves}}
\end{figure}

The final light curves of our 21 variable objects are shown in Figure \ref{fig_variable_curves}.  The high cadence and accuracy of our {\it Spitzer} photometry allows us to confidently establish that some of our variables have regular short-term periodicities, while others are irregular or have long periods.  Based on the preceding discussion (Section~\ref{sec_variability_fitting}), we categorize our variables as follows.


\begin{figure}
\plottwo{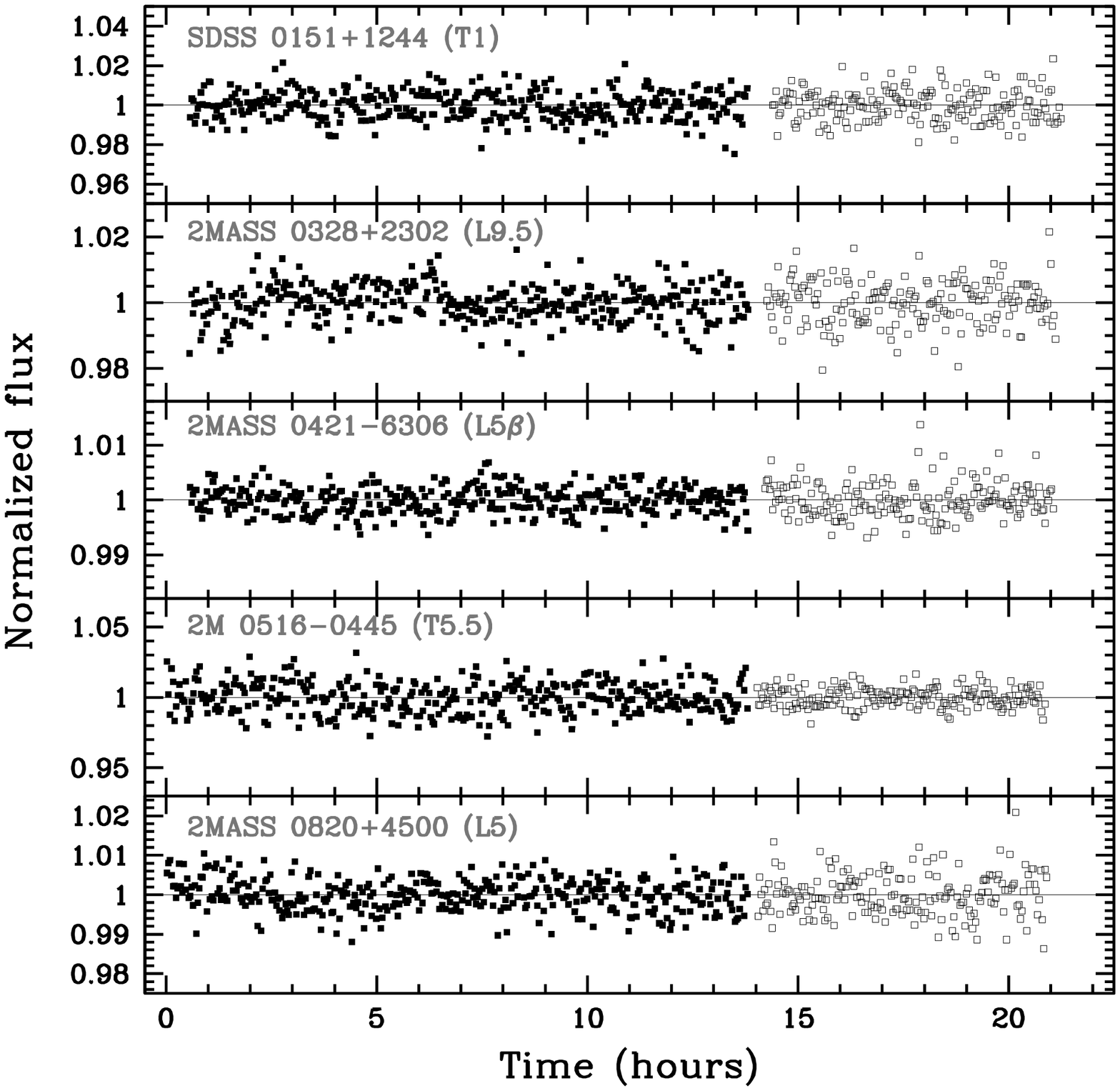}{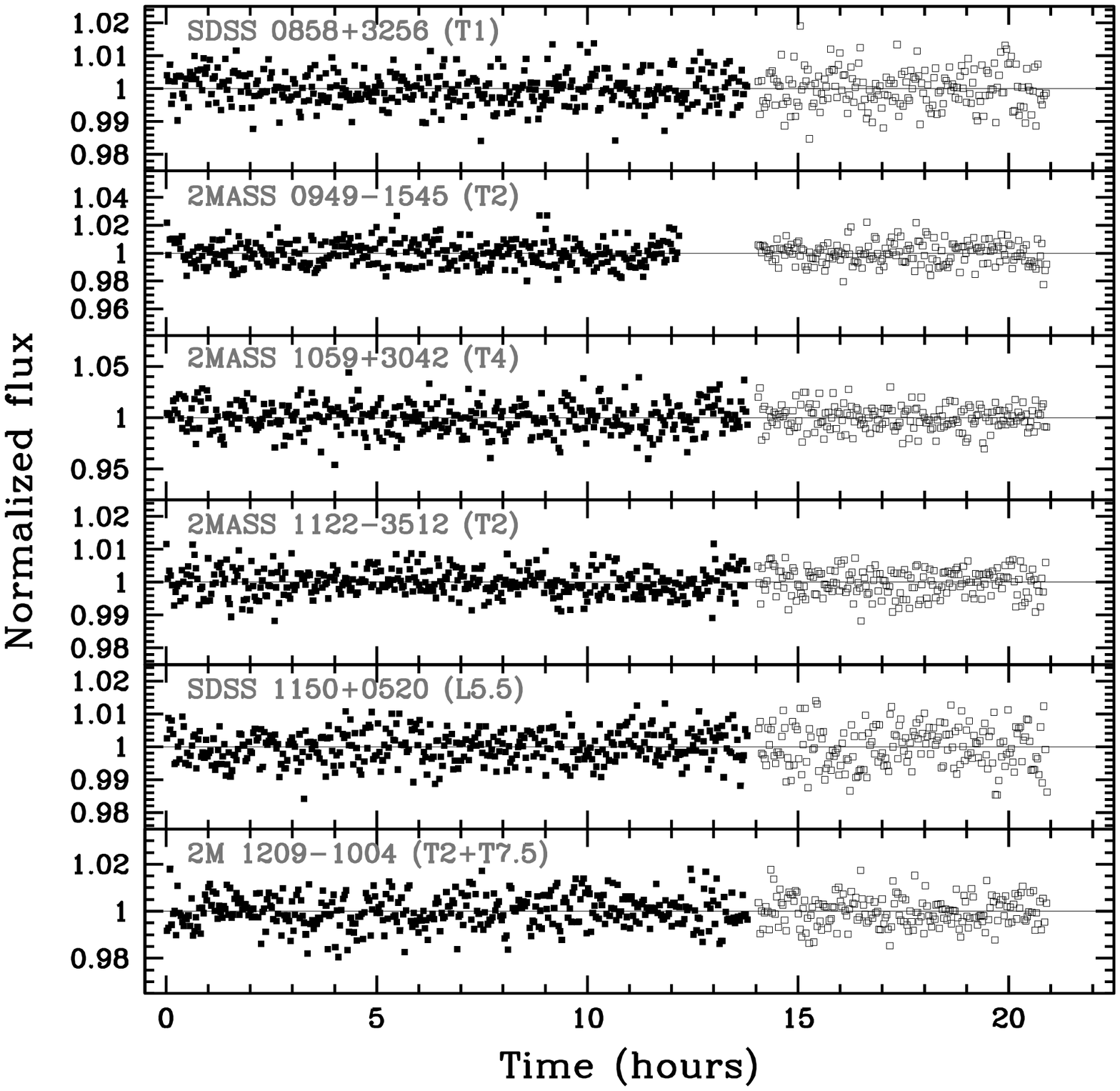} \\
\plottwo{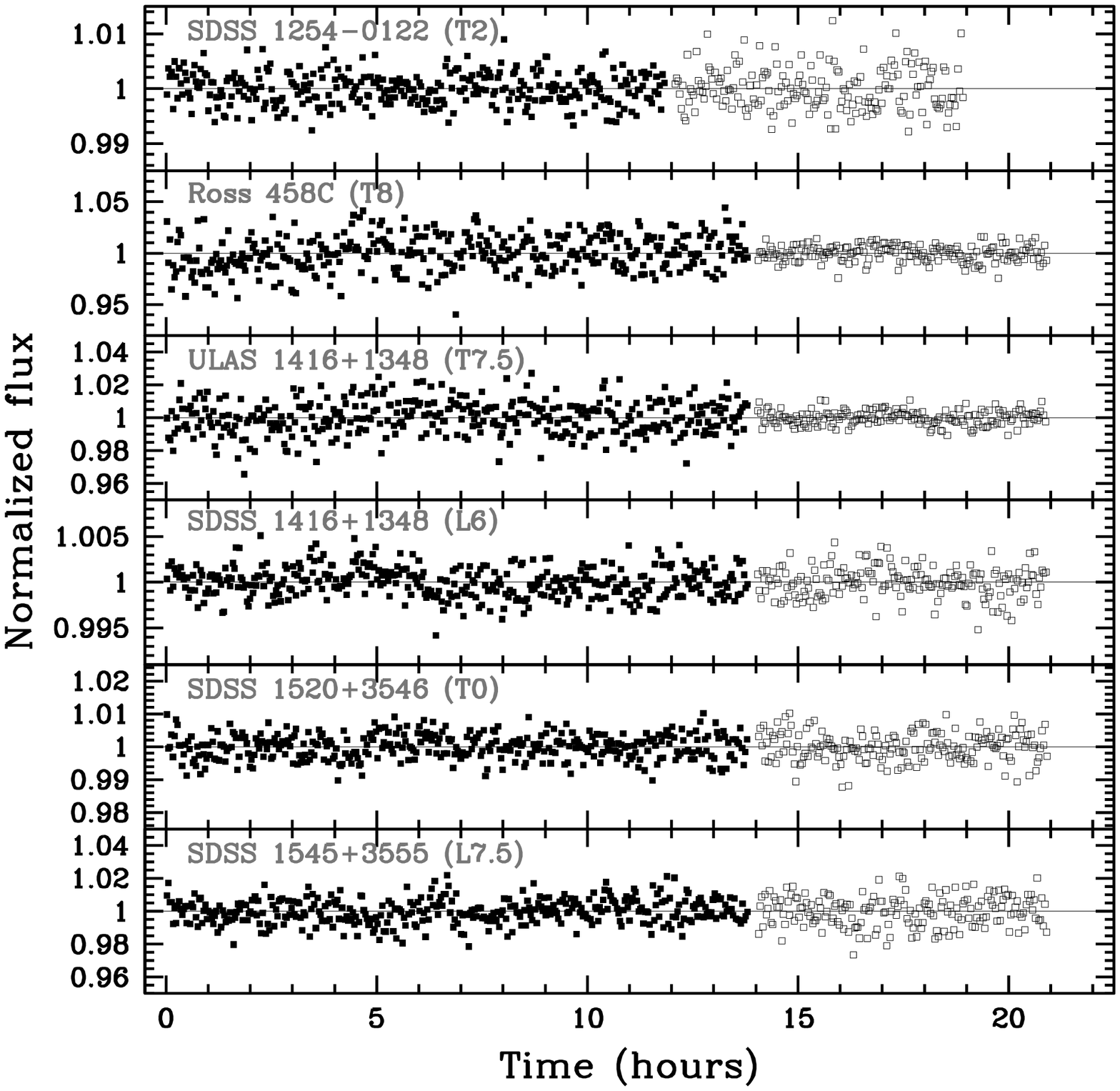}{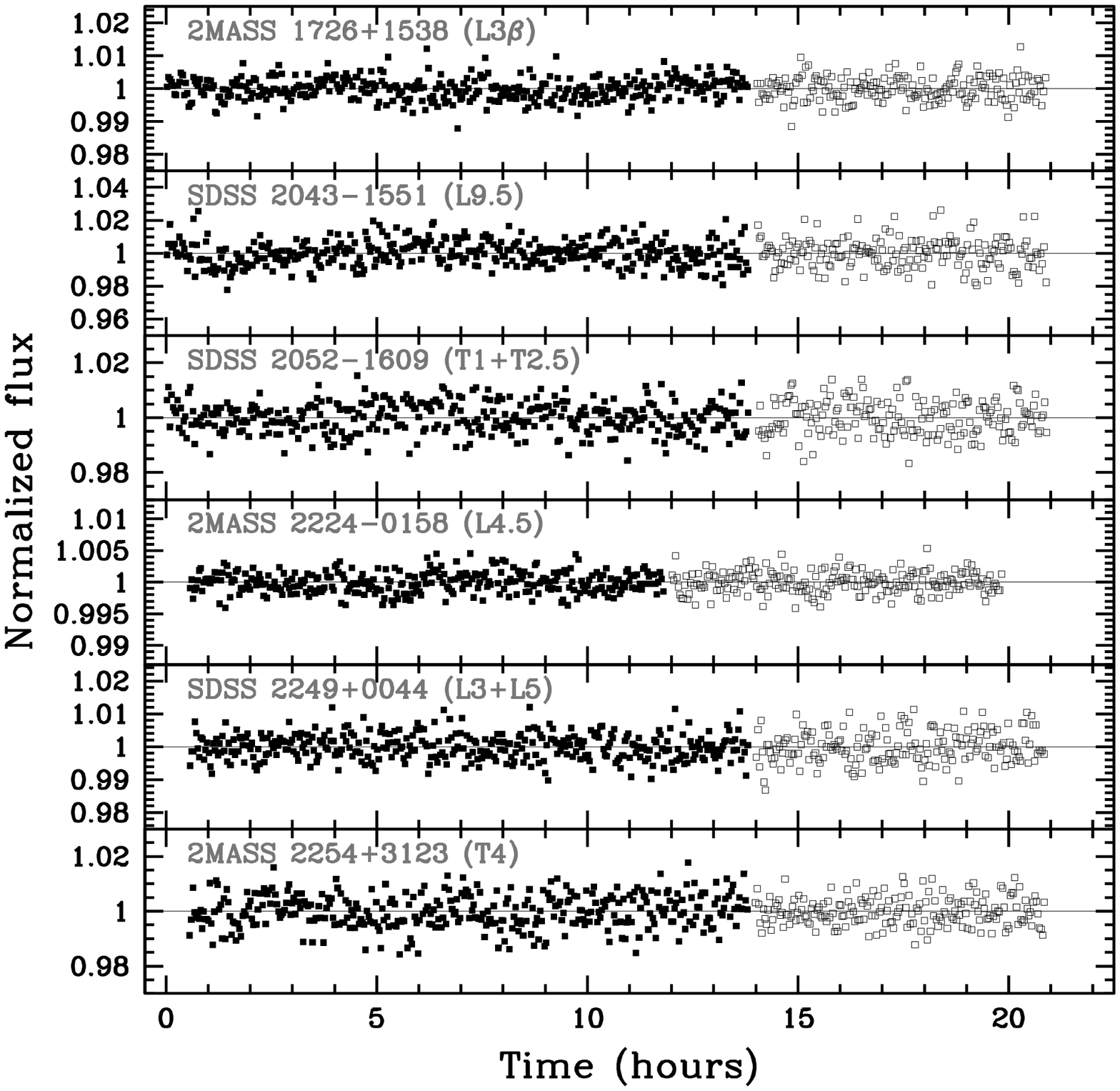}
\figcaption{\footnotesize Normalized {\it Spitzer} IRAC [3.6] (filled symbols) and [4.5] (open symbols) light curves of the 23 non-variable targets in our sample.
\label{fig_nonvariable_curves}}
\end{figure}




\paragraph{Regular variables}
are those for which we identify at least two complete rotations in the total (usually 21-hour) IRAC channel 1 and 2 observation, and the number of recorded rotations is greater than the number of Fourier terms required to fit the light curve.  That is, the fits to the light curves of the regular variables are well-constrained.  Eight of our 21 variables are regular, and are noted with ``reg'' in the Periodicity column of Table~\ref{tab_results}.

The regular variables have reliable estimates of periods and period uncertainties in Table~\ref{tab_results}, obtained from the range of period values produced by the respective single-band fits and by the two-band fit.  The [4.5]/[3.6] amplitude ratios are obtained from the fitted light curves.  These ratios can be different from the ratios of the individual [4.5] and [3.6] amplitudes, which are fit independently. Two of the regular variables (DENIS J1058.7--1548 [L3] and 2MASS J11263991--5003550 [L4.5]) have significant variability only at [3.6], and one (2MASS J00501994--3322402 [T7]) only at [4.5].  In these cases we fit only the variable-band data, and set upper limits on the amplitudes in the non-variable bands.  
However, we do list the best-fit [4.5]/[3.6] amplitude ratios from the joint fits on the [3.6]-only regular variables.

\paragraph{Irregular variables}
are those with at least two recorded rotations whose fits require more Fourier terms than the number of recorded rotations.  We believe that in these cases the photospheric brightness distribution on the brown dwarf was changing during our observations, and therefore the astrophysical variations were not strictly periodic.  Four of our variables are irregular: marked with ``irreg'' in Table~\ref{tab_results}.

The fits for irregular variables are effectively non-periodic because the nominal periods of the Fourier series approach the length of our monitoring.  The fits represent the simplest Fourier model that was able to account for all the data.  They also likely represent the lowest-order fits that solve accurately for the pixel-phase parameters, rather than producing pixel-phase results that are biased by astrophysical variations not captured by the Fourier model.  The photometry in Figure~\ref{fig_variable_curves} has been corrected based on the pixel-phase parameters produced by these final fits.   

The periods, and in particular the amplitudes and [4.5]/[3.6] amplitude ratios of the irregular variables, are often less well constrained.  We estimate the rotation periods 
using periodograms of the final, corrected data, and by identifying commonalities in terms of frequency components among the single-band fits and among fits using different numbers of Fourier terms.  
All four irregular variables do show dominant periodicities on a time scale shorter than half of the observing sequence, and we are able to identify uncertainties for these periods.  The quasi-periodic behavior of the irregular variables is in agreement with the expectation of rotational modulations---as for the regular variables---and we assume that these correspond to the objects' rotation periods.

The amplitudes of the irregular variables correspond to the maximum peak-to-peak variation observed in each band.  Because variations outside of our observing window could have even greater amplitudes, the quoted amplitudes are effectively lower limits.  Because of the rapid changes in the light curves of the irregular variables, we refrain from using the joint two-band fits to determine the [4.5]/[3.6] amplitude ratios.  Instead, we report the ratio of the amplitudes of the unconstrained single-band fits.  
 

\paragraph{Long-period variables} are those for which our observations cover less than two rotations.  Six of our 21 variables have such long periods, and are marked with ``long'' in the Periodicity column of Table~\ref{tab_results}.  The four for which we see one full rotation have estimates of the period uncertainties, and [4.5]/[3.6] amplitude ratios.  One of the four, the L5.5 + T5 close binary SDSS J151114.66+060742.9, is significantly variable only at [3.6], and we provide only an upper limit to the [4.5] amplitude from the stand-alone channel 2 data.

One of the other two long-period variables, 2MASS J16154255+4953211 (L4$\beta$), has a period somewhat longer than the 21~h channel 1 and 2 AOR sequence.  We do not estimate an uncertainty on its period, and the [4.5] amplitude can not be estimated from the channel 2 data alone.  However, we are able to tentatively estimate the [4.5]/[3.6] amplitude ratio from a joint fit to the [3.6] and [4.5]-band light curves, if we allow a significant offset---larger than the [4.5]-band amplitude upper limit---in the [4.5]-band curve.

The remaining long-period variable, 2MASS J175334518--6559559 (L4), does not show any periodicity, but only a trend.  The trend is significant only in the [3.6] data, where the object is more variable than 99\% of the 636 comparison stars from our entire {\it Spitzer} campaign.  While the [4.5] data in Figure~\ref{fig_variable_curves} are shown systematically below the [3.6] data, this is a consequence of our assumption that the [3.6] and [4.5] variability are phased, which requires that the [3.6] and [4.5] light curves intersect at unity.  The [4.5] data alone show no evidence of variability, and we do not estimate a [4.5]/[3.6] amplitude ratio for this object.  As will be detailed in a forthcoming publication \citep{heinze_etal15}, we suspect that 2MASS 175334518--6559559 may be viewed close to pole-on, and that we may be seeing spot pattern evolution on the visible hemisphere.  As such, the light curve probably does not reflect the spin period of the object.

In principle, it may be possible to explain most of the long-period light curves in our sample through cloud evolution and near pole-on viewing geometry.  Nonetheless, we note that none of the long-period variables show evidence of rotational variations on shorter time scales.  Where such evidence is present, we classify the variables in the remaining category of ``irregular/long-period variables.''

\paragraph{Irregular/long-period variables} are three variables--marked ``irreg/long'' in Table~\ref{tab_results}---that show significant periodicities on multiple scales, with a marked improvement in the quality of the fit for periods longer than 10 hours.  For these we have adopted the shortest period at which there is a highly significant peak in the periodogram.
Thus, both 2MASSI J0825196+211552 (L7.5) and SDSS J151643.01+305344.4 (T0.5) show substantial power in periods that are approximately half of the best-fit period with three Fourier terms.  These periods, 7.6~h and 6.7~h are the ones that we have adopted for these objects.  However, we have not quoted period uncertainties since the actual period may be significantly longer.  

The remaining irregular/long-period variable, SDSSp J010752.33+004156.1 (L8), is well fit by a five-Fourier term solution with a 13.0~h period, which effectively matches the beginnings of the channel 1 and 2 light curves.  Because of the erratic appearance of this light curve, we believe that we may be witnessing rapid evolution of the spot pattern that may be obscuring the actual rotation period.  In seeking a dominant time scale that would potentially reflect the spin of the object, we observe that single-Fourier term fits to the channel 2 or combined channel 1+2 data reveal periodicities of 5.0~h to 5.5~h, while a single-term fit to the channel 1 data alone reveals a periodicity that is approximately twice as long: 10.2~h.  Neither of these single-term fits are even remotely satisfactory.  However, noting the $\approx$5~h multiples in the single-term periods, we adopt 5~h as our best guess for the period of SDSSp J010752.33+004156.1.
As for the irregular variables, the amplitude estimates for the irregular/long variables are effectively lower limits.


\subsection{Non-Variables and Amplitude Upper Limits}
\label{sec_nonvariables}

For objects that are not variable (Fig.~\ref{fig_nonvariable_curves}) we use a Monte Carlo method to calculate upper limits on the amplitude of any undetected variability.  For each given object, we create a large ensemble of simulated data sets, each with the same sampling as the real data.  Each simulated data set contains a signal of fixed period and amplitude, a random phase, and a distinct realization of Gaussian noise matched to the RMS scatter of the real data.  The amplitude upper limits were determined assuming fixed ten-hour periods.  As
our sensitivity is better for shorter periods, this is a conservative choice.

To account for possible suppression of signal by the pixel phase correction, we correct the synthetic data sets using the same prescription as for the real data.  We find the periodogram FAP of each synthetic data set in the ensemble, and determine in what fraction of simulated cases the FAP is lower (that is, more significant variations were detected) than in the real data set.  We adjust the amplitude of the simulated signal until the FAP becomes lower than for the real data set in 95\% of cases.  
This threshold sets our 95\% confidence level upper limit on the amplitude of sinusoidal variations in each non-variable brown dwarf (Table~\ref{tab_results}).

As we already noted, the periodogram is not optimally sensitive to non-sinusoidal variations.  We performed additional tests with a different input signal: the sum of two equal-amplitude identically-phased sinusoids differing by a factor of two in period.  We considered this a reasonable representation of some of the extreme amplitude behavior observed in the lightcurves our variables.
Such an input signal aims to model cases where the lightcurve of a variable spends most of its time near the mean, and has only one narrow peak and one narrow trough per cycle.  Such variables would be more easily missed compared to perfectly sinusoidal variables with the same amplitude because of the leakage of periodogram power out of the main peak: resulting partly from the presence of a second period, partly from the small number of periods covered by our observation.
We do not list these more conservative ``non-sinusoidal'' upper limits, although note that they are on average 50\% higher at [3.6] and 30\% higher at [4.5].

\begin{figure}[ht]
\plottwo{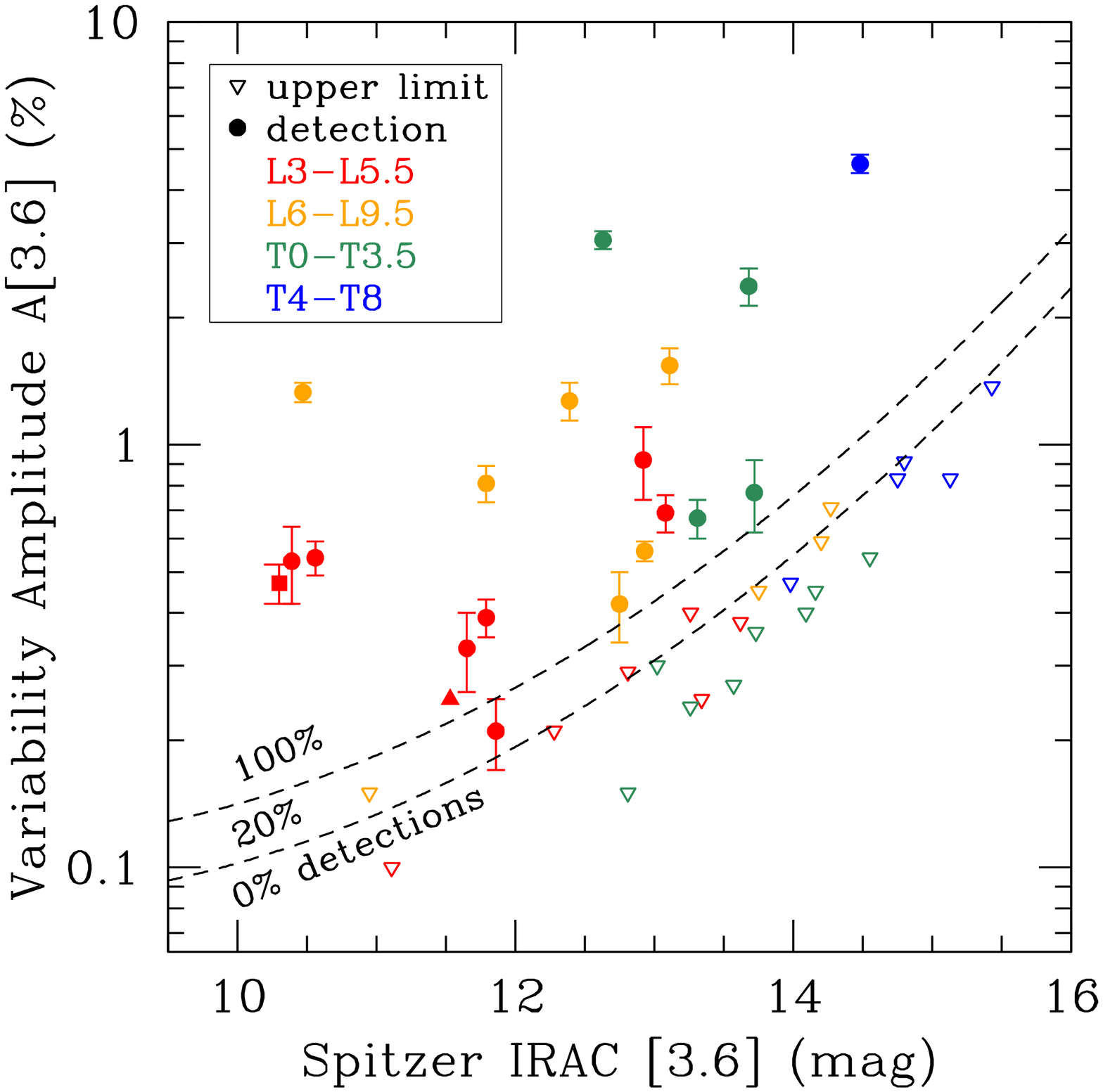}{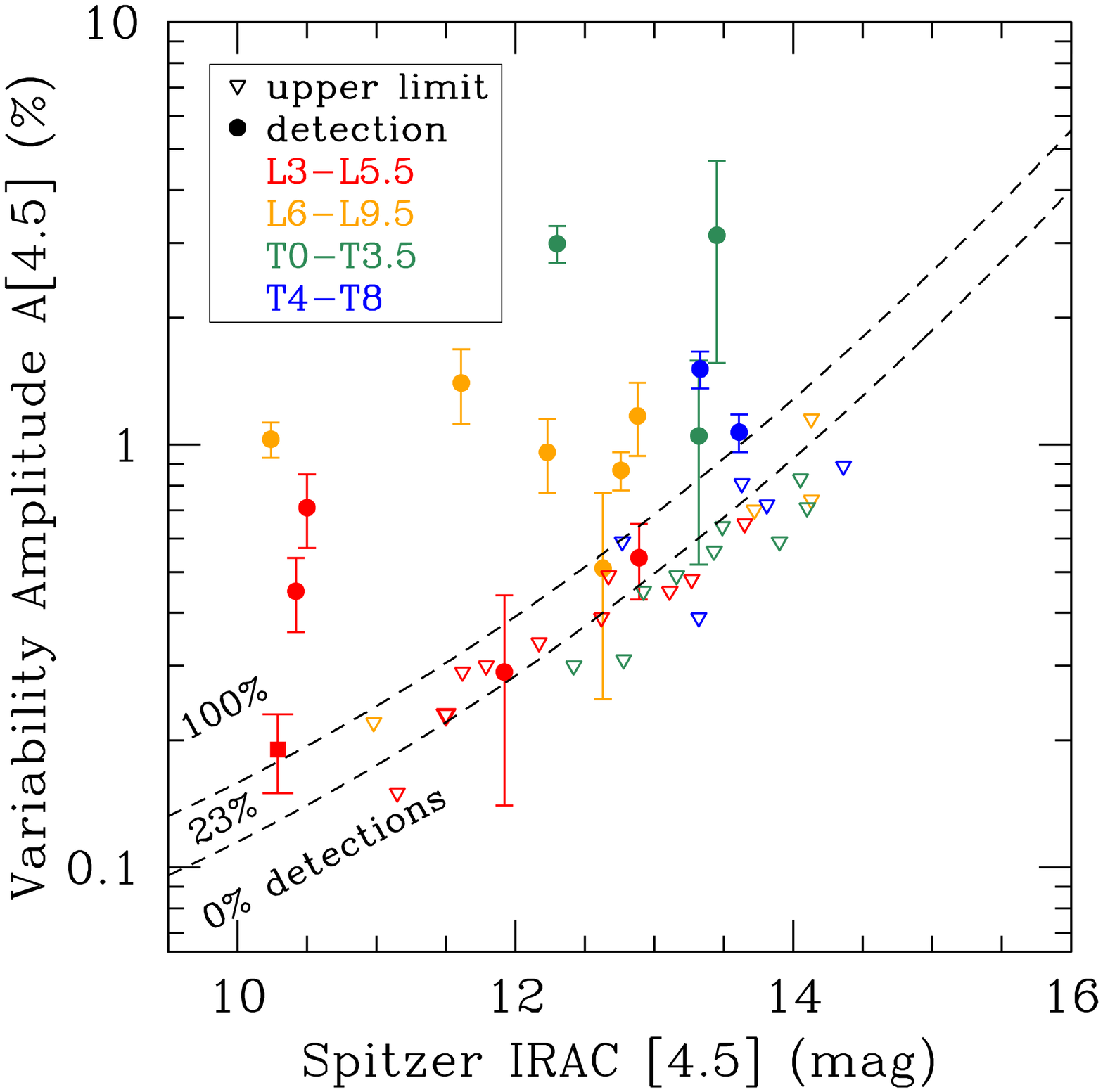}
\figcaption{\footnotesize Variability amplitudes vs.\ target brightness for L3--T8 dwarfs at [3.6] ({\it left}) and [4.5] ({\it right}).   The L4 dwarf 2MASS J175334518--6559559 shows only a linear trend at [3.6], and we have plotted the lower limit on its [3.6] amplitude with a solid red upward-pointing triangle.  
The known magnetically active L3.5 dwarf 2MASSW J0036159+182110
is shown with a filled red square. The dashed curves 
delineate regions of detection completeness, and are scaled linearly from the photometric  precision limits in Figure~\ref{fig_photometric_precision}.  The variability detection rates 
in each region correspond to the fraction of detected variables.  These detection completeness rates are used in the Monte Carlo simulations to determine the overall survey incompleteness (Sec.~\ref{sec_monte_carlo}).
\label{fig_Avsmag}}
\end{figure}

The 95\% upper limits on sinusoidal [3.6] and [4.5] variability amplitudes are plotted along with the amplitudes of the detected variables in the two panels of Figure~\ref{fig_Avsmag}.  Most of our non-detections are T dwarfs, consistent with the relative faintness of T dwarfs in our sample compared to L dwarfs.  The dashed curves plotted in each panel of Figure~\ref{fig_Avsmag} are scaled versions of the respective photometric precision limits from Figure~\ref{fig_photometric_precision}, and separate the majority of the detections from the majority of non-detections. 

We note that the loci of detections and upper limits partially overlap in the ``20\% detections'' and ``23\% detections'' bands in the [3.6]- and [4.5]-band panels.  In these regions we have both low-amplitude (but significant) detections, and upper limits that scatter higher than some of the detections.  The scatter is caused by several factors, most significant among which is the amount of periodogram power in the data for a given object.  Some objects can not be classified as variables although they have considerable periodogram power and FAP values near the variability threshold (Fig.~\ref{fig_fap}).  Such objects may exhibit real astrophysical variations, albeit below our detection limit.  For these objects we are not able to rule out amplitudes as small as those for targets of similar brightness that show almost no periodogram power. Other factors contributing to the scatter in amplitude upper limits include differences among the photometric properties of each set of reference stars, different pixel phase effect systematics in the observations, and in a few cases, shorter AORs.
 

We note that the transition between detections and non-detections of variability on either side of the detection limits curve in Figure~\ref{fig_Avsmag} is smooth and continuous.
This further demonstrates that while we do not detect low-amplitude variables among the cooler brown dwarfs, that is likely because of our
poorer sensitivity on fainter targets.  

\section{THE VARIABILITY OF L3--T8 DWARFS AT 3--5~$\micron$
\label{sec_variability}}

\begin{figure}
\plotone{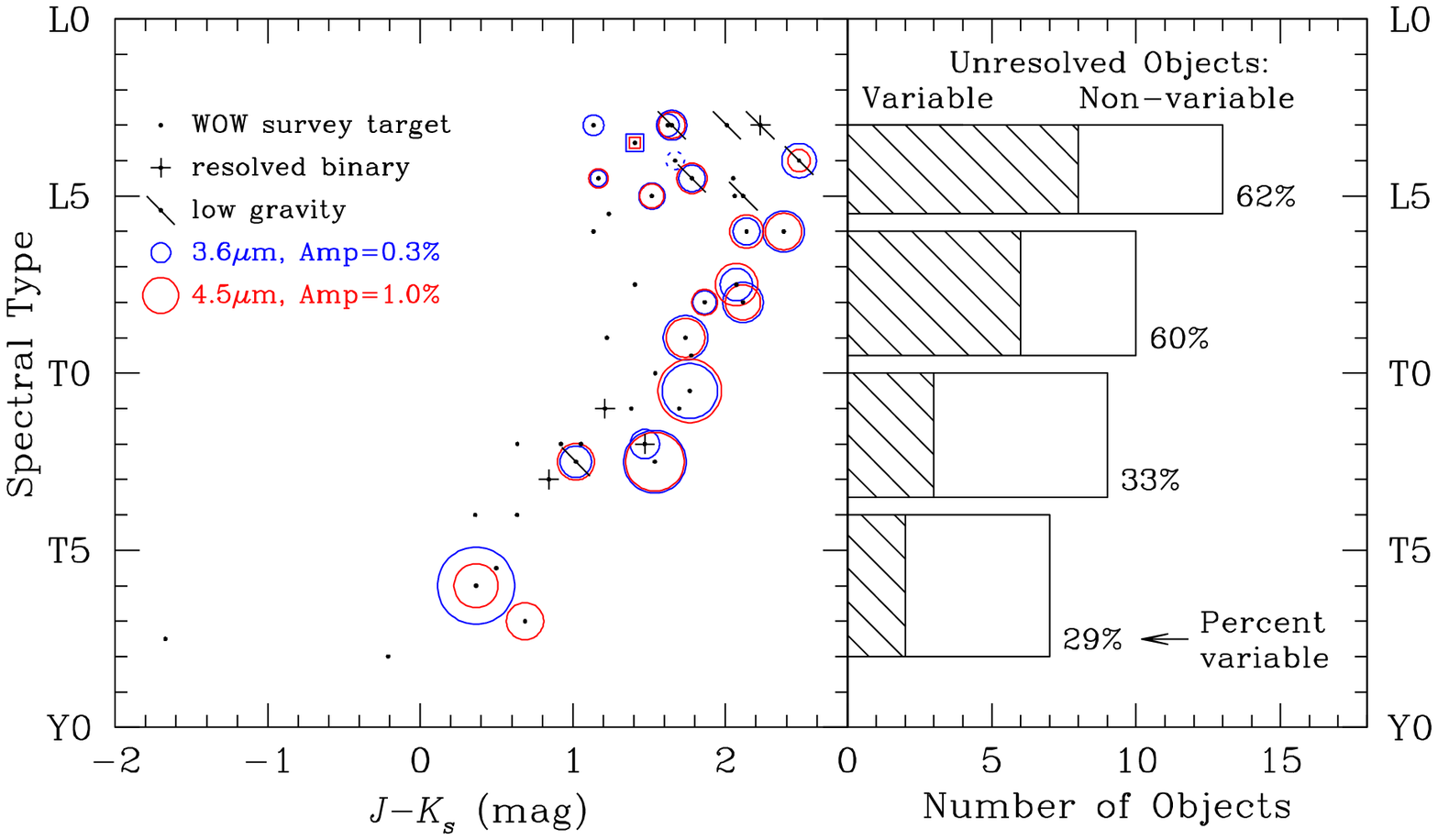}[hb]
\figcaption{\footnotesize {\bf Left:} Color, spectral type, and variability distribution of our 44 L3--T8 targets.  Circles enclose the variable targets, with the area of the circle proportional to the variability amplitude in the IRAC [3.6] band (blue) or [4.5] band (red).   The dashed blue circle encloses object 2MASS J175334518--6559559 (L4), which displays only a linear trend at [3.6], and does not have a well defined amplitude.  The previously known magnetically active L3.5 dwarf 2MASSW J0036159+182110
is variable and shown with concentric squares.  Known tight binaries are marked with $+$, and are plotted at their systemic spectral type and color.  Inclined bars denote low-gravity objects, including six L3--L5 dwarfs (one a close binary) and the T2.5 dwarf HN~Peg~B.  
{\bf Right:} Distribution and frequency of [3.6] or [4.5] variability of 
the 39 objects in our unresolved sample, excluding the previously known magnetically active L3.5 variable 2MASSW J0036159+182110.
\label{fig_spt_jk_hist}}
\end{figure}

Our {\it Spitzer} program detected 21 variables at $>$95\% confidence among 44 L3--T8 targets, including 19 variables among the 39 in the unresolved sample.  Seventeen of these are newly-detected variables.  The spectral type and $J-K_s$ color distribution of our sample, including both variables and non-variables, is shown in Figure~\ref{fig_spt_jk_hist}.  In the following we present the key results on L and T dwarf variability from the program.

\subsection{Variability Is Observed throughout the L3--T8 Spectral Type Range
\label{sec_var_spt}}

We detect photometric variations at virtually all spectral subtypes, with the warmest variables being L3's, and the latest a T7.  Variability is detected twice as frequently among L3--L9.5 dwarfs, with 14 out of 23 ($61\%_{-20\%}^{+17\%}$, 95\% binomial confidence interval) L dwarfs being variable, than among T0--T8 dwarfs, where 5 out of 16 ($31\%_{-17\%}^{+25\%}$) are variable (Fig.~\ref{fig_spt_jk_hist}).  The inclusion of the four binaries, one of which is variable, does not affect these results significantly.


The lower fraction of detected variables among the T dwarfs is fully consistent with the decreasing apparent brightness of cooler objects in our sample (Sec.~\ref{sec_nonvariables}): our average L dwarf is 1.8~mag brighter at [3.6] than our average T dwarf.   While T dwarfs have redder $[3.6]-[4.5]$ colors than L dwarfs, that does not compensate for their relative faintness and the $\approx$0.2~mag poorer photometric precision at [4.5] compared to [3.6] (Fig.~\ref{fig_photometric_precision}).  An incompleteness-corrected estimate of the fraction of variable L and T dwarfs is discussed in Section~\ref{sec_spots_ubiquitous}.

\begin{figure}
\epsscale{0.7}
\plotone{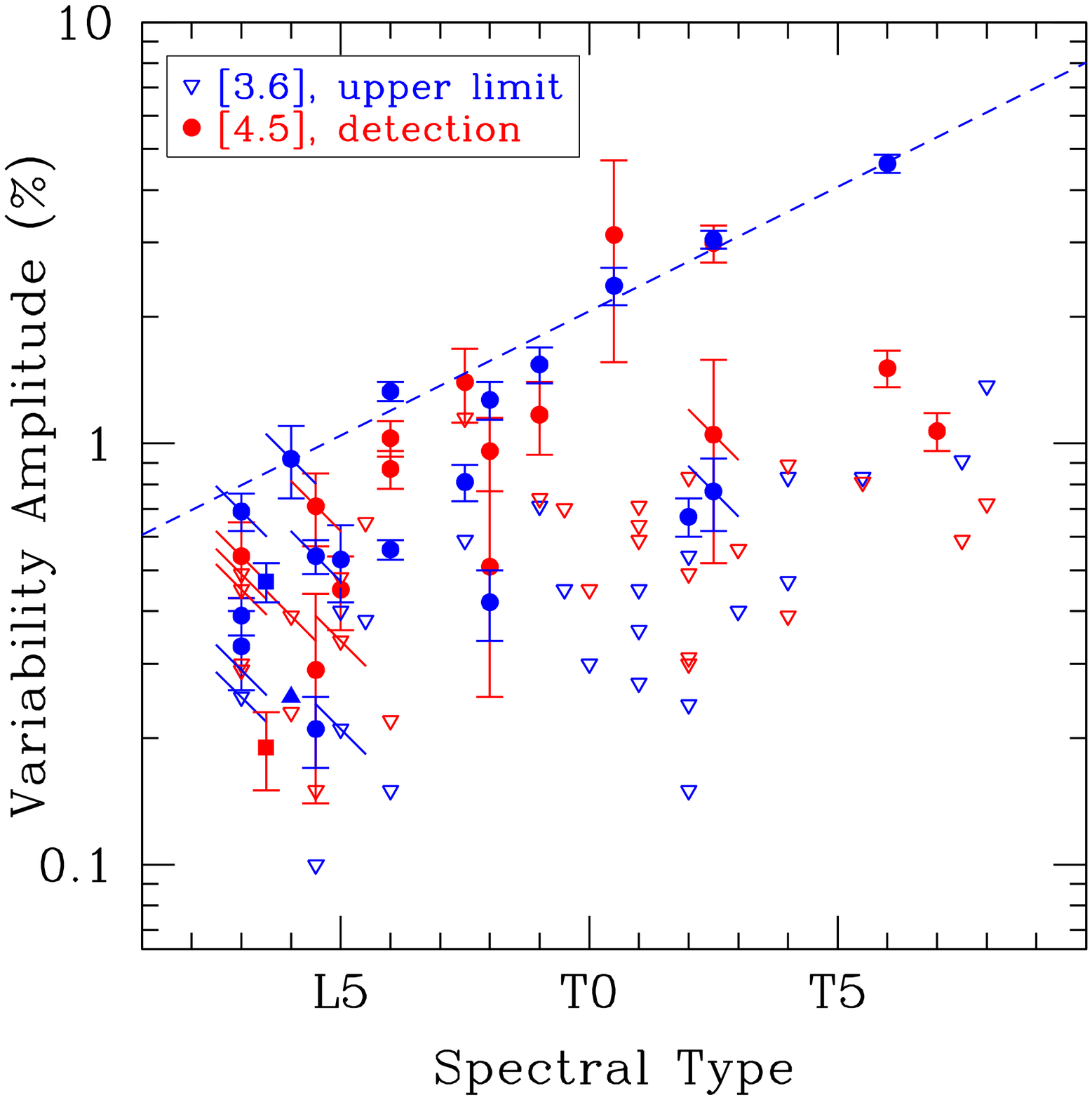}
\figcaption{\footnotesize Variability amplitude as a function of spectral type.  Blue symbols represent [3.6] data while red symbols show [4.5] data.  Open downward-pointing triangles show 95\% confidence upper limits on the amplitudes of non-variables. 
Inclined bars denote low or moderately low gravity objects.
The filled blue upward-pointing triangle marks the lower limit on the [3.6] amplitude of 2MASS J175334518--6559559 (L4), which shows only a linear trend in channel 1.  The solid squares mark the [3.6] and [4.5] amplitudes of the deliberately added known variable 2MASSW J0036159+182110 (L3.5).  The blue dashed line is a fit to the upper envelope of [3.6]-band amplitudes, using the eight closest [3.6] amplitude measurements. 
\label{fig_Avsspt}}
\end{figure}

Similarly to RLJ14 and R14, we find large-amplitude ($>$2\%) variables near the L/T transition: in the L9--T3.5 spectral type range.  
Two of our three $>$2\% amplitude variables are at the L/T transition.  However, we find that they are not exceptional in the context of the overall variability frequency or amplitude distribution  
(Sec.~\ref{sec_maximum_amplitudes}; Fig.~\ref{fig_Avsspt}).  A further comparison between the findings at near-IR wavelengths and our 3--5~$\micron$ {\it Spitzer} results is rendered in Section~\ref{sec_comparison}.

\subsection{The Maximum Variability Amplitude Steadily Increases from L to T Dwarfs}
\label{sec_maximum_amplitudes}

The amplitudes from the independent fits to the channel 1 and 2 light curves range between 0.2\%--4.6\% in the [3.6] band and between 0.2\%--3.2\% in the [4.5] band (Fig.~\ref{fig_Avsspt}).  Four objects vary significantly only at [3.6], for three of which we are able to fit [4.5] amplitudes if constraining the fit jointly with [3.6]. These channel 2 amplitudes range between 0.1\%--0.3\%.  However, because they can not be confirmed independently from the [4.5] data, we do not consider them in our amplitude distribution analysis.
Only one object, 2MASS J00501994--3322402 (T7), is observed to vary significantly only in channel 2, with an upper limit on the channel 1 amplitude, and so a lower limit on the [4.5]/[3.6] ratio.   

An interesting result that emerges from our survey is that the maximum variability amplitude in either {\it Spitzer} IRAC channel steadily increases over the L and T spectral range.  The expression
\begin{equation}
\log (A[3.6]_{\rm max}) = (0.059\pm0.002)\times {\rm SpT} - 0.28\pm0.03,
\label{eqn_Amax_spt}
\end{equation}
where SpT = 0 at L0 and SpT=18 at T8, represents the upper [3.6]-band variability envelope well, with the line fit to the eight nearest [3.6]-band amplitudes in Figure~\ref{fig_Avsspt}.  
The trend is well supported in the L3--T2.5 range, which contains seven of the eight defining data points.  The two highest-amplitude objects in the L3--L5.5 bin have low surface gravities that may have enhanced their variations (Sec.~\ref{sec_low_gravity}).  Regardless, this does not alter the observation that the maximum amplitudes in the L3--L5.5 bin are smaller than in any of the later-type bins.

The projection of an increasing maximum variability amplitude beyond spectral type T3 is more speculative, as in that range it is substantiated by only a single data point: the [3.6]-band amplitude of the T6 dwarf 2MASS J22282889--4310262. The trend is also not confirmed in the [4.5]-band amplitudes of $>$T3 dwarfs. Nonetheless, we note that 2MASS J22282889--4310262 represents half of the variability detections in the T4--T8 bin, so its high [3.6]-band amplitude may not be entirely random.  Besides, the trend only marks the maximum observed amplitude, rather than typical amplitudes.  It does not imply that late-T dwarfs or even Y dwarfs will generally have such large amplitudes, but only that increasingly larger amplitudes are possible at cooler effective temperatures.  Incidentally, this agrees with the $\sim$20\% projected integrated variability of Jupiter at 4.78~$\micron$ \citep{gelino_marley00}.  Overall, the trend for increasing maximum amplitudes at later spectral types 
indicates a propensity for greater brightness contrasts than in warmer brown dwarfs.  


Conversely, the lack of large-amplitude variations in the early-L dwarfs points to greater homogeneity in the appearance of their 3--5~$\micron$ photospheres.  One one hand, this could be caused by smaller temperature differences associated with multiple molecular species condensing at slightly different temperatures and forming multiple cloud decks.  On the other hand, a constant or a slowly varying temperature differential with spectral type, e.g., 
as a result of a temperature perturbation \citep{robinson_marley14} from atmospheric wave breaking \citep[e.g.,][]{young_etal97}, would also lead to smaller flux variations at earlier spectral types, since the temperature perturbation would be smaller in a relative sense.

\subsection{Amplitude Ratios over 3--5 $\micron$ Are Not Correlated with Spectral Type}
\label{sec_amplitude_ratios}

Measuring the dependence of the [4.5]/[3.6] amplitude ratio on spectral type was one of the main goals of our {\it Weather on Other Worlds} program.  Given the strong wavelength dependence of the brightness temperatures of molecule-rich ultra-cool atmospheres, the amplitude ratio can be used as a probe of the temperature gradient among cloud layers or between regions of thick and thin clouds.  In \citet{heinze_etal13} we argued that a ratio of $A[4.5]/A[3.6]<1$ for the variable L3 dwarf DENIS-P J1058.7--1548 indicated fractional coverage by warm spots.  We concluded that we were most likely observing a two-component $\Delta T\sim 100$~K cloud deck with holes in the upper deck revealing the warmer deck underneath.

\begin{figure}
\plotone{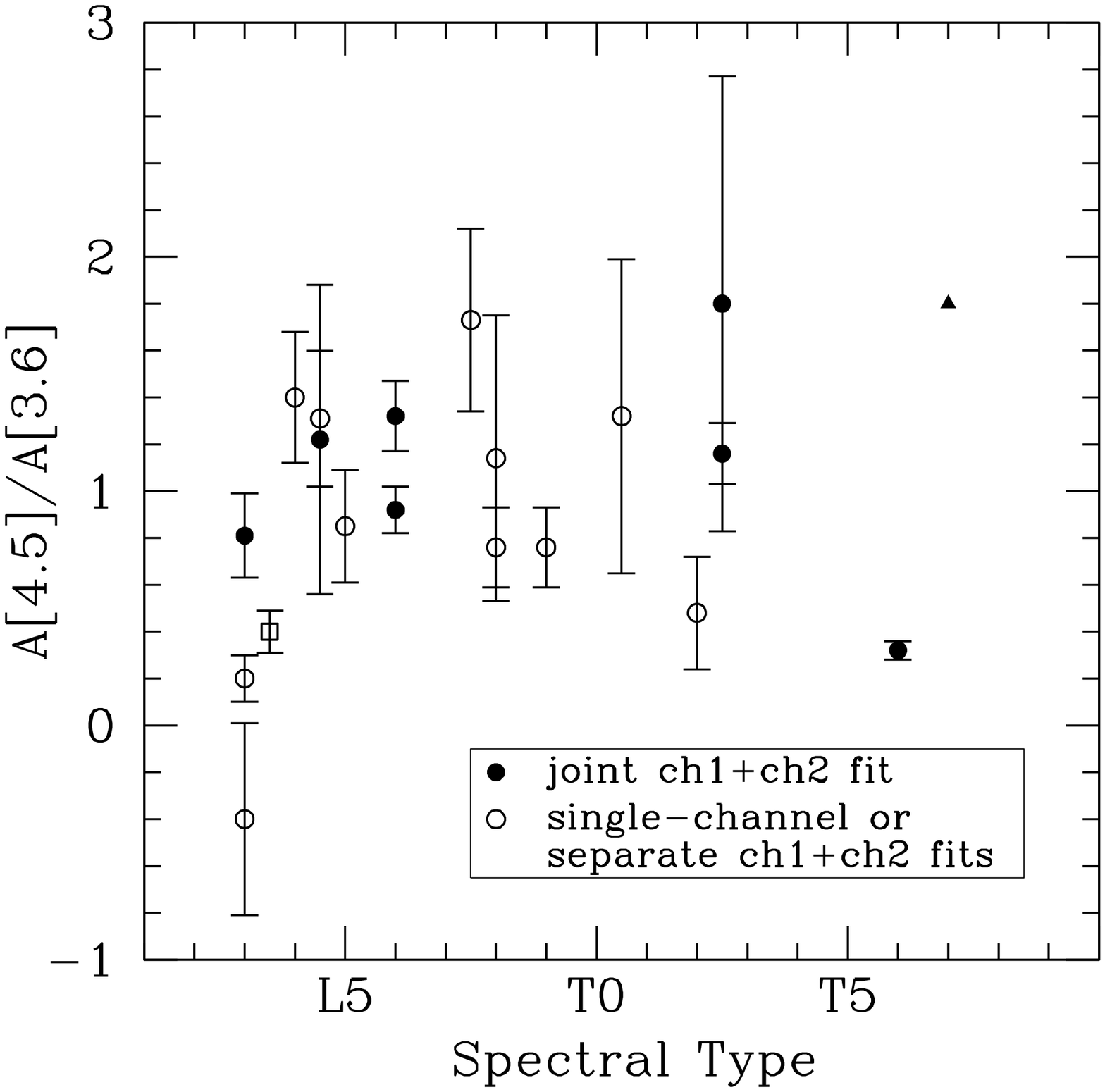}
\figcaption{\footnotesize Variability amplitude [4.5]/[3.6] ratios as a function of spectral type.  Solid symbols show reliably estimated amplitude ratios of objects with periodic variations in both bands, obtained from simultaneous fits to the [3.6] and [4.5] data under the constraints of identical period and phase.  Open symbols denote cases where a simultaneous fit was not possible because of irregular or long-term variations (Note 1 in Table~\ref{tab_results}), or when variability was only detected at [3.6] (Note 3 in Table~\ref{tab_results}).  The amplitude ratios in these cases may not be representative of the true amplitude ratios under simultaneous or more sensitive observations.
The amplitude ratio of the magnetically active irregular variable 2MASSW J0036159+182110 (L3.5) is shown with an open square.  A lower limit on the amplitude ratio of the [4.5]-only regular variable 2MASS J00501994--3322402 (T7) is shown with a solid upward pointing triangle.
\label{fig_amplitude_ratios}}
\end{figure}

Figure~\ref{fig_amplitude_ratios} shows the [4.5]/[3.6] amplitude ratios from Table~\ref{tab_results} for most of the variable objects in our survey as a function of spectral type.  Only one object has been excluded from this analysis, 2MASS J17534518--6559559 (L4), for which the observed variability is solely a linear trend in channel 1.  We find no obvious correlation, except only that all three variable L3--L3.5 dwarfs in the unresolved sample, and the known magnetically active L3.5 dwarf 2MASSW J0036159+182110, have [4.5]/[3.6] amplitude ratios below unity, like DENIS-P J1058.7--1548.  Formally, the mean [4.5]/[3.6] amplitude ratio over the L3--T8 domain is 1.0, with a standard deviation of 0.7.  In the context of cloudy models, this suggests that either small warm holes in dominant cold cloud decks or small cold patches of high-altitude clouds above a prevailing warm photosphere are equally likely on L and T dwarfs.

\subsubsection{The Variable Component of the L5.5 + T5 Binary SDSS~J151114.66+060742.9 Is the Primary}
\label{sec_sdss1511}

The consideration of [4.5]/[3.6] amplitude ratios is appropriate for discerning which of the two components in the close L5.5 + T5 binary SDSS J151114.66+060742.9 varies.  While it may be logical to assume that the brighter component is responsible for the observed variability, in the context of increasing maximum amplitude with spectral type (Sec.~\ref{sec_maximum_amplitudes}), it is worth considering whether the cooler secondary may have unusually large amplitude that drives the combined flux variations.

There are two arguments that favor variability in the brighter component.  First, given typical 3--5~$\micron$ absolute magnitudes and colors of L5 and T5 dwarfs from {\it WISE} \citep{kirkpatrick_etal11}, the secondary is $\approx$1.5~mag redder in $[3.6]-[4.5]$ than the primary and also $\approx$1.5~mag fainter at [4.5].  Being altogether
$\approx$3.0~mag fainter than the primary at [3.6], the secondary would have to vary by $\sim$10\% to account for the observed 0.67\% [3.6]-band amplitude in combined light.  No 3--5~$\micron$ amplitudes this high are observed in any of the other variable L or T dwarfs in our sample.  Such large-amplitude variability seems to so far be contained only to shorter wavelengths \citep{radigan_etal12, gillon_etal13, heinze_metchev15}.

Second, no similarly large [4.5]-band amplitude could be deduced for the T5 secondary, even if a [4.5]-band detection is favored given the red $[3.6]-[4.5]$ colors of mid-T dwarfs.  Rather, SDSS J151114.66+060742.9 is one of our [3.6]-only variables, although a simultaneous fit to the [3.6]- and [4.5]-band light curves gives a [4.5]/[3.6] amplitude ratio of $0.5\pm0.2$ in combined light.  Because of the $[3.6]-[4.5]$ color differential between the primary and the secondary, should the variability be originating only from the secondary, its actual [4.5]/[3.6] amplitude ratio would be only $\sim$0.15. This again contravenes the behavior of the other five variable T dwarfs, all of which show significant [4.5]-band variations, with [4.5]/[3.6] amplitude ratios $>$0.3.  Conversely, if the variability arose from the L5.5 primary, which dominates the total flux, the 0.5 amplitude ratio in combined light would be normal for a dwarf in the L3--L5.5 bin.

We therefore conclude that the observed variability in the close L5.5 + T5 binary SDSS J151114.66+060742.9 likely originates from the brighter L5.5 component.  We have nonetheless retained the systemic T2 spectral type for plotting purposes in Figures~\ref{fig_spt_jk_hist}--\ref{fig_p_spt} until a resolved spectroscopic characterization of the binary is available \citep{bardalez_gagliuffi_etal15}.


\subsection{Irregular Variables Are Common among L Dwarfs
\label{sec_irregular}}

Our sample contains seven irregular or irregular/long variables, including the known magnetically active L3.5 dwarf 2MASSW J0036159+182110.  Six of these seven irregular variables are L dwarfs, and the seventh is the T0.5 dwarf SDSS J151643.01+305344.4.  Among the 14 variable L dwarfs in the unresolved sample, five are irregular.  None of the irregular variables are known or candidate close binaries.  Hence, unresolved multiplicity within the {\it Spitzer} PSF can not account for the large number of Fourier terms required to fit their light curves.

A possible reason for the high incidence of irregular variability in L dwarfs is that we are detecting rapid changes in the distribution of photospheric spots, potentially across multiple cloud layers.  The light curve of the L5 dwarf 2MASSW J1507476--162738 (Fig.~\ref{fig_variable_curves}) provides a clear example of spot evolution: an oscillation appears around 7 hours into the channel 1 observing sequence, and continuously grows in amplitude until the end of the channel 2 sequence, five 2.5~h rotations later.  The behavior of the L4.5 dwarf 2MASS J18212815+1414010 is similar, on a longer, 4.2~h period.  

We will analyze the properties of the irregular variables in more detail in upcoming publications \citep{flateau_etal15, heinze_etal15}.  At present, we only note that the almost exclusive appearance of irregular variables among the L dwarfs points to more complex and rapidly evolving spot configurations in $\gtrsim$1400~K atmospheres.  This may be an indication that the ratio of the convective-overturn time scale to the spin period is smaller in L dwarfs than in T dwarfs \citep{zhang_showman14}.  Given that the set of irregular variables includes the previously known magnetically active L3.5 dwarf 2MASSW J0036159+182110, it is also possible that we are witnessing low-level magnetic activity, or a combination of cloud- and magnetically-induced photometric variations.

\subsection{Low-Gravity 
L3--L5.5 Variables May Have Enhanced Amplitudes
\label{sec_var_gravity}}

Our unresolved sample contains six individual objects that have been characterized as low- or moderately low-surface gravity dwarfs (Sec.~\ref{sec_low_gravity}).  Five of these are in the L3--L5.5 bin, and the remaining is the $\sim$500~Myr-old T2.5 dwarf HN~Peg~B.  The tight L3 + L5 binary SDSSp J224953.45+004404.2---not part of the unresolved sample---also has low surface gravity \citep{allers_etal10}: for a total of eight low-gravity objects in our complete sample.  

The variability fraction among the putative low-gravity L3--L5.5 dwarfs is 3/7 or 3/5, depending on whether the individual components of the non-varying binary SDSSp J224953.45+004404.2 are counted separately, or whether it is altogether excluded from the sample.  
Within the statistical uncertainties, this is indistinguishable from the fraction of variables among the high-gravity objects in the L3--L5.5 bin: 5/8.  HN Peg B is one of three single variable dwarfs in the T0--T3.5 bin.  That is, variability among T0--T3.5 dwarfs is detected both in moderate- and in high-gravity objects.  Combining the results for the L3--L5.5 and the T0--T3.5 bins, we do not see an enhanced variability frequency among low-gravity objects, although our sample is too small to confidently exclude a correlation.

Instead, we do detect a tentative correspondence between amplitude and surface gravity among the set of eight L3--L5.5 dwarfs that are variable.  The three L3--L5.5 variables that show signatures of low gravity also have the highest [3.6]-band amplitudes in the L3--L5.5 bin (Fig.~\ref{fig_Avsspt}): 2MASS J16154255+4953211 (L4$\beta$), 2MASSW J2208136+292121 (L3$\gamma$), and 2MASS J18212815+1414010 (L4.5).  The latter two objects have the highest bin amplitudes also at [4.5].  We only provide an upper limit to the [4.5]-band variability of 2MASS J16154255+4953211 because its inferred period ($\sim$24~h) is much longer than the 7~h channel 2 AOR.  In reality, the combined channel $1+2$ light curve fit gives a [4.5]/[3.6] amplitude ratio of 1.4 (Table~\ref{tab_results}), which would make 2MASS J16154255+4953211 the strongest [4.5]-band L3--L5.5 variable, and all three low-gravity L3--L5.5 variables would have the highest amplitudes also at [4.5].

A consideration of all possible ways to choose three objects from eight shows that the three low-gravity variables would have the highest amplitudes among the eight variables in the L3--L5.5 bin in $\frac{3}{8}\frac{2}{7}\frac{1}{6}=1.8\%$ of cases.  That is, the result might appear 98.2\% significant.  

More generally, we would have likely considered any outcome that includes the amplitudes of the three low-gravity L3--L5.5 dwarfs among the top half in the bin.  
We also need to incorporate the four low-gravity L3--L5.5 dwarfs---including the individual near-equal flux components of the L3 + L5 binary SDSSp J224953.45+004404.2---that are not detected as variables.  Otherwise, the exclusion of censored data could bias our conclusion.  We test the significance of the result by combining all detections and non-detections in a Monte Carlo approach (see Section~\ref{sec_monte_carlo}).  To account for the diminished sensitivity to variations from either of the components of the L3 + L5 binary, we count it as two individual objects that are half as bright.  We consider as positive any outcome that includes at least three detected low-gravity L3--L5.5 variables, with [3.6] or [4.5] amplitudes all in the top half of the L3--L5.5 bin.  
We find that this scenario arises at random in 8\% of our simulations.  That is, the association between low surface gravity and enhanced variability amplitude is 92\% significant.

In arriving at the above conclusion, we have assumed that 2MASS J18212815+1414010 (L4.5) has low surface gravity.  However, as already discussed in Section~\ref{sec_low_gravity}, while a moderately low surface gravity is the favored explanation for its spectroscopic appearance and galactic space motion, it is not unique.
If we exclude 2MASS J18212815+1414010 from the above analysis, the association between low surface gravity and enhanced variability amplitude would not be significant. 

In summary, while we can not conclude that low surface gravity leads to higher incidence of detectable variability, we find that low gravity 
may be correlated with higher 3--5~$\micron$ amplitudes among variable L3--L5.5 dwarfs.







\subsection{L and T Dwarf Periods Range from 1 h to $>$20 h
\label{sec_periods}}
A natural by-product of the {\it Weather on Other Worlds} program is the determination of rotation periods for L and T dwarfs.  Indeed, the program is the most sensitive campaign to measure L and T dwarf rotations.

Our survey was designed to cover at least two $<$10~h rotation periods per object.  The 10~h upper limit was partly motivated by the lack of $v\sin i<10$~km~s$^{-1}$ measurements among L and T dwarfs \citep[e.g.,][]{bailer_jones04, blake_etal10}, which imply spin periods of $<$12~h for one Jupiter-radius objects.
Separately, all L and T dwarf photometric periods measured in high-cadence, intensive monitoring campaigns have been shorter than 9~h (\citealt{clarke_etal02, clarke_etal08, koen_etal05, artigau_etal09}, RLJ14).  However, we note that both the $v \sin i\gtrsim10$~km~s$^{-1}$ and the $P\lesssim$9~h constraints from previous surveys may well be selection effects: either related to the maximum resolving power ($R\sim30000$) of sensitive near-IR spectrographs (e.g., NIRSPEC on Keck), or to the diurnal cycle.  While some variations with time scales $>$10~h have been reported in \citet{bailer-jones_mundt01} and \citet{gelino_etal02}, the sparse sampling of the light curves in these observations---once a night for $\sim$1~h over several nights---leaves a high probability that the detected frequencies may be aliases of shorter periods or even that the variability may be spurious.  For example, \citet{bailer-jones_mundt01} note that despite a peak in the periodogram of the L5 dwarf SDSSp J053951.99--005902.0 at 13.3~h, no clear pattern is seen in the object's light curve.


Our observations are most sensitive to periods shorter than 7~h---half of the 14~h channel 1 sequence---most of which we measure to 5\% accuracy on the period or better.  However, the 21~h in continuous channel $1+2$ observations permit a probe of much longer periods for the first time.

Between six and nine of our 20 L3--T8 variables, i.e., approximately a third, have $>$10~h periodicities: a result uniquely enabled by our long uniterrupted observations.  This set of objects comprises the six long-period variables, and possibly some of the three irregular/long-period variables identified in Section~\ref{sec_variable_classification}.  
In the three most extreme cases, the light curves follow only slowly changing trends in our 14~h [3.6] AORs (2MASS J16154255+4953211 [L4$\beta$], 2MASS 175334518--6559559 [L4], and 2MASS J21481628+4003593 [L6]; Fig.~\ref{fig_variable_curves}).  
The case of 2MASS J175334518--6559559 is particularly unusual, as all that we observe over the entire 14~h [3.6] sequence is a linear trend.  The estimated $>$50~h time scale for the trend
is very uncertain.   The unusually long time scale suggests a variability mechanism other than rotation.  As we surmised in Section~\ref{sec_variable_classification}, we may be observing the effect of cloud evolution on an object that is vewied nearly pole-on.

\begin{figure}
\plotone{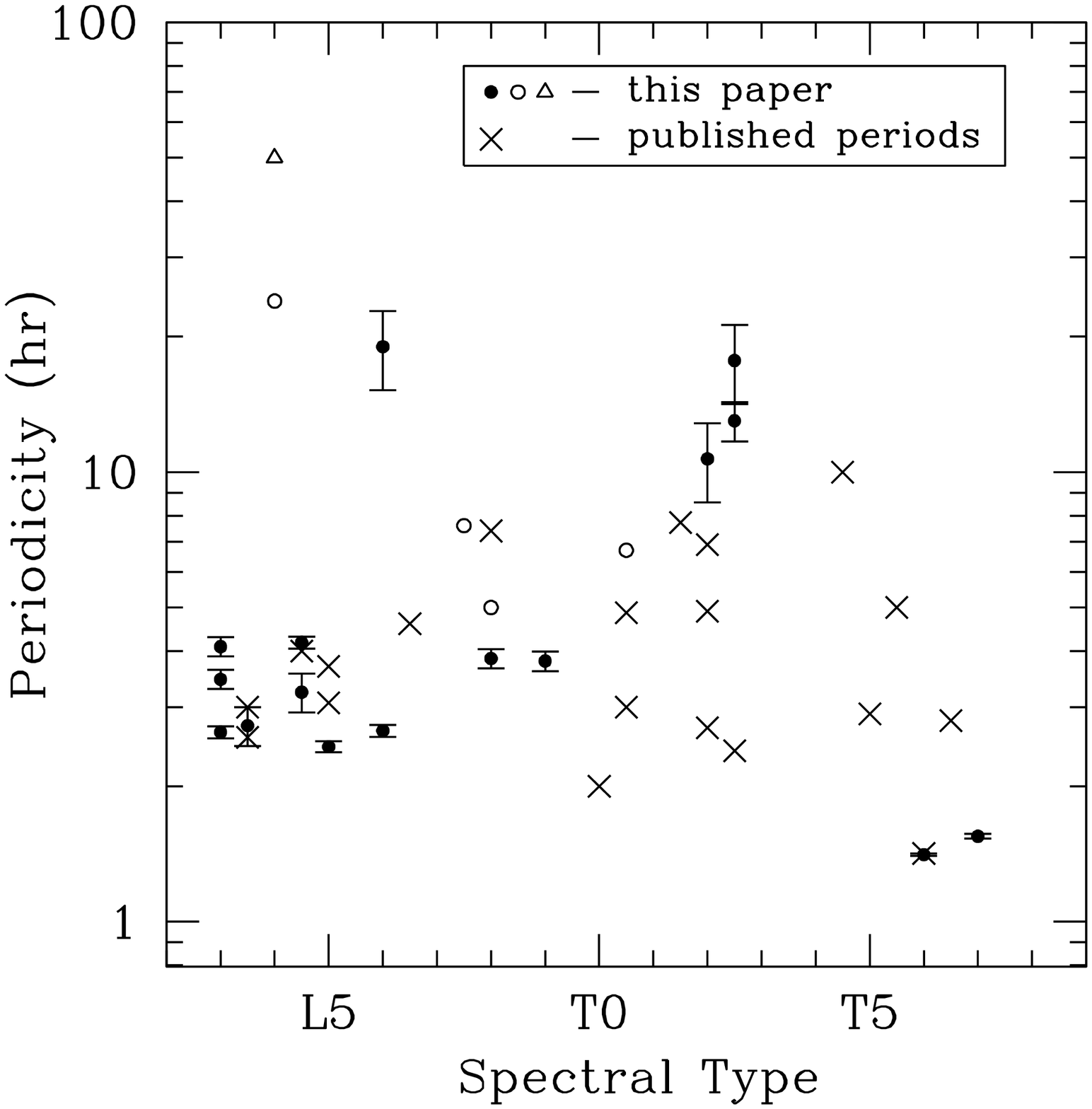}
\figcaption{\footnotesize Estimated period as a function of spectral type.  Solid circles with errorbars show objects with well-determined periodicities, for which we believe that we have the rotation period.  Open symbols are objects with unreliable periods, with uncertainties of $\geq$50\%.  
An upward-facing triangle denotes the 50-hr lower limit on the periodicity of 2MASS 175334518--6559559 (L4).  
Reliably measured L3--T8 dwarf rotation periods from 
\citet{koen04, koen13b, berger_etal05, clarke_etal08, artigau_etal09, gillon_etal13, girardin_etal13}, and RLJ14 are shown with the `$\times$' symbol.  We have not included all L3--T8 periods compiled in \citet{crossfield14}, as these contain variables that have not withstood independent confirmation. 
\label{fig_p_spt}}
\end{figure}

The periods of our variables are shown as a function of spectral type in Figure~\ref{fig_p_spt}.  Sixteen of our 21 variables have reliably determined rotation periods, including all eight regular, four irregular, and four of the long-period variables with high-SNR variations and $<$21~h periods.
The periodic variables are shown as solid symbols with errorbars on Figure~\ref{fig_p_spt}.  Thirteen of the periods are for objects that are newly discovered to be variable.  We also confirm the previously established periods for  2MASSW J0036159+182110 \citep[L3.5;][]{berger_etal05}, 2MASS J22282889--4310262 \citep[T6;][]{clarke_etal08, buenzli_etal12}, and 2MASS J11263991--5003550 (L4.5; RLJ14).  
We note that because of the limited time span of our observations, we technically can not exclude longer rotation periods for some of these objects.  This is relevant especially to the irregular and to the long-period variables, even though much longer periods for the latter would be even more surprising.  Our analysis does exclude any significant power at shorter periods for all of our objects.  Hence, our periods and their quoted uncertainties may be strictly regarded as one-sided error bars giving lower, but not upper, limits to the periods.

The variability time scales of objects with uncertain periods, including the two longest-period variables and the three irregular/long-period variables, are shown either with open symbols without error bars, or as an upward pointing triangle (for 2MASS J175334518--6559559 [L4]) in Figure~\ref{fig_p_spt}.  
We retain $>$95\% confidence in the existence of variability in these objects by comparison to the pool of $>$600 reference stars taken at random from the entire program (Section~\ref{sec_variable_identification}).  However, we are unable to constrain the periods of these variables to better than a factor of $\sim$2.

Altogether, we have doubled the number of L3--T8 dwarfs with reliably measured rotation periods.
In addition, our newly discovered L4--T2.5 population of slow (or pole-on) substellar rotators is ideal for establishing a grid of high-dispersion low-$v\sin i$ substellar standards.

\section{THE OCCURRENCE OF SPOTS ON L AND T DWARFS
\label{sec_spots}}

We combine the results on the spectral type distribution of L3--T8 variables with the limits on photometric sensitivity, and use Monte Carlo simulations (Sec.~\ref{sec_monte_carlo}) to estimate the overall fraction of spotted L3--T8 dwarfs (Section~\ref{sec_spots_ubiquitous}).  We discuss our results in the context of published near-IR surveys in Section~\ref{sec_comparison}.

\subsection{Correcting for Incompleteness with Monte Carlo Simulations
\label{sec_monte_carlo}}

We simulate the effect of our photometric precision limits on detecting [3.6]- and [4.5]-band variability as a function of target brightness.  Our main assumptions are that: (1) the maximum [3.6]-band variability amplitude increases monotonically as described by Equation~\ref{eqn_Amax_spt} and evidenced in Figure~\ref{fig_Avsspt}, and (2) small-amplitude variations are likely at all spectral types.  Several independent factors support the second assumption.  First, we already noted that there is no significant empty phase space between most variability detections and the majority of non-detections on the amplitude vs.\ magnitude diagrams in Figure~\ref{fig_Avsmag} (Sec.~\ref{sec_nonvariables}).   Second, the distribution of the logarithm of the [3.6]-band amplitudes ($\log A[3.6]$) 
on Figure~\ref{fig_Avsmag} is approximately uniform at each spectral type bin.  Finally, the fraction of variables toward later spectral type bins decreases along with the relative decrease in $\log(A)$-magnitude phase space above the detection limits in Figure~\ref{fig_Avsmag}.  Hence, we conclude that low-amplitude variables likely exist even at late spectral types, where they may have been below our sensitivity threshold. 

Figure~\ref{fig_Avsspt} offers an independent assessment of the assumed amplitude vs.\ spectral type relation.  The approximately uniform distribution in $\log A[3.6]$ is again evident in the L dwarfs in either of the L3--L5.5 or L6--L9.5 spectral type bins.  A gap at $\sim$2\% amplitudes might exist among the T dwarfs, but with only six T variables (five at [3.6] and a different set of five at [4.5]), the existence of such a gap is not significant in either of the IRAC bands.  
Overall, the data are consistent with an increase in the number of variables toward lower amplitudes at a fixed spectral type.  An inverse proportionality in the frequency of variables as a function of amplitude at a fixed spectral type is the lowest order approximation of this trend.  Our sample statistics are insufficient to seek a higher-order description.

The bright targets in our sample are L dwarfs, among which we observe that the presence of low-amplitude variations is independent of spectral subtype.  Therefore, we treat all L dwarfs the same, and correct for incompleteness down to the lowest detected amplitudes: 0.2\%.  We can not confirm whether low-amplitude variations exist throughout the T spectral type because of poorer sensitivity.  We therefore limit our incompleteness correction in the T dwarfs to higher variability amplitudes: $>$0.4\% at [3.6] or [4.5].  This threshold corresponds to the 95\% upper limits for approximately half of our T dwarfs.

Our Monte Carlo simulations aim to reproduce the observed variability characteristics of the unresolved sample of 39 objects---23 L dwarfs and 16 T dwarfs---in our {\it Weather on Other Worlds} program.  We use the ensemble detection rates from Figure~\ref{fig_Avsmag} (dashed lines) to gauge whether a simulated variable is detected.   We set the probability of detecting a simulated variable from the fraction of observed variables in the corresponding amplitude vs.\ magnitude phase space between the detection rate curves.  For example, if a target's simulated [3.6]-band magnitude and variability amplitude fall in the 20\% detection rate band (left panel of Figure~\ref{fig_Avsmag}), the simulated target is given a 20\% chance of being detected as a variable.  Simulated L3--T1 variables are given a finite probability of being irregular, in which case the more conservative, non-sinusoidal upper limits discussed in Section~\ref{sec_nonvariables} apply.  


We note that the detection rates may slightly underestimate the actual completeness in the intermediate (20\% or 23\%) or non-detection (0\%) regions, inasmuch as not all of the upper limits track the upper boundaries of these regions.  That is, it is possible for a simulated variable to have an amplitude slightly larger than the upper limit for the corresponding sample target, and that it yet falls below the 100\% completeness region.  Such a variable would be given the nominal completeness-dependent probability of being detected, i.e., 0\% or 20\%/23\%, when it should have been much more readily detected if its amplitude was above the 95\% upper limit of the sample target.  Nonetheless, we note that the 95\% upper limits themselves are derived under object-specific variability assumptions, and that various data-dependent factors induce a scatter in the upper limits at similar object magnitudes (Sec.~\ref{sec_nonvariables}).  Conversely, the ensemble detection rates are expected to be representative of our magnitude-dependent completeness, under the assumption of a continuous distribution of amplitudes.  Hence, we retain the 100\%, 20\%/23\%, and 0\%  detection rates as actual completeness estimates for our simulations.


The [3.6] and [4.5] magnitudes of the simulated variables are allowed to vary in a Gaussian fashion within 1$\sigma$ errors of 0.05~mag.  Spectral types are assumed to be uncertain by 0.5 (1$\sigma$) subtypes.  As we find no clear trend in the [4.5]/[3.6] amplitude ratio vs.\ spectral type (Sec.~\ref{sec_amplitude_ratios}; Fig.~\ref{fig_amplitude_ratios}), we adopt a mean [4.5]/[3.6] amplitude ratio of 1.0 with a Gaussian standard deviation of 0.7, based on the available data.


The input rates of variability in each of the L and T spectral types and the fraction of irregular L3--T1 dwarfs were treated as free parameters that were adjusted until the simulations matched the detected variability frequencies: 14/23 (61\%) in the L dwarfs and 5/16 (31\%) in the T dwarfs, with 6/15 (40\%) of L3--T1 variables being irregular.  Our observational results offer additional validation checks, such as the fraction of single-band variables (3/19 at [3.6] and 1/19 at [4.5]), the fraction of high-amplitude ($>$1\%) variables at each band (6/18 at [3.6], 8/15 at [4.5]) and at each spectral type (4/14 at L, 5/5 at T), etc.  The simulations were able to reproduce all of the observed properties of our sample to satisfactory approximation.  We found that the input fraction of irregular L3--T1 variables only weakly affects the outcome of the simulations, mostly because the L dwarfs in our sample are relatively bright, and the majority have variability amplitudes well above the 0.2\% minimum threshold.  Input L3--T1 irregular fractions near 50\% produced results that were most consistent with the various aspects of the data.

\subsection{Spots Are Ubiquitous on L3--T8 Dwarfs
\label{sec_spots_ubiquitous}}

Our Monte Carlo simulations show that 
80\% of L dwarfs are variable at $>$0.2\% between 3--5$\micron$, with a 95\% confidence interval on the variability fraction of 
53\%--100\%.
The result is fully consistent with the variability frequency of the brightest subset of our targets---all L dwarfs---for which we are nearly complete to 0.2\% [3.6]-band amplitudes: among the ten L dwarfs brighter than $[3.6]=12$~mag, eight are variable.  This agreement independently validates our incompleteness correction.  


The detection of spot-induced brightness variations depends on viewing geometry.  The median spin axis inclination of an object is $i=60^\circ$.  A single spot between 60$^\circ$--90$^\circ$ latitude will appear to rotate in and out of sight, and so cause significant rotationally-modulated variations, only when $i>60^\circ$: i.e., in less than half of the cases.  More generally, if spots can occur with equal probability per unit area anywhere on the surface of a brown dwarf, a spot will be always out of view on average in 11\% of cases, and always in view in another 11\% of cases.  We would not detect any variations from a brown dwarf with a single dominant spot in the former 11\% of cases. In the latter 11\% of cases, when a spot is permanently in view, variations may be detectable only if there is substantial periodic modulation of the visible cross-section of the spot, or if the spot itself varies in intensity.  Given such geometric incompleteness, our finding that 
80\% of L dwarfs are variable is fully consistent with all L dwarfs having spots.  Some spotted L dwarfs may simply not produce rotational variations because the spots are never visible or always in view.

The incompleteness-corrected variability fraction for T dwarfs at $>$0.4\% amplitudes is a factor of $\sim$2 lower: 
36\%, with a 95\% confidence interval of 19\% to 62\%.
For comparison, if we only consider L dwarfs with $>$0.4\% amplitudes, 
53\% are variable, with a 95\% confidence interval of 35\% to 69\%.
Overall, at amplitudes of $>$0.4\% the variability fractions among L and T dwarfs are consistent.

We have no reason to suspect that T dwarfs do not exhibit smaller, $<$0.4\% amplitude variations in the {\it Spitzer} bands.  The low-amplitude ($\lesssim$1\%) T dwarf detections from our program merge smoothly with the non-detections, suggesting continuity of amplitudes even below our detection limits.  In addition, the $\sim$50~min light curves of several of the T dwarfs in the B14 {\it HST} spectroscopic survey show very shallow gradients over certain wavelength ranges, while at other wavelength ranges the gradients are stronger.  That is, the B14 data indicate that low-amplitude 1.1--1.7~$\micron$ variations exist among the T dwarfs, too.  It is reasonable to assume that low amplitudes can extend to the 3--5~$\micron$ region, and that we have simply missed them because of our poorer sensitivity on T dwarfs.

Summarizing the above evidence, we conclude that spots are present on virtually 100\% of L3--L9.5 dwarfs, and probably also on most T0--T8 dwarfs.

\subsection{Comparison between the 3--5~$\micron$ and the 1.1--1.7~$\micron$ Variability Trends
\label{sec_comparison}}

Summarizing the relevant results from our {\it Spitzer} program: (1) after correcting for incompleteness, we find that 
$80\%_{-27\%}^{+20\%}$
of L3--L9.5 dwarfs vary with peak-to-peak amplitudes $>$0.2\% 
and 
and $36\%_{-17\%}^{+26\%}$ of T0--T8 dwarfs vary at $>$0.4\%, (2) there is no evidence for enhanced occurrence of variables at the L/T transition, and (3) there is a continuous trend of increasing maximum amplitude throughout the L3--T8 sequence. 

The result that variability-inducing spots are common has already been suggested from the two most sensitive 1.1--1.7~$\micron$ ground (RLJ14) and space-based (B14) surveys.  RLJ14 find that 7/41 (17\%) of their L4--T9 dwarfs vary, with an enhanced variability fraction, 5/16 (31\%), for spectral types between L9--T3.5.  After marginalizing over spin-axis orientations, RLJ14 find that 53\% of their L/T-transition (L9--T3.5) dwarfs would be variable with $>$2\% amplitudes at $J$ band.  Similar conclusions are echoed in R14's combined analysis of the RLJ14 and \citet{wilson_etal14} surveys. B14 correspondingly find that at least 27\% of L5--T6 dwarfs are variable, independent of spectral subtype, and estimate that the intrinsic variability rate may be as high as 50\%.  Our {\it Weather on Other Worlds} {\it Spitzer} program is more sensitive than either survey because of the factor of $\sim$10 better photometric precision compared to RLJ14 and the factor of $\sim$30 longer continuous on-target integrations than in B14.  Hence, our higher variability fractions are consistent with the previous findings, and conclusively demonstrate that spots are ubiquitous on L dwarfs, and probably also on T dwarfs.

The second result is in marginal disagreement with the findings for an enhanced fraction of large-amplitude ($>$2\%) $J$-band variables at the L/T transition by RLJ14 and R14.  While two of our three $>$2\% amplitude variables are at the L/T transition, the occurrence rate of such 3--5~$\micron$ amplitudes in the L9--T3.5 spectral type range is 2/15 (13\%): lower than, even if formally consistent with the 25\% and $24\%^{+11\%}_{-9\%}$ frequencies in RLJ14 and R14.  If we decreased the threshold defining a large amplitude in the {\it Spitzer} IRAC bands to 1\%, then the occurrence of $>$1\% variables between L9--T3.5 becomes 4/12, or 33\%: in closer agreement with the RLJ14 and R14 fundings for large-amplitude variables.  However, we note that L9--T3.5 dwarfs are not unusual as $>$1\% variables at 3--5~$\micron$ compared to L6--L8 dwarfs (3/7; 43\%) or T4--T8 dwarfs (2/7; 29\%).  That is, L/T-transition dwarf variability does not stand out at 3--5~$\micron$ as it does at $J$ band.

It is tempting to interpret this discrepancy in the context of cloud models, 
since different wavelengths probe different pressure levels and depths in opacity-dependent fashion.  The $J$-band flux arises from deeper regions with higher atmospheric pressures ($\sim$10~bar), while the [3.6]- and [4.5]-band flux originates on average at lower pressures ($\sim$1~bar) and higher altitudes.  Any difference between the two sets of results might suggest that $J$-band observations are more sensitive to silicate cloud break-up, expected to occur just above the 10~bar pressure level in L/T transition objects, while the 3--5~$\micron$ photometry is sensitive mostly to changes in the higher-altitude atmospheric structure.  Brightness temperature variations at these higher altitudes may also be affected by other processes, such as temperature fluctuations in the outermost convective layer of the atmosphere \citep{robinson_marley14, zhang_showman14}.

More generally, different wavelengths probe not only different pressure levels and temperatures, but a convolution of these factors with the cloud cover and the source function. Spectroscopic observations over a larger set of distinct wavelengths may therefore reveal a more nuanced picture.  Thus, the smaller (22-object) {\it HST} survey of B14 finds a more uniform frequency of 1.1--1.7~$\micron$ variables between L5 and T6 spectral types. 
In particular, the broader set of wavelengths and the higher precision of the B14 {\it HST} spectroscopic measurements contribute a large fraction of mid-L and mid-T 0.5\%--1.0\% variables compared to RLJ14. 
Unfortunately, the much shorter duration of their observations, only 40 min per target, precludes B14 from measuring the variability amplitudes.
Overall, the existence of significant 1.1--1.7~$\micron$ variations from brown dwarfs outside of the L/T transition, as observed by both RLJ14 and B14, agrees with our 3--5~$\micron$ findings.  

The third result, of an increasing maximum amplitude throughout the L and T spectral types is unique to the 3--5~$\micron$ region.  The RLJ14 and R14 $J$-band results reveal a significant peak in detected amplitudes in the L9--T3.5 range.  The B14 1.1--1.7~$\micron$ {\it HST} snap shot survey is not sensitive to amplitudes; only to the time derivatives of the variations.   Once again, the discrepancy with the RLJ14 and R14 results is suggestive of a difference in the atmospheric processes at the respective pressure levels and altitudes.  However, to the extent that the RLJ14 and R14 analyses can not rule out a gradual rise of $J$-band amplitudes into the L/T transition, the $J$-band and the 3--5~$\micron$ data are consistent with each other.   

Comparisons with variability studies at other wavelengths are a powerful tool to derive the cloud structure of brown dwarfs.  The two variables that our survey shares in common with RLJ14 and R14 have differing amplitudes.  The T6 dwarf 2MASS J22282889--4310262 has peak-to-peak amplitudes of $A[J]=1.6\%$, 
$A[3.6]=4.6\%$, and $A[4.5]=1.6\%$. The L4.5 dwarf 2MASS J11263991--5003550 has amplitudes of $A[J]=1.2\%$, $A[3.6]=0.21\%$, and $A[4.5]=0.29\%$.  A dependence of amplitude on wavelength is expected in the context of cloud models.  However, results featured in \citet{artigau_etal09}, \citet{metchev_etal13}, and \citet{gillon_etal13} clearly demonstrate that the amplitude of variations can change significantly on the time scale of several rotation periods.  The long time span between the RLJ14 $J$-band and our 3--5~$\micron$ observations precludes joint constraints on the cloud structure.  The two-band {\it Spitzer} observations alone can be used on individual objects as in \citet{heinze_etal13}.  As we discussed in Section~\ref{sec_amplitude_ratios}, the ensemble of results from our program does not paint a simple uniform picture of small hot spots on a prevailing colder surface vs.\ small cold spots on a prevailing hotter surface.    

The only conclusion that we draw at present from the maximum 3--5~$\micron$ amplitudes vs.\ spectral type trend is that the maximum flux contrast between warm and cold regions gradually increases toward cooler brown dwarfs.

\section{CONCLUSION
\label{sec_conclusion}}

The {\it Weather on Other Worlds} {\it Spitzer} Science Exploration program is the most sensitive large survey for photometric variability in brown dwarfs.  Our sample comprised 
44 L3--T8 dwarfs: 25 L and 19 T dwarfs.  These included seven systems---eight unique targets---with
low or moderately low surface gravities selected in order to seek a correlation with variability.   A known L3.5 radio-emitting dwarf was included as a test case for the effect of magnetic activity at the warm end of our sample.  A subsample of 23 L and 16 T dwarfs---all spatially unresolved---was used to infer the variability properties of L3--T8 dwarfs.  We summarize the new findings 
below.  

\begin{enumerate}
\item Photometric variability is common among L3--T8 dwarfs, with 
19 of the 39 ($49\%\pm15\%$, 95\% binomial confidence interval) single
objects detected as variables at the $>$95\% confidence level.

\item The rate of variability detection is approximately twice as high among L dwarfs than among T dwarfs: 
$61\%_{-20\%}^{+17\%}$ vs.\ $31\%_{-17\%}^{+25\%}$.
However, the difference is likely a consequence of our poorer photometric precision on the fainter T dwarfs.

\item After applying a moderate incompleteness correction for photometric sensitivity, we conclude that 
$80\%_{-27\%}^{+20\%}$ (95\% confidence interval)
of L3--L9.5 dwarfs are variable at $>$0.2\% amplitudes in either of the {\it Spitzer} [3.6] or [4.5] bands, and 
$36\%_{-17\%}^{+26\%}$ 
of T0--T8 dwarfs are variable at $>$0.4\%.  If only amplitudes $>$0.4\% are considered for the L dwarfs, the variability fraction among them is 
$53\%_{-18\%}^{+16\%}$,
comparable to that of the T dwarfs at the same amplitude.

\item A further consideration of the randomness of spin-axis orientations demonstrates that spots are likely ubiquitous on L dwarfs, even if they do not always produce detectable variability.  
Given the similar fraction of $>$0.4\% amplitude variables among L and T dwarfs, spots are likely present on most T dwarfs, too.

\item The observed variability amplitudes range from 0.2\%--4.6\%, with the smallest amplitudes found among L3--L5.5 dwarfs: a selection effect because of their greatest apparent brightness.  The maximum observed amplitude increases monotonically as a function of spectral type 
through the mid-T dwarfs.  Few L dwarfs have $>$1\% amplitudes and only T dwarfs have $>$2\% amplitudes.

\item We find tentative (92\% confidence) evidence that among L3--L5.5 dwarfs that are variable, the low-gravity ones 
may have higher 3--5~$\micron$ amplitudes than their field-aged counterparts.  However, we can not confirm that surface gravity also affects the {\it frequency} of variability among L3--L5.5 dwarfs.

\item A significant fraction of the variables have irregular light curves that require multiple Fourier terms to fit adequately.  At least three, and potentially as many as six of the 19 variables in the unresolved sample are irregular.  The known radio-emitting L3.5 dwarf that was deliberately added to our survey is also an irregular variable.  All irregular variables have spectral types of T0.5 or earlier.  
The high occurrence of irregular variables in the L dwarfs points to complex spot patterns, including multiple and/or rapidly changing spots.

\item We have doubled the number of L3--T8 dwarfs with reliably measured rotation periods.  L3--T8 dwarf variability time scales range from 1.4~h to $>$20~h, where the upper end of the range is limited by the 21~h duration of our uninterrupted observations.  

\item  Between six and nine of our total sample of 21 variables, i.e., approximately a third, show $>$10~h periods.  Likely not all of these periodicities reflect rotation, as at least one curve suggests spot evolution on a pole-on rotator.  This new population of L4--T2.5 slow (or pole-on) substellar rotators is ideal for establishing a grid of high-dispersion spectra of low-$v\sin i$ substellar standards.

\item We find a notable absence of correlation between the [3.6]- and [4.5]-band amplitudes. The [4.5]/[3.6] amplitude ratios scatter randomly around unity, and are not correlated with spectral type.  In the context of cloudy models, this indicates that either small warm spots on cooler atmospheres or small cold spots on warmer atmospheres are equally likely.

\end{enumerate}

We compare our results to the largest and most sensitive 1.1--1.7~$\micron$ L and T dwarf variability surveys \citep{radigan_etal14, buenzli_etal14}.  Our findings extend the results from these studies with variability detection fractions that are factors of 1.5--3 higher: because of the greater sensitivity to both amplitudes and longer periods in our uninterrupted 21~h observations.  Overall, all three surveys point to variability being common in L and T dwarfs.  We do not find increased variability fractions or uniquely high amplitudes at the L/T transition, as reported in \citet{radigan_etal14} and \citet{radigan14}.  Part of the discrepancy may be attributed to the difference in sensitivity limits and to sample stochastics.  However, the comparison is also suggestive of differences in the cloud structure appearance at 1.1--1.7~$\micron$ and 3--5~$\micron$ wavelengths.



The unprecedented sensitivity of our survey further allows us to conclude that, upon correcting for incompleteness, spotted brown dwarfs are not only common, but probably ubiquitous.  This reinforces the utility of variability-based studies to characterize the cloud and atmospheric properties of brown dwarfs.  In particular, our data set will be very suitable for Doppler imaging observations of brown dwarf clouds as performed by \citet{crossfield_etal14} on Luhman 16B.  Our sample provides both accurate periods needed for the phase folding of high-resolution spectroscopic observations, and slowly rotating brown dwarfs for least-squares deconvolution.   Furthermore, our finding that variability amplitudes may be enhanced at low surface gravities reveals a tantalizing potential for variability studies of directly imaged exoplanets \citep{kostov_apai13}, e.g., with the Gemini Planet Imager on the Gemini South telescope or with SPHERE on the Very Large Telescope.  High-contrast variability monitoring will also be an important tool for exoplanet characterization with the {\it James Webb Space Telescope}.  The present 3--5~$\micron$ findings will serve as a basis for the interpretation of these future observations.

\acknowledgments {\bf Acknowledgments.}

{\it Weather on Other Worlds} was a Cycle 8 Exploration Science program with the {\it Spitzer} Space Telescope.  We acknowledge team support through NASA grants to S.M., D.A., M.M., and P.P.  This research has benefitted from the M, L, T, and Y dwarf compendium housed at DwarfArchives.org. 

{\it Facility:} \facility{Spitzer (IRAC)}

\clearpage

\begin{deluxetable}{lccccccccrrc}
\rotate
\tablewidth{0pt}
\tabletypesize{\scriptsize}
\tablecaption{Results on L3--T8 Dwarf Variability \label{tab_results}}
\tablehead{ & & \colhead{$A[3.6]$} & \colhead{$A[4.5]$} & & \colhead{Fit $P$} & \colhead{Adopted $P$} & \colhead{$\sigma_P$} & & 
\colhead{$\lg({\rm FAP})$\tablenotemark{\ddag}} & 
\colhead{$\lg({\rm FAP})$\tablenotemark{\ddag}} & \\
	\colhead{Object} & \colhead{SpT} & \colhead{(\%)} & \colhead{(\%)} & \colhead{$A[4.5]/A[3.6]$\tablenotemark{\dag}} &
	\colhead{(h)} & \colhead{(h)} & \colhead{(h)} & \colhead{Periodicity} & \colhead{[3.6]} & \colhead{[4.5]} & \colhead{Note}}
\startdata
\input{table2.tex}
\enddata
\tablenotetext{\dag}{The $A[4.5]/A[3.6]$ ratios of the peak-to-peak amplitudes are from simultaneous, phased [3.6] and [4.5] fits, unless noted.  They may not correspond to the ratios of the independently fit [3.6]- and [4.5]-band amplitudes.  See Sections \ref{sec_variability_fitting}--\ref{sec_variable_classification} for explanation.}
\tablenotetext{\ddag}{The periodogram FAP thresholds below which we claim variability at the 95\% confidence level are $\lg({\rm FAP})=-3.4$ at [3.6] and $\lg({\rm FAP})=-1.5$ at [4.5].}

\tablecomments{1. For all short-period irregular and irregular/long variables, $A[4.5]/A[3.6]$ is simply the ratio of the unconstrained [4.5] and [3.6] amplitude fits.  
2. $A[3.6]$ upper limit obtained form Markov Chain Monte Carlo simulations. 
3. No significant variation is seen in the [4.5] data alone.  $A[4.5]$ is determined by fixing the [4.5] period to the [3.6] period, and fitting the joint data set.  
4. Because of a period longer than the observing sequence, the amplitude in one or both {\it Spitzer} channels can not be determined.}
\end{deluxetable}

\twocolumn

\input{ms_arxiv.bbl}
\onecolumn

\end{document}

%% file: table1.tex
\begin{deluxetable}{llccrcccrr}
\tablewidth{0pt}
\tabletypesize{\scriptsize}
\tablecaption{Sample and Observations \label{tab_sample}}
\tablehead{ & & & \colhead{$J$} & \colhead{$J-K_s$} & \colhead{[3.6]} & \colhead{[4.5]} & \colhead{$v\sin i$} & \colhead{$t_{3.6}$} & \colhead{$t_{4.5}$}  \\
	\colhead{Object} & \colhead{Spectral Type}  & \colhead{Ref} & \colhead{(mag)} & \colhead{(mag)} & \colhead{(mag)} & \colhead{(mag)} & \colhead{(km~s$^{-1}$)} & \colhead{(hr)} & \colhead{(hr)}}
	
\startdata
\object{	2MASSW J0036159+182110	}\tablenotemark{i}	&	L3.5		& 1 &	12.47	&	1.41	&	10.30	&	10.29	&	35.1	\tablenotemark{a}	&	8	&	6	\\
\object{	2MASS J00501994-3322402	}		&	T7		& 2 &	15.93	&	0.69	&	14.97	&	13.61	&	\nodata		&	14	&	7	\\
\object{	2MASSI J0103320+193536	}		&	L6		& 1 &	16.29	&	2.14	&	12.93	&	12.76	&	\nodata		&	14	&	7	\\
\object{	SDSSp J010752.33+004156.1}		&	L8		& 3 &	15.82	&	2.12	&	12.39	&	12.23	&	\nodata		&	14	&	7	\\
\object{	SDSS J015141.69+124429.6	}		&	T1		& 2 &	16.57	&	1.38	&	14.16	&	13.90	&	\nodata		&	14	&	7	\\
\object{	2MASSI J0328426+230205	}		&	L9.5		& 4 &	16.69	&	1.78	&	13.75	&	13.72	&	\nodata		&	14	&	7	\\
\object{	2MASS J04210718-6306022	}		&	L5$\beta$ & 5 &	15.57	&	2.12	&	12.28	&	12.17	&	\nodata		&	14	&	7	\\
\object{	2MASS J05160945-0445499	}		&	T5.5		& 2 &	15.98	&	0.50	&	14.75	&	13.63	&	\nodata		&	14	&	7	\\
\object{	2MASSW J0820299+450031	}		&	L5		& 1 &	16.28	&	2.06	&	13.26	&	13.27	&	\nodata		&	14	&	7	\\
\object{	2MASSI J0825196+211552	}\tablenotemark{{ii}}	&	L7.5		& 1 &	15.10	&	2.07	&	11.79	&	11.61	&	19.0	\tablenotemark{b}	&	12	&	9	\\
\object{	SDSS J085834.42+325627.7	}		&	T1		& 6 &	16.45	&	1.70	&	13.57	&	13.49	&	\nodata		&	14	&	7	\\
\object{	2MASS J09490860-1545485	}		&	T2 (T1+T2?) & 2,7 &	16.15	&	0.92	&	14.55	&	14.05	&	\nodata		&	14	&	7	\\
\object{	SDSS J104335.08+121314.1	}		&	L9		& 8 &	16.00	&	1.74	&	13.11	&	12.88	&	\nodata		&	14	&	7	\\
\object{	DENIS-P J1058.7-1548	}\tablenotemark{{iii}}	&	L3		& 9 &	14.16	&	1.62	&	11.79	&	11.79	&	37.5	\tablenotemark{c}	&	8	&	6	\\
\object{	2MASS J10595185+3042059}		&	T4		& 10 &	16.20	&	0.64	&	15.13	&	14.36	&	\nodata		&	14	&	7	\\
\object{	2MASS J11220826-3512363	}		&	T2		& 2 &	15.02	&	0.64	&	13.26	&	12.78	&	\nodata		&	14	&	7	\\
\object{	2MASS J11263991-5003550	}\tablenotemark{{ii}}	&	L4.5		& 11 &	14.00	&	1.17	&	11.86	&	11.92	&	\nodata		&	14	&	7	\\
\object{	SDSS J115013.17+052012.3	}		&	L5.5		& 12 &	16.25	&	1.24	&	13.62	&	13.65	&	\nodata		&	14	&	7	\\
\object{	2MASS J12095613-1004008	}		&	T3 (T2+T7.5) & 2,13 &	15.91	&	0.84	&	14.09	&	13.43	&	\nodata		&	14	&	7	\\
\object{	SDSSp J125453.90-012247.4}		&	T2		& 2 &	14.89	&	1.05	&	12.81	&	12.42	&	27.0	\tablenotemark{d}	&	12	&	7	\\
\object{	Ross 458C	}\tablenotemark{{iv}}	&	T8		& 14 &	16.67	&	-0.21	&	15.43	&	13.81	&	\nodata		&	14	&	7	\\
\object{	2MASS J13243559+6358284}		&	T2.5 (L8+T3.5?) & 8,7 &	15.60	&	1.54	&	12.63	&	12.30	&	\nodata		&	14	&	7	\\
\object{	ULAS J141623.94+134836.3}\tablenotemark{v}	&	T7.5	& 15 &	17.26	&	-1.67	&	14.80	&	12.77	&	\nodata		&	14	&	7	\\
\object{	SDSS J141624.08+134826.7	}\tablenotemark{v}	&	L6	& 16 &	13.15	&	1.14	&	10.95	&	10.98	&	\nodata		&	14	&	7	\\
\object{	2MASSW J1507476-162738	}		&	L5		& 1 &	12.83	&	1.52	&	10.39	&	10.42	&	21.3	\tablenotemark{a}	&	12	&	8	\\
\object{	SDSS J151114.66+060742.9	}		&	T2 (L5.5+T5)	& 17,7,18 &	16.02	&	1.47	&	13.31	&	13.16	&	\nodata		&	14	&	7	\\
\object{	SDSS J151643.01+305344.4	}		&	T0.5		& 6 &	16.85	&	1.77	&	13.68	&	13.45	&	\nodata		&	14	&	7	\\
\object{	SDSS J152039.82+354619.8	}		&	T0		& 6 &	15.54	&	1.54	&	13.02	&	12.92	&	\nodata		&	14	&	7	\\
\object{	SDSS J154508.93+355527.3	}		&	L7.5		& 6 &	16.83	&	1.41	&	14.20	&	14.13	&	\nodata		&	14	&	7	\\
\object{	2MASS J16154255+4953211}		&	L4$\beta$ & 19 &	16.79	&	2.48	&	12.92	&	12.62	&	\nodata		&	14	&	7	\\
\object{	2MASSW J1632291+190441	}		&	L8		& 9 &	15.87	&	1.86	&	12.75	&	12.63	&	30.0	\tablenotemark{e}	&	8	&	6	\\
\object{	2MASSI J1721039+334415	}		&	L3		& 20 &	13.63	&	1.14	&	11.65	&	11.62	&	\nodata		&	14	&	7	\\
\object{	2MASSI J1726000+153819	}		&	L3$\beta$ & 5 &	15.67	&	2.01	&	12.81	&	12.67	&	\nodata		&	14	&	7	\\
\object{	2MASS J17534518-6559559	}		&	L4::		& 21 &	14.10	&	1.67	&	11.53	&	11.50	&	\nodata		&	14	&	7	\\
\object{	2MASS J18212815+1414010}		&	L4.5 	& 22 &	13.43	&	1.78	&	10.56	&	10.50	&	28.9	\tablenotemark{a}	&	10	&	6	\\
\object{	SDSS J204317.69-155103.4	}		&	L9		& 6 &	16.63	&	1.22	&	14.27	&	14.13	&	\nodata		&	14	&	7	\\
\object{	SDSS J205235.31-160929.8	}		&	T1 (T1+T2.5) & 6,23 &	16.33	&	1.21	&	13.73	&	14.10	&	\nodata		&	14	&	7	\\
\object{	HN PegB	}\tablenotemark{{vi}}	&	T2.5		& 24 &	16.06	&	1.02	&	13.72	&	13.32	&	\nodata		&	14	&	7	\\
\object{	2MASS J21481628+4003593}		&	L6		& 8 &	14.15	&	2.38	&	10.47	&	10.24	&	\nodata		&	14	&	7	\\
\object{	2MASSW J2208136+292121	}		&	L3$\gamma$	& 5 &	15.80	&	1.65	&	13.08	&	12.89	&	\nodata		&	14	&	7	\\
\object{	2MASSW J2224438-015852	}		&	L4.5		& 1 &	14.07	&	2.05	&	11.11	&	11.15	&	25.5	\tablenotemark{a}	&	12	&	8	\\
\object{	2MASS J22282889-4310262	}\tablenotemark{{ii}}	&	T6		& 2 &	15.66	&	0.37	&	14.48	&	13.33	&	\nodata		&	14	&	7	\\
\object{	SDSSp J224953.45+004404.2}		&	L3$\beta$ (L3+L5) & 3,25 &	16.59	&	2.23	&	13.34	&	13.11	&	\nodata		&	14	&	7	\\
\object{	2MASSI J2254188+312349	}		&	T4		& 2 &	15.26	&	0.36	&	13.98	&	13.32	&	\nodata		&	14	&	7	\\

\enddata
\tablecomments{
{\bf Notes on individual objects:} i.~Known radio emitter \citep{berger02}; not a part of the statistical sample of L3--T8 dwarfs. ii.~Previously known variable from \citet{radigan_etal14}, \citet{clarke_etal08}, or \citet{buenzli_etal14}. iii.~H$\alpha$ emission detected by \citet{tinney_etal97}, \citet{kirkpatrick_etal99}, \citet{martin_etal99}, or \citet{gelino_etal02}. iv.~Companion to Ross 458AB \citep{scholz10b}. v.~A 9$\farcs$5 L6+T7.5 binary \citep{scholz10a}. vi.~A 43$\arcsec$ companion to HN Peg; moderately low gravity \citep{luhman_etal07b}. \\
{\bf References for spectral types:} 1.~\citet{kirkpatrick_etal00}, 2.~\citet{burgasser_etal06b}, 3.~\citet{hawley_etal02}, 4.~\citet{knapp_etal04}, 5.~\citet{cruz_etal09}, 6.~\citet{chiu_etal06}, 7.~\citet{burgasser_etal10b}, 8.~\citet{kirkpatrick_etal10}, 9.~\citet{kirkpatrick_etal99}, 10.~\citet{sheppard_cushing09}, 11.~\citet{burgasser_etal08}, 12.~\citet{zhang_etal09}, 13.~\citet{liu_etal10}, 14.~\citet{burgasser_etal10}, 15.~\citet{burgasser_etal10c}, 16.~\citet{bowler_etal10b}, 17.~\citet{albert_etal11}, 18.~\citet{bardalez_gagliuffi_etal15}, 19.~\citet{cruz_etal07}, 20.~\citet{cruz_etal03}, 21.~\citet{reid_etal08b}, 22.~\citet{looper_etal08}, 23.~\citet{stumpf_etal11}, 24.~\citet{luhman_etal07b}, 25.~\citet{allers_etal10}. \\
{\bf References for $v\sin i$ measurements:} a.~\citet{blake_etal10}, b.~\citet{reiners_basri08}, c.~\citet{basri_etal00}, d.~\citet{zapatero-osorio_etal06}, e.~\citet{mohanty_basri03}.}

\end{deluxetable}

%% file: table2.tex
\object{	2MASSW J0036159$+$182110	}	&	L3.5	&		0.47	$\pm$	0.05	&		0.19	$\pm$	0.04	&		0.40	$\pm$	0.09	&	2.7	&	2.7	&	0.3	&	irreg	&	$-$4.9	&	$-$3.7	&	1	\\
\object{	2MASS J00501994$-$3322402	}	&	T7	&	$<$	0.59			&		1.07	$\pm$	0.11	&	$>$	1.8			&	1.55	&	1.55	&	0.02	&	reg	&	$-$1.4	&	$-$9.1	&	2	\\
\object{	2MASSI J0103320$+$193536	}	&	L6	&		0.56	$\pm$	0.03	&		0.87	$\pm$	0.09	&		1.32	$\pm$	0.15	&	2.7	&	2.7	&	0.1	&	reg	&	$-$20.1	&	$-$10.0	&		\\
\object{	SDSSp J010752.33$+$004156.1	}	&	L8	&		1.27	$\pm$	0.13	&		1.0	$\pm$	0.2	&		0.8	$\pm$	0.2	&	13	&	5	&	unc.	&	irreg/long	&	$-$25.5	&	$-$5.6	&	1	\\
\object{	SDSS J015141.69$+$124429.6	}	&	T1	&	$<$	0.45			&	$<$	0.59			&		\nodata			&	\nodata	&	\nodata	&	\nodata	&		&	$-$0.3	&	0.0	&		\\
\object{	2MASSI J0328426$+$230205	}	&	L9.5	&	$<$	0.45			&	$<$	0.70			&		\nodata			&	\nodata	&	\nodata	&	\nodata	&		&	$-$1.5	&	$-$0.4	&		\\
\object{	2MASS J04210718$-$6306022	}	&	L5$\beta$	&	$<$	0.21			&	$<$	0.34			&		\nodata			&	\nodata	&	\nodata	&	\nodata	&		&	$-$1.6	&	$-$0.5	&		\\
\object{	2MASS J05160945$-$0445499	}	&	T5.5	&	$<$	0.83			&	$<$	0.81			&		\nodata			&	\nodata	&	\nodata	&	\nodata	&		&	$-$1.0	&	$-$0.8	&		\\
\object{	2MASSW J0820299$+$450031	}	&	L5	&	$<$	0.40			&	$<$	0.48			&		\nodata			&	\nodata	&	\nodata	&	\nodata	&		&	$-$3.0	&	$-$0.2	&		\\
\object{	2MASSI J0825196$+$211552	}	&	L7.5	&		0.81	$\pm$	0.08	&		1.4	$\pm$	0.3	&		1.7	$\pm$	0.4	&	11.7	&	7.6	&	unc.	&	irreg/long	&	$-$31.3	&	$-$26.9	&	1	\\
\object{	SDSS J085834.42$+$325627.7	}	&	T1	&	$<$	0.27			&	$<$	0.64			&		\nodata			&	\nodata	&	\nodata	&	\nodata	&		&	$-$0.2	&	$-$0.5	&		\\
\object{	2MASS J09490860$-$1545485	}	&	T2 (T1+T2?)	&	$<$	0.54			&	$<$	0.83			&		\nodata			&	\nodata	&	\nodata	&	\nodata	&		&	$-$0.3	&	$-$0.5	&		\\
\object{	SDSS J104335.08$+$121314.1	}	&	L9	&		1.54	$\pm$	0.15	&		1.2	$\pm$	0.2	&		0.8	$\pm$	0.2	&	3.8	&	3.8	&	0.2	&	irreg	&	$-$14.4	&	$-$7.7	&	1	\\
\object{	DENIS-P J1058.7$-$1548	}	&	L3	&		0.39	$\pm$	0.04	&	$<$	0.30			&		0.20	$\pm$	0.10	&	4.1	&	4.1	&	0.2	&	reg	&	$-$9.9	&	$-$0.3	&	3	\\
\object{	2MASS J10595185$+$3042059	}	&	T4	&	$<$	0.83			&	$<$	0.89			&		\nodata			&	\nodata	&	\nodata	&	\nodata	&		&	$-$0.2	&	0.0	&		\\
\object{	2MASS J11220826$-$3512363	}	&	T2	&	$<$	0.24			&	$<$	0.31			&		\nodata			&	\nodata	&	\nodata	&	\nodata	&		&	$-$0.3	&	$-$0.1	&		\\
\object{	2MASS J11263991$-$5003550	}	&	L4.5	&		0.21	$\pm$	0.04	&		0.29	$\pm$	0.15	&		1.2	$\pm$	0.7	&	3.2	&	3.2	&	0.3	&	reg	&	$-$6.5	&	$-$2.9	&		\\
\object{	SDSS J115013.17$+$052012.3	}	&	L5.5	&	$<$	0.38			&	$<$	0.65			&		\nodata			&	\nodata	&	\nodata	&	\nodata	&		&	$-$1.3	&	$-$0.4	&		\\
\object{	2MASS J12095613$-$1004008	}	&	T3 (T2+T7.5)	&	$<$	0.40			&	$<$	0.56			&		\nodata			&	\nodata	&	\nodata	&	\nodata	&		&	$-$0.3	&	$-$0.2	&		\\
\object{	SDSSp J125453.90$-$012247.4	}	&	T2	&	$<$	0.15			&	$<$	0.30			&		\nodata			&	\nodata	&	\nodata	&	\nodata	&		&	$-$0.1	&	$-$0.1	&		\\
\object{	Ross 458C	}	&	T8	&	$<$	1.37			&	$<$	0.72			&		\nodata			&	\nodata	&	\nodata	&	\nodata	&		&	$-$1.2	&	$-$0.3	&		\\
\object{	2MASS J13243559$+$6358284	}	&	T2.5 (L8+T3.5?)	&		3.05	$\pm$	0.15	&		3.0	$\pm$	0.3	&		1.16	$\pm$	0.13	&	13	&	13	&	1	&	long	&	$<-40$	&	$-$17.4	&		\\
\object{	ULAS J141623.94$+$134836.3	}	&	T7.5	&	$<$	0.91			&	$<$	0.59			&		\nodata			&	\nodata	&	\nodata	&	\nodata	&		&	$-$1.9	&	$-$1.3	&		\\
\object{	SDSS J141624.08$+$134826.7	}	&	L6	&	$<$	0.15			&	$<$	0.22			&		\nodata			&	\nodata	&	\nodata	&	\nodata	&		&	$-$2.3	&	$-$1.3	&		\\
\object{	2MASSW J1507476$-$162738	}	&	L5	&		0.53	$\pm$	0.11	&		0.45	$\pm$	0.09	&		0.8	$\pm$	0.2	&	2.5	&	2.5	&	0.1	&	irreg	&	$-$7.5	&	$-$15.6	&	1	\\
\object{	SDSS J151114.66$+$060742.9	}	&	T2 (L5.5+T5)	&		0.67	$\pm$	0.07	&	$<$	0.49			&		0.5	$\pm$	0.2	&	11	&	11	&	2	&	long	&	$-$9.3	&	0.0	&	3	\\
\object{	SDSS J151643.01$+$305344.4	}	&	T0.5	&		2.4	$\pm$	0.2	&		3.1	$\pm$	1.6	&		1.3	$\pm$	0.7	&	14	&	6.7	&	unc.	&	irreg/long	&	$-$27.4	&	$-$20.5	&	1	\\
\object{	SDSS J152039.82$+$354619.8	}	&	T0	&	$<$	0.30			&	$<$	0.45			&		\nodata			&	\nodata	&	\nodata	&	\nodata	&		&	$-$1.3	&	$-$0.4	&		\\
\object{	SDSS J154508.93$+$355527.3	}	&	L7.5	&	$<$	0.59			&	$<$	1.15			&		\nodata			&	\nodata	&	\nodata	&	\nodata	&		&	$-$1.3	&	$-$1.1	&		\\
\object{	2MASS J16154255$+$4953211	}	&	L4$\beta$	&		0.9	$\pm$	0.2	&	$<$	0.39			&		1.4	$\pm$	0.3	&	24	&	24	&	unc.	&	long	&	$-$4.1	&	$-$0.4	&	3,4	\\
\object{	2MASSW J1632291$+$190441	}	&	L8	&		0.42	$\pm$	0.08	&		0.5	$\pm$	0.3	&		1.1	$\pm$	0.6	&	3.9	&	3.9	&	0.2	&	reg	&	$-$4.7	&	$-$3.0	&	3	\\
\object{	2MASSI J1721039$+$334415	}	&	L3	&		0.33	$\pm$	0.07	&	$<$	0.29			&		$-$0.4	$\pm$	0.4	&	2.6	&	2.6	&	0.1	&	reg	&	$-$7.6	&	$-$1.1	&	3	\\
\object{	2MASSI J1726000$+$153819	}	&	L3$\beta$	&	$<$	0.29			&	$<$	0.49			&		\nodata			&	\nodata	&	\nodata	&	\nodata	&		&	$-$2.0	&	$-$1.4	&		\\
\object{	2MASS J17534518$-$6559559	}	&	L4	&	$>$	0.25			&		\nodata			&		\nodata			&	$>$50	&	$>$50	&	unc.	&	long	&	$-$6.5	&	$-$0.6	&	4	\\
\object{	2MASS J18212815$+$1414010	}	&	L4.5	&		0.54	$\pm$	0.05	&		0.71	$\pm$	0.14	&		1.3	$\pm$	0.3	&	4.2	&	4.2	&	0.1	&	irreg	&	$-$13.3	&	$-$13.5	&	1	\\
\object{	SDSS J204317.69$-$155103.4	}	&	L9	&	$<$	0.71			&	$<$	0.74			&		\nodata			&	\nodata	&	\nodata	&	\nodata	&		&	$-$2.5	&	$-$0.2	&		\\
\object{	SDSS J205235.31$-$160929.8	}	&	T1 (T1+T2.5)	&	$<$	0.36			&	$<$	0.71			&		\nodata			&	\nodata	&	\nodata	&	\nodata	&		&	$-$0.4	&	$-$0.6	&		\\
\object{	HN PegB	}	&	T2.5	&		0.77	$\pm$	0.15	&		1.1	$\pm$	0.5	&		1.8	$\pm$	1.0	&	18	&	18	&	4	&	long	&	$-$3.4	&	$-$2.1	&		\\
\object{	2MASS J21481628$+$4003593	}	&	L6	&		1.33	$\pm$	0.07	&		1.03	$\pm$	0.10	&		0.92	$\pm$	0.10	&	19	&	19	&	4	&	long	&	$-$14.9	&	$-$18.6	&		\\
\object{	2MASSW J2208136$+$292121	}	&	L3$\gamma$	&		0.69	$\pm$	0.07	&		0.54	$\pm$	0.11	&		0.8	$\pm$	0.2	&	3.5	&	3.5	&	0.2	&	reg	&	$-$19.3	&	$-$3.7	&		\\
\object{	2MASSW J2224438$-$015852	}	&	L4.5	&	$<$	0.10			&	$<$	0.15			&		\nodata			&	\nodata	&	\nodata	&	\nodata	&		&	$-$0.2	&	$-$0.5	&		\\
\object{	2MASS J22282889$-$4310262	}	&	T6	&		4.6	$\pm$	0.2	&		1.51	$\pm$	0.15	&		0.32	$\pm$	0.04	&	1.41	&	1.41	&	0.01	&	reg	&	$<-40$	&	$-$18.0	&		\\
\object{	SDSSp J224953.45$+$004404.2	}	&	L3$\beta$ (L3+L5)	&	$<$	0.25			&	$<$	0.45			&		\nodata			&	\nodata	&	\nodata	&	\nodata	&		&	$-$0.4	&	$-$0.2	&		\\
\object{	2MASSI J2254188$+$312349	}	&	T4	&	$<$	0.47			&	$<$	0.39			&		\nodata			&	\nodata	&	\nodata	&	\nodata	&		&	$-$1.0	&	0.0	&		\\